\documentclass{article}
\usepackage{amssymb}

\input{tcilatex}

\begin{document}

\bigskip

\bigskip

Article

{\Large Maxwell-Dirac isomorphism revisited:}

{\Large from foundations of quantum mechanics }

{\Large to geometrodynamics and cosmology}

{\Large \bigskip }

\bigskip\ \ Arkady L.Kholodenko

\bigskip\ \ \ \ \ \ \ \ \ \ \ \ \ \ \ \ \ \ \ \ \ \ \ \ \ 375 H.L.Hunter
Laboratories, Clemson University\textit{,}

\ \ \ \ \ \ \ \ \ \ \ \ \ \ \ \ \ \ \ \ \ \ \ \ \ Clemson, SC 29634,USA;
string@clemson.edu\textit{\bigskip }

\ \ \ \ \ \ \ \ \ \ \ \ \ \ \ \ \ \ \ \ \ \ \ \ \ \ \ \ \ \ \ \ 

\ \ \ \ \ \ \ \ \ \ \ \ \ \ \ \ \ \ \ \ \ \ \ 

\ \ \ \ \ \ \ \ \ \ \ \ \ \ \ \ \ \ \ \ \ \textbf{Abstract}\bigskip :
Although electrons (fermions)and photons

\ \ \ \ \ \ \ \ \ \ \ \ \ \ \ \ \ \ \ \ \ (bosons) produce the same
interference patterns in the

\ \ \ \ \ \ \ \ \ \ \ \ \ \ \ \ \ \ \ \ \ two-slit experiments, known in
optics for photons since

\ \ \ \ \ \ \ \ \ \ \ \ \ \ \ \ \ \ \ \ \ 17th century, the description of
these patterns for

\ \ \ \ \ \ \ \ \ \ \ \ \ \ \ \ \ \ \ \ \ electrons and photons thus far
were markedly different.

\ \ \ \ \ \ \ \ \ \ \ \ \ \ \ \ \ \ \ \ \ Photons are spin one, relativistic
and massless particles,

\ \ \ \ \ \ \ \ \ \ \ \ \ \ \ \ \ \ \ \ \ while electrons are spin 1/2
massive particles producing

\ \ \ \ \ \ \ \ \ \ \ \ \ \ \ \ \ \ \ \ \ the same\ interference patterns
irrespective of their speed.

\ \ \ \ \ \ \ \ \ \ \ \ \ \ \ \ \ \ \ \ \ Experiments with other massive
particles demonstrate

\ \ \ \ \ \ \ \ \ \ \ \ \ \ \ \ \ \ \ \ \ the same kind of interference
patterns. Despite of these

\ \ \ \ \ \ \ \ \ \ \ \ \ \ \ \ \ \ \ \ \ differences, already in early
30ies of 20ieth century the

\ \ \ \ \ \ \ \ \ \ \ \ \ \ \ \ \ \ \ \ \ isomorphism between the
source-free Maxwell and Dirac

\ \ \ \ \ \ \ \ \ \ \ \ \ \ \ \ \ \ \ \ \ equations was established. In this
work it permitted us

\ \ \ \ \ \ \ \ \ \ \ \ \ \ \ \ \ \ \ \ \ to replace Born probabilistic
interpretation of quantum

\ \ \ \ \ \ \ \ \ \ \ \ \ \ \ \ \ \ \ \ \ mechanics with optical. In 1925
Rainich combined

\ \ \ \ \ \ \ \ \ \ \ \ \ \ \ \ \ \ \ \ \ source-free Maxwell equations with
Einstein's equations

\ \ \ \ \ \ \ \ \ \ \ \ \ \ \ \ \ \ \ \ \ for gravity. His results were
rediscovered in the late 50ies

\ \ \ \ \ \ \ \ \ \ \ \ \ \ \ \ \ \ \ \ \ by Misner and Wheeler who
introduced word

\ \ \ \ \ \ \ \ \ \ \ \ \ \ \ \ \ \ \ \ \ "geometrodynamics" for description
of the unified field

\ \ \ \ \ \ \ \ \ \ \ \ \ \ \ \ \ \ \ \ \ theory of gravity and
electromagnetism. Absence of

\ \ \ \ \ \ \ \ \ \ \ \ \ \ \ \ \ \ \ \ \ sources \ remained a problem in
this unified theory until

\ \ \ \ \ \ \ \ \ \ \ \ \ \ \ \ \ \ \ \ \ Ranada's work of late 80ies. But
his results required the

\ \ \ \ \ \ \ \ \ \ \ \ \ \ \ \ \ \ \ \ \ existence of null electromagnetic
fields. These were absent

\ \ \ \ \ \ \ \ \ \ \ \ \ \ \ \ \ \ \ \ \ in Rainich's- Misner's-Wheeler's
geometrodynamics. They

\ \ \ \ \ \ \ \ \ \ \ \ \ \ \ \ \ \ \ \ \ were added to it in 60ies by
Geroch. Ranada's solutions of

\ \ \ \ \ \ \ \ \ \ \ \ \ \ \ \ \ \ \ \ \ source-free Maxwell's equations
came out as knots and

\ \ \ \ \ \ \ \ \ \ \ \ \ \ \ \ \ \ \ \ \ links. In this work we establish
that due to their topology

\ \ \ \ \ \ \ \ \ \ \ \ \ \ \ \ \ \ \ \ \ these knots/links \ acquire masses
and charges. They live

\ \ \ \ \ \ \ \ \ \ \ \ \ \ \ \ \ \ \ \ \ on the Dupin cyclides-the
invariants of Lie sphere

\ \ \ \ \ \ \ \ \ \ \ \ \ \ \ \ \ \ \ \ \ geometry. Symmetries of Minkowski
space-time also

\ \ \ \ \ \ \ \ \ \ \ \ \ \ \ \ \ \ \ \ \ belong to this geometry. Using
these symmetries

\ \ \ \ \ \ \ \ \ \ \ \ \ \ \ \ \ \ \ \ \ Varlamov recently demonstrated
group-theoretically,

\ \ \ \ \ \ \ \ \ \ \ \ \ \ \ \ \ \ \ that experimentally known mass
spectrum for all mesons

\ \ \ \ \ \ \ \ \ \ \ \ \ \ \ \ \ \ \ and baryons is obtainable with one
formula containing

\ \ \ \ \ \ \ \ \ \ \ \ \ \ \ \ \ \ \ electron mass as an input. In this
work, using some facts

\ \ \ \ \ \ \ \ \ \ \ \ \ \ \ \ \ \ \ from polymer physics and differential
geometry new

\ \ \ \ \ \ \ \ \ \ \ \ \ \ \ \ \ \ \ proof of knotty nature of electron is
established. The

\ \ \ \ \ \ \ \ \ \ \ \ \ \ \ \ \ \ \ obtained result perfectly blends with
the description

\ \ \ \ \ \ \ \ \ \ \ \ \ \ \ \ \ \ \ of rotating and charged black hole.

\ \ \ \ \ \ \ \ \ \ \ \ \ \ \ \ \ \ \ 

\bigskip

\ \ \ \ \ \ \ \ \ \ \ \ \ \ \ \ \ \ \ \ 


\ \ \ \ \ \ \ \ \ \ \ \ \ \ \ \ \ \ \ quantum mechanics; differential
geometry; knots and links;

\ \ \ \ \ \ \ \ \ \ \ \ \ \ \ \ \ \ \ Chladni patterns; Dupin cyclides;
polymer physics;

\ \ \ \ \ \ \ \ \ \ \ \ \ \ \ \ \ \ \ neutrinos; axions; geometrodynamics

\ \ \ \ \ \ \ \ \ \ \ \ \ \ \ \ \
------------------------------------------------------------------------

\bigskip \bigskip

\ \ \ \ \ \ \ \ \ \ \ \ \ \ \ \ \ \ \ \ \ \textbf{1.Introduction\bigskip }

\bigskip\ 

If electrons and photons and other massive particles produce the same
interference patterns in the two-slit experiments [1-5], why then the optics
formalism cannot be applied unchanged to electrons and heavier particles?
What makes use of Born's probabilistic interpretation of the wave function
in quantum mechanics superior to the intensity interpretation of the
Maxwellian wave functions in optics? For description of the two-slit
experiments in which the results are visibly the same for light and massive
particles? \ Study of this issue was initiated by David Bohm in his
monograph on quantum mechanics [6], pages 97-98. From these pages it
follows, that the differences do exist but in just few places. The updated
theoretical comparison more recently was made by Sanz and Miret-Artes, in
chapters 4 and 7 of their book [7]. From the results of these chapters it
follows that all objections made by Bohm in the remaining few places of his
analysis can be removed. Fortunately, the results of [7] along with those
originating from references on which these results are based (and even more
recent ones), can still be improved. The first purpose of this paper to do
just this. We can develop the unifying quantum mechanical-optical formalism
for photons and massive particles with or without spin. Since nowadays the
sophisticated quantum mechanical experiments are typically done optically
[8,9], the results of our paper may provide additional guidelines for
interpretation of these optical experiments and vice versa. In the light of
correspondence just stated, it makes sense to claim that our understanding
of subtleties of quantum mechanics is contingent upon our understanding of
subtleties of the developed optical formalism adapted for quantum mechanical
needs. In this paper we discuss several features of this formalism which
were not yet discussed (to our knowledge) in the context of quantum
mechanics or quantum field theories.

\ Even though the two-slit interference experiments visibly produce the same
results in formalisms of both optics and quantum mechanics, it is not
immediately possible to adopt word-for-word the conventional optical
formalism to quantum mechanics. This is so because Maxwell's equations
contain vector quantities, like \textbf{E} and \textbf{H}. Besides, the
polarization in optics, whose quantum analog is spin [10], causes
difficulties since at the nonrelativistic level at which Schr\"{o}dinger's
equation is being used there is no mention of spin. At the same time, the
two-slit interference fringe patterns for monochromatic light depend
strongly upon the light polarization. For spinless particles in
nonrelativistic quantum mechanics used in the two-slit experiments this
effect of polarization is absent while in optics, depending on polarization,
there are four distinct cases of study. They were discovered by Fresnel and
Arago at the beginning of 19th century [11,12]. \ For the record, these are:
1. Two rays of light polarized in the same plane. They interfere like rays
of ordinary (unpolarized) light. 2. Two rays polarized at the right angles
to each other. They do not interfere. 3. Two rays originally polarized at
the right angles and then brought into the same plane of polarization. They
do not produce interference pattern. 4 Two rays originally polarized at the
right angles, if derived from the same linearly polarized wave and
subsequently brought into the same plane, can interfere.

These and other facts, e.g. masses and spins, relativistic effects, etc.
complicate comparison between the two-slit interference\ experiments in
optics and in quantum mechanics. Fortunately, there are ways out of these
difficulties. In this work we discuss these ways. Surprisingly, the solution
of this difficulties has resulted in solution to many other things presented
in this paper. Thus far they have been considered as unrelated. \ Already
the solution of the first task required us to take several unexpected steps.
The first step is described in Section 2. Its purpose is to introduce some
known results\ to be used in the rest of this paper. Next, in Section 3 we
discuss the condition (absence of sources/charges) under which the vector
electromagnetic field can be replaced by the complex scalar field [13,14].
This feature was known but not used for many years. The existence of
source-free Maxwellian fields is of central importance for all results of
this paper. In Section 4 we use this complex scalar field to reobtain known
results for nonrelativistic Schr\"{o}dinger's quantum mechanics. Use of
complex scalar fields (seemingly, of electromagnetic origin) in quantum
mechanics allows us to extend known quantum mechanical formalism in Section
5. In it we introduce the Chladni patterns and discuss the existence of
knotted and linked configurations for the wave functions. The existence of
these configurations in Schrodinger's version of quantum mechanics makes
Schrodinger's and Heisenberg's interpretations of quantum mechanics not
equivalent. Section 6 is devoted to the considerable extension of results of
Ranada [15,16]. They are linked to the results of previous sections because
already in his 1st paper [15] Ranada used the complex scalar fields for
description of the electromagnetic field. \ He did it with the purpose of
designing knotted and linked structures out of electromagnetic fields. In
his second paper [16] he demonstrated that these knotty structures are
acting like particles having masses, spins and charges. Additional helpful
technical details of this particle-like properties of knotty \
electromagnetic fields are presented in [17]. In the light of these results
and in the spirit of Einstein's quantization program sketched in the same
section, the Maxwell-Dirac isomorphism is introduced and treated. Then,
using the notion of electromagnetic duality the axions are introduced. These
are particles believed to be linked with the \textsl{dark matter.} \ The
evidence is provided in favor of the statement that the axions are
responsible for the Chladni patterns in the sky (that is in space -time of
the Universe). Section 7 provides \bigskip additional support (culminating
in Section 9) of Einstein's quantization program. It discusses the
Maxwell-Dirac isomorphism from various angles- all based on known rigorous
mathematical results which have not been collected yet in one place thus
far. In particular, in this section we discuss peculiar results coming from
treating the mass in the Dirac equation. Although mathematically to make the
mass go to zero is perfectly permissible, so that it could be made as small
as possible, physically such limiting process loses its meaning.\ This fact
is not logical but experimental. After a certain threshold (theoretically
not known yet), the Dirac equation must be replaced by the Majorana
equation. It is commonly believed that neutrinos are described by the
Majorana equations with vanishingly small but nonzero masses. This fact
causes some serious difficulties in uses of the Maxwell-Dirac
correspondence. To resolve these difficulties, the neutrino theory of light
was proposed (e.g., read Wikipedia). This theory is not working, though,
because neutrinos have masses(very small but nonzero) while the light is
massless. \ Story with neutrino also teaches us about the notion of
chirality. This phenomenon exists not only for elementary particles (the
Standard Model of particle physics is built on chiralities), though. It also
exists in the macroscopic world. How this property is showing itself in the
macroscopic nature and how to account for chirality in the Maxwell-Dirac
(Majorana) formalism is also discussed in Section 7. In section 8 we develop
our own mathematical proof of the Maxwell-Dirac isomorphism having in mind
physical applications discussed in Section 9. Surprisingly, the arguments
presented in this section are based on some results from polymer physics
which were developed by the author in the past. Also, in this section we use
some not so widely known results by Dirac which belong to the domain of
projective relativity. This field of study emerged before
Rainich-Misner-Wheeler\ geometrodynamics and is also aimed at designing of
the unified field theory of gravity and electromagnetism. But it does have a
significant merit on its own, especially because it is technically connected
with the Lie sphere geometry (subsection 8.1). The Dupin cyclides-invariants
of this geometry, are also mentioned in Section 8. Although Section 9 is a
discussion section, it is of independent value of major importance. We
advise our readers to read it immediately after this section since, in a
way, it could be looked upon as an extension of this section. At the same
time, it should be read subsequently, after Section 8. This is especially
true in the light of the fact that Varlamov's mass formula, presented in
Section 9, is obtained with help of the group-theoretic methods applied to
the Lorentz group. For its use it requires at least the mass of the electron
as an input. The model of electron obtained in Section 8 fits Varlamov's
formula \ since both the ground and the excited states which this formula
describes have knot-theoretic interpretation. All knots are divided into two
groups with respect to knot-theoretic operation of the pass equivalence. The
Arf invariant of a knot/link (e.g., read Wikipedia) is taking two values 0
and 1, with respect to the pass -equivalence operation. Those knots which
are pass -equivalent to the unknot have the Arf invariant 0, while those
which are pass-equivalent to the trefoil knot have the Arf invariant equal
to 1. As demonstrated in Section 8, the description of electron utilizes the
trefoil knot. It is the simplest torus knot and the torus is the simplest
Dupin cyclide. The mass excitation formula by Varlamov generates unstable
particles, all having Arf invariant 1, and all eventually decaying back to
the trefoil knot. The pass-equivalence operation can be interpreted
physically in terms of the equidistant spectrum of harmonic oscillator.
Indeed, Varlamov's mass formula involves a harmonic oscillator-like spectrum
as an input.

Although the Kaluza-Klein, projective relativity and the geometrodynamics
serve the same purpose-to unify gravity with electromagnetism, only
Rainich's-Misner's-Wheeler's geometrodynamics could be considered as the
theory in which such unification was finally achieved. The major difficulty
of the initial Misner's-Wheeler's version of geometrodynamics associated
with the mathematically rigorous description of the sources of
electromagnetic fields was solved by Ranada in late 80ies when he discovered
particle-like behaving knotty solutions of the source-free electromagnetic
fields. Neither Ranada, nor the rest of researchers -developers of his
results, the author included, were aware of a connection between Ranada's
results and geometrodynamics. It is made only now, in this paper. Much later
on it was realized that for existence and stability of Ranada's \
knots/links \ it is required that these should come out as solutions
involving the null fields, that is the electromagnetic fields for which $%
\mathbf{E}\cdot \mathbf{H=}0$ and $\left\vert \mathbf{E}\right\vert
^{2}=\left\vert \mathbf{H}\right\vert ^{2}.$ The existence of null fields in
geometrodynamics was a problem in the Rainich,Misner-Wheeler version of
geometrodynamics. It was solved subsequently by Geroch (Section 9) in the
mid of 60ies. The problem of stability of Ranada's knots (or their ability
to brake) was seriously treated mathematically only a few years ago. The
problem of existence \ of sources, especially for the extended objects, in
gravitation and non Abelian gauge theories, to our knowledge is not solved
satisfactorily thus far. At the same time, from the very beginnings of
development of quantum field theory Einstein objected to the existing
methods of quantization of sources and fields in the quantum field theory
separately. With the results of Section 8, in which the electron is
represented as the trefoil knot and living on the Dupin cyclide,
superimposed with Varlamov's mass formula, the source problem is solved in
geometrodynamics in accordance with Einstein's vision of quantization of
quantum fields. Furthermore, in the same section it is demonstrated that the
developed description of electron blends seamlessly with the description of
\ charged rotating black hole. With these accomplishments, the description
of neutinos requires uses of the Majorana equation (since neutrinos are
neutral particles of very small masses). This experimental fact apparently
restricts uses of the Maxwell-Dirac isomorphism. The first steps toward
solving this problem were made only in 2023.\bigskip

\textbf{2. \ Derivation of the Schr\"{o}dinger equation for a single photon}

\textbf{\bigskip }

We begin with writing down Maxwell's equations without sources and currents
in the vacuum. These are:

\begin{eqnarray}
\mathbf{\nabla }\cdot \mathbf{E} &=&0,\mathbf{\nabla }\times \mathbf{E}=-%
\frac{1}{c}\frac{\partial \mathbf{H}}{\partial t},  \nonumber \\
\mathbf{\nabla }\cdot \mathbf{H} &=&0,\mathbf{\nabla }\times \mathbf{H}=%
\frac{1}{c}\frac{\partial \mathbf{E}}{\partial t}.  \TCItag{1}
\end{eqnarray}%
In the above equations we keep only the speed of light $c$ while the rest of
constants we put equal to unity. These always can be restored if needed.
Incidentally, the situation described by (1) is in accord with that known in
standard quantum mechanics. This was emphasized by Bohm [6] and Bohm and
Hiley [$18]$. The De Brogile and Schr\"{o}dinger's waves in quantum
mechanics do not have the sources or sinks (charges) and, accordingly, the
currents. In the absence of knotted/linked configurations the Maxwellian
fields also do not have sources or sinks. In section 5 ,following Ranada and
others,we argue that the knotted/linked \ Maxwellian field configurations
could be reinterpreted as charges. Thus, if the charges in electrodynamics
are of topological origin, their movements then could be interpreted as the
currents. But charges do have masses! Thus, such defined masses are of
topological origin and could be looked upon as linked and knotted
configurations of electromagnetic fields. Details will be given below, in
sections 5-7. Reciprocally, these results are possible to apply to Schr\"{o}%
dinger's quantum mechanics.

Next, we introduce the Riemann-Silberstein complex vector, 
\begin{equation}
\mathbf{F}_{\lambda }=\frac{1}{\sqrt{2}}(\mathbf{E}_{\lambda }+i\lambda 
\mathbf{H}_{\lambda }),  \tag{2}
\end{equation}%
where $\lambda =1(-1)$ corresponds to the positive (negative) helicity. The
concept of helicity is related to the concept of polarization which\ for the
light is equivalentto spin [16,17]. All this is described in detail in [10].
Following Kobe [$19]$, and Smith and Raymer [$20,21]$,we temporarily
suppress the subscript $\lambda $ \ and introduce the energy density $%
\QTR{sl}{\varepsilon }$ as follows 
\begin{equation}
\mathbf{F}^{\ast }\cdot \mathbf{F}=\frac{1}{2}(\text{\textbf{E}}^{2}+\mathbf{%
H}^{2})\equiv \QTR{sl}{\varepsilon }.  \tag{3}
\end{equation}%
This result then allows us to introduce the photon wave function,%
\begin{equation}
\Psi _{i}=\sqrt{\frac{1}{\mathcal{E}}}F_{i},  \tag{4}
\end{equation}%
Here \textit{i}=1,2,3 labels the Euclidean coordinates while $\mathcal{E}$
is defined in (6), below . \ Using (4) we require%
\begin{equation}
\dsum\limits_{i=1}^{3}\dint d^{3}x\Psi _{i}^{\ast }\Psi _{i}=1,  \tag{5}
\end{equation}%
with the total energy $\mathcal{E}$ defined by 
\begin{equation}
\dint d^{3}x\QTR{sl}{\varepsilon }=\mathcal{E}.  \tag{6}
\end{equation}%
By design, the wave function $\Psi _{i}$ is satisfying the Schr\"{o}dinger's
equation for the photon

\begin{eqnarray}
i\frac{\partial }{\partial t}\mathbf{F} &=&c\mathbf{\nabla }\times \mathbf{F,%
}\text{ }  \TCItag{7} \\
\text{or, }i\hbar \frac{\partial }{\partial t}\mathbf{F} &=&ic\mathbf{%
p\times F,}\text{ }\mathbf{p=-i\hbar \nabla ,}  \TCItag{8}
\end{eqnarray}%
provided that

\begin{equation}
\mathbf{\nabla }\cdot \mathbf{F}=0.  \tag{9}
\end{equation}%
Equations (7) and (9) are equivalent to the set of Maxwell's equations (1)
as required. The continuity equation for the probability now reads,%
\begin{equation}
\frac{\partial \rho }{\partial t}+\text{div}\cdot \mathbf{j=}0\mathbf{,}%
\text{ }\mathbf{j}=\alpha \mathbf{E}\times \mathbf{H=}\text{ }\alpha \mathbf{%
\mathbf{F}^{\ast }\times \mathbf{F},}\text{ }\rho \text{ }\mathbf{=}%
\dsum\limits_{i=1}^{3}\Psi _{i}^{\ast }\Psi _{i},\alpha =\frac{c}{\mathcal{E}%
}.  \tag{10}
\end{equation}%
These results are perfectly fine as far as quantization of electromagnetic
field is of interest only. They are not exhibiting the universal connection
with quantum mechanics of particles, though. \ They were initially designed
for photons only. This gives us an opportunity to describe such a connection
in the next section.\bigskip \bigskip \bigskip

\textbf{3. \ Electromagnetic field in the absence of currents and sources as}

\ \ \ \textbf{\ \ complex scalar field}\bigskip \bigskip

Following Green and Wolf [13,14], we notice that in a region of space free
of currents and charges the electromagnetic field is fully specified by the
single vector potential \textbf{A}. For such a case we can write: $\mathbf{E}%
=-\frac{1}{c}\frac{\partial }{\partial t}\mathbf{A\equiv -}\frac{1}{c}%
\mathbf{\dot{A},}$ $\mathbf{H=\nabla \times A.}$ \ This observation then
allows us to reobtain the vector potential \textbf{A} from a single (in
generally complex) scalar field potential $V(\mathbf{x},t)$. It can be shown
that the total energy $\mathcal{E}$, defined in (3) and (6), can be
rewritten in terms of $V(\mathbf{x},t)$ as 
\begin{equation}
\dint d^{3}x\QTR{sl}{\varepsilon }\equiv \frac{1}{2}\dint d^{3}x\text{ }(%
\frac{1}{c^{2}}\dot{V}\dot{V}^{\ast }+\mathbf{\nabla }V\cdot \mathbf{\nabla }%
V^{\ast })=\mathcal{E}  \tag{11}
\end{equation}%
Accordingly, the Poynting flux $\mathbf{j}$\textbf{\ \ }of \ electromagnetic
energy density\textbf{\ }is given by 
\begin{equation}
\mathbf{j}=-\frac{1}{2}(\dot{V}^{\ast }\mathbf{\nabla }V+\dot{V}\mathbf{%
\nabla }V^{\ast }).  \tag{12}
\end{equation}%
Therefore, the\ analog of the continuity equation (10) now reads as 
\begin{equation}
\frac{\partial \QTR{sl}{\rho }}{\partial t}+\text{div}\cdot \mathbf{j=}%
0,\rho =V^{\ast }V.  \tag{13}
\end{equation}%
It should be clear then that the obtained results are equivalent to those
presented in section 2. The advantage of having these results rewritten with
help of the scalar potential $V$ lies in the opportunity of bringing them
into correspondence with quantum mechanics and of introducing knots and
links into discussion. In fact, as we soon demonstrate, such a scalar form
of Maxwell's equations allows us to accomplish much more. To begin, we
represent $V(\mathbf{x},t)$ in the form

\begin{equation}
V(\mathbf{x},t)=\dint d^{3}k[\alpha (\mathbf{k},t)\cos (\mathbf{k}\cdot 
\mathbf{x})+\beta (\mathbf{k},t)\sin (\mathbf{k}\cdot \mathbf{x})]\equiv
\dint d^{3}kV(\mathbf{k},t),  \tag{14}
\end{equation}%
at the same time, 
\begin{equation}
\mathbf{A}(\mathbf{x},t)=\dint d^{3}k[\mathbf{a}(\mathbf{k},t)\cos (\mathbf{k%
}\cdot \mathbf{x})+\mathbf{b}(\mathbf{k},t)\sin (\mathbf{k}\cdot \mathbf{x}%
)]=\dint d^{3}k\mathbf{A}(\mathbf{k},t).  \tag{15}
\end{equation}%
From electrodynamics it is known that the vector potential \textbf{A} is
satisfying the wave equation for the vector field \textbf{A:} 
\begin{equation}
\frac{1}{c^{2}}\frac{\partial ^{2}}{\partial ^{2}t}\mathbf{A}-\nabla ^{2}%
\mathbf{A}=0.  \tag{16}
\end{equation}%
Use of (15) in (16) leads to%
\begin{equation}
\frac{1}{c^{2}}\frac{\partial ^{2}}{\partial ^{2}t}\mathbf{A}(\mathbf{k},t)+%
\mathbf{k}^{2}\mathbf{A}(\mathbf{k},t)=0.  \tag{17}
\end{equation}%
By comparing equations (14),(15), it follows that 
\begin{equation}
\frac{1}{c^{2}}\frac{\partial ^{2}}{\partial ^{2}t}V(\mathbf{k},t)+\mathbf{k}%
^{2}V(\mathbf{k},t)=0  \tag{18}
\end{equation}%
and, therefore, 
\begin{equation}
\frac{1}{c^{2}}\frac{\partial ^{2}}{\partial ^{2}t}V(\mathbf{x},t)-\nabla
^{2}V(\mathbf{x},t)=0.  \tag{19}
\end{equation}%
Next, suppose that the solution of (19) can be presented in the form,%
\begin{equation}
V(\mathbf{x},t)=\mathfrak{v}(\mathbf{x})\exp \{i\phi (\mathbf{x},t)\},\text{ 
}\phi (\mathbf{x},t)=k\Phi (\mathbf{x})-\omega t.  \tag{20}
\end{equation}%
It should be noted that such a representation is physically motivated by its
direct connection with Huygens' principle. Details are best explained in
Luneburg [22]. \ Some of these ideas were subsequently developed by Maslov
[23]. Different/alternative interpretation of (20), adopted for formalism of
quantum mechanics, was independently developed by Bohm in his Bohmian
mechanics [6,18]. Using (20) in (19) the following two equations are
obtained: 
\begin{equation}
\left( \nabla \Phi \right) ^{2}-\frac{1}{k^{2}\mathfrak{v}}\nabla ^{2}%
\mathfrak{v}=n^{2},  \tag{21}
\end{equation}%
\begin{equation}
\nabla \Phi \cdot \nabla \mathfrak{v+}\frac{1}{2}\left( \nabla ^{2}\Phi
\right) \mathfrak{v}=0.  \tag{22}
\end{equation}%
Here $k=c/\omega ,$ $n^{2}=1.$ In the case of quantum mechanics, typically, $%
n^{2}=n^{2}(\mathbf{x}).$ This fact \ will be discussed further below. For
now, however, consider both of these equations in the limit $k^{2}$ $%
\rightarrow \infty .$ Such a limit is typical for phenomena described by the
methods of geometrical optics [23], [24]. In this limit (21) is converted
into : 
\begin{equation}
\left( \nabla \Phi \right) ^{2}=n^{2}\text{ .}  \tag{23}
\end{equation}%
The physics behind equations (21) and (22) is predetermined by the
properties of geometric optics limit. Specifically, in this limit the
surfaces $\Phi =const$ represent the \textsl{wave fronts} while their duals
represent the \textsl{\ light rays. }These are orthogonal to \ the
wavefronts. Following [23],[24], we notice that: a) (23) is known as the 
\textsl{eikonal equation}, b) (22) is known as the \textsl{transport equation%
}. This equation can be simplified with the help of the following arguments.
If $\frac{\partial }{\partial \tau }$ denotes the differentiation along a
particular ray, then according to [23],[24] we write, 
\begin{equation}
\frac{\partial }{\partial \tau }\cdot \cdot \cdot =\nabla \Phi \cdot \nabla
\cdot \cdot \cdot .  \tag{24}
\end{equation}%
Displayed identity allows us to rewrite (22) as 
\begin{equation}
\frac{\partial \mathfrak{v}}{\partial \tau }+\frac{1}{2}\left( \mathfrak{v}%
\nabla ^{2}\Phi \right) =0.  \tag{25}
\end{equation}%
Integration of the last equation is straightforward and is yielding the
result,%
\begin{equation}
\mathfrak{v}(\tau )=\mathfrak{v}(\tau _{0})\exp [-\frac{1}{2}%
\dint\limits_{\tau _{0}}^{\tau }d\tau ^{\prime }\nabla ^{2}\Phi ].  \tag{26}
\end{equation}%
Next, following [24], chr.7, we notice that the ray trajectory $\mathbf{x}$($%
\tau )$ can be derived from the equations of motion, 
\begin{equation}
\frac{d\mathbf{x}}{d\tau }=\mathbf{\nabla }\Phi .  \tag{27}
\end{equation}%
By combining equations(25) and (27) the solution, equation (26), can be
rewritten as 
\begin{equation}
\mathfrak{v}(\mathbf{x}(\tau ))=\mathfrak{v}(\mathbf{x}(0))\exp [-\frac{1}{2}%
\dint\limits_{\tau _{0}}^{\tau }d\tau ^{\prime }\nabla ^{2}\Phi (\mathbf{x}%
(\tau ^{\prime }))].  \tag{28}
\end{equation}%
Notice next that $\mathbf{x}(\tau )=\mathbf{x}(\mathbf{x}_{0},\tau )$ so
that, by definition, $\mathbf{x}(\mathbf{x}_{0},\tau _{0})=\mathbf{x}_{0}.$
This observation allows us to introduce the Jacobian$\ J$ as 
\begin{equation}
J=\det (\frac{\partial x^{i}(\mathbf{x}_{0},\tau )}{\partial x_{0}^{j}}). 
\tag{29}
\end{equation}%
In view of (27), we then obtain as well%
\[
\frac{1}{J}\frac{dJ}{d\tau }=\nabla \cdot \nabla \Phi =\nabla ^{2}\Phi (%
\mathbf{x}(\tau )) 
\]%
implying%
\begin{equation}
J(\mathbf{x}_{0},\tau )=\exp [\dint\limits_{\tau _{0}}^{\tau }d\tau ^{\prime
}\nabla ^{2}\Phi (\mathbf{x}(\tau ^{\prime }))]\equiv J(\mathbf{x}(\tau )). 
\tag{30}
\end{equation}%
By combining (28) and (30) we finally obtain: 
\begin{equation}
\mathfrak{v}(\mathbf{x}(\tau ))=\frac{\mathfrak{v}(\mathbf{x}(\tau _{0}))}{%
\sqrt{J(\mathbf{x}(\tau ))}}.  \tag{31}
\end{equation}%
In view of this result, it is always possible to normalize $\mathfrak{v}(%
\mathbf{x}(\tau )),$ that is to require%
\begin{equation}
\dint_{\Delta }d^{3}x\mathfrak{v}^{2}=1  \tag{32}
\end{equation}%
where $\Delta $ is the domain of integration determined by a particular
problem to be solved. According to [13], the energy density $\QTR{sl}{%
\varepsilon ,}$ defined in (11), can be represented in view of (23) as%
\begin{equation}
\varepsilon =\frac{1}{2}[k^{2}\mathfrak{v}^{2}+\frac{k^{2}}{n^{2}}\mathfrak{v%
}^{2}(\left( \nabla \Phi \right) ^{2}+\frac{1}{k^{2}}(\nabla \ln \mathfrak{v}%
)^{2})]\rightarrow _{k\rightarrow \infty }k^{2}\mathfrak{v}^{2}.  \tag{33}
\end{equation}%
In the limit $k\rightarrow \infty $ the analog of the wavefunction density $%
\rho $, (10), is given now by 
\begin{equation}
\rho =\frac{\varepsilon }{\mathcal{E}}\rightarrow _{_{k\rightarrow \infty }}%
\frac{\mathfrak{v}^{2}}{\dint d^{3}x\mathfrak{v}^{2}}\rightarrow \Psi ^{\ast
}\Psi ,\text{ }\Psi =\mathfrak{v}(\mathbf{x}(\tau ))\exp \{i\phi (\mathbf{x}%
,t)\}.  \tag{34}
\end{equation}%
Substitution of the ansatz (20) into expression for the energy current (12)
results in 
\begin{equation}
\frac{\mathbf{j}}{\mathcal{E}}=\frac{1}{2}\mathfrak{v}^{2}(\mathbf{x})%
\mathbf{\nabla }\Phi .  \tag{35}
\end{equation}%
The associated Hamilton-Jacobi (H-J) equation is given by (23). In the limit 
$k\rightarrow \infty $ its solutions are those of the Schr\"{o}dinger
equation. This fact is well known from the WKB theory, where $k\rightarrow
\infty $ limit\ is the same as $\hbar \rightarrow 0$ limit (the classical
limit). We shall recover below Schr\"{o}dinger's equation without recourse
to the $k\rightarrow \infty $ (or $\hbar \rightarrow 0)$ limit.

\bigskip .

\bigskip

\textbf{4. \ Schr\"{o}dinger's quantum mechanics in terms of complex scalar}

\ \ \ \textbf{\ \ fields of electromagnetic origin}

\ \ \ \ 

To our knowledge, interpretation of electromagnetic fields in terms of
complex scalar fields was not in use in physics literature till late 50ies
[13,14]. Roman-one of the founders of modern quantum field theory- had
written a paper [$25$] on this topic in 1959 and made the results of [13,14]
perfectly rigorous. However, his efforts- to write source-free
electrodynamics- were left unnoticed until work by Ranada [$15$] who
rediscovered this possibility without actually being guided by[13,14,$25$].
It is quite remarkable that Max Born- the designer of probabilistic
interpretation of quantum mechanics, later on in life had written a book
"Principles of Optics" [$26$] (along with Emil Wolf) in which only
implicitly (that is without any mention of quantum mechanics) on page 430,
in section 8.4, the sclalar field theory interpretation of electromagnetic
field can be found. In recent literature this mention is briefly presented
in [$27$]. While the papers [13,14,$25$] were left unnoticed, much more
cumbersome Duffin-Kemmer (D-K) formalism has become in vogue for
electromagnetic and other bosonic fields [$28-30$]. A consistent
relativistic quantum mechanics/quantum field theory for spin 0 and 1 bosons
had been developed with help of the D-K formalism [$28-31$]. \ In it the
Dirac looking the 1st order equation is used in which the Dirac $\gamma $
matrices are replaced with $\beta $ matrices obeying commutation relations
different from those known for the Dirac matrices. In dealing with the
(anti) self -dual electromagnetic fields the twistorial interpretation of
equations for these fields is also possible [$32$]. The comparison between
the vector (Maxwell), the complex scalar, the Dirac and the twistorial
fields is seemingly possible only for the massless versions of these fields.
The inclusion of masses into results just mentioned can be done either by
hands or topologically, e.g. as it was done by Ranada [16,17]. The role of
masses will be further studied in Sections 5-7.

In standard field-theoretic notations[$10,28$] (making for a moment all
constants being equal to unity) the Lagrangian $\mathcal{L}$ for the complex
massless scalar field is given by, e.g. see [$28$], page 32, 
\begin{equation}
\mathcal{L}[\varphi ,\varphi ^{\ast }]=\dsum\limits_{n}\frac{\partial
\varphi }{\partial x^{n}}\frac{\partial \varphi ^{\ast }}{\partial x_{n}}. 
\tag{30}
\end{equation}%
By varying the fields $\varphi $ and $\varphi ^{\ast }$in the action $%
\mathcal{A}$, 
\begin{equation}
\mathcal{A}=\dint d^{4}x\mathcal{L}[\varphi ,\varphi ^{\ast }],  \tag{31}
\end{equation}%
while assuming that these fields are independent and decaying at infinity,
leads to the following equations of motion,%
\begin{eqnarray}
\square \varphi &=&0,  \TCItag{32} \\
\square \varphi ^{\ast } &=&0.  \TCItag{33}
\end{eqnarray}%
Here the d'Alembertian $\square $ is defined as: $\square =\frac{\partial
^{2}}{\partial ^{2}t}-\nabla ^{2},$ with $\nabla ^{2}$ being the
3-dimensional Laplacian$.$ In his work [25] Roman was not interested in the
standard field-theoretic analysis of $\mathcal{L}[\varphi ,\varphi ^{\ast }]$%
, e.g. that which is done in [28] on page 32. He was interested in proving
that the gauge transformations of the Maxwellian fields when rewritten in
the formalism of complex scalar fields will keep the action $\mathcal{A}$
form (or gauge)-invariant (up to the total divergences vanishing at the
boundary of space-time). With this result proven, the task of this paper is
different. Given that the action $\mathcal{A}$ is \ gauge-invariant, we
apply the standard field-theoretic treatment to (30) with purposes which
will become clear upon reading. From [28] we find the time component $T^{00}$
of the energy-momentum tensor (that is the energy density). It is given by, 
\begin{equation}
T^{00}=\frac{\partial \varphi }{\partial t}\frac{\partial \varphi ^{\ast }}{%
\partial t}+\mathbf{\nabla }\varphi \cdot \mathbf{\nabla }\varphi ^{\ast }. 
\tag{34}
\end{equation}%
Notice now that $T^{00}$ coincides with $\QTR{sl}{\varepsilon }$ defined in
(11), where we temporarily put $c=1$. Accordingly, the momentum density $%
T^{0i}$ given by 
\begin{equation}
T^{0i}=-(\frac{\partial \varphi ^{\ast }}{\partial t}\nabla _{i}\varphi +%
\frac{\partial \varphi }{\partial t}\nabla _{i}\varphi ^{\ast }),i=1,2,3, 
\tag{35}
\end{equation}%
is up to a constant coincides with the flux $\mathbf{j}$ defined in (12).
Evidently, in this setting the continuity equation (13) is nothing but the
law of conservation of the energy-momentum tensor [28]%
\begin{equation}
\frac{\partial }{\partial x^{\mu }}T_{\nu }^{\mu }=0.  \tag{36}
\end{equation}%
For the record, we use the Minkowski- space metric tensor $g_{\mu \nu }$ of
signature $(+,-,-,-)$ so that $g_{\mu \nu }=g^{\mu \nu }$. For the 4-vector $%
a_{\mu }$ we have $a_{\mu }=g_{\mu \nu }a^{\nu },$ etc.

Because we are dealing with the complex scalar field, there is also the
current vector $J_{\mu }$ responsible for carrying the charge (recall that
the description of charges and currents associated with them is always
associated with the existence of the global gauge symmetry inseparably
linked with uses of complex scalar fields [$28$]). Mathematically, the
conservation of current $J_{\mu }$ is expressible by analogy with (36) as%
\begin{equation}
\frac{\partial }{\partial x^{\mu }}J^{\mu }=0.  \tag{37}
\end{equation}%
Explicitly, the current $J_{\mu }$ is given by 
\begin{equation}
J_{\mu }=i(\varphi ^{\ast }\frac{\partial \varphi }{\partial x^{\mu }}-\frac{%
\partial \varphi ^{\ast }}{\partial x^{\mu }}\varphi ).  \tag{38}
\end{equation}%
Let $Q=\dint d^{3}xJ_{0}$ be the total charge. Then (37) is the continuity
equation associated with the charge conservation. It is well known
[10,28,33] that $J_{0}$ is not always positively defined quantity. This fact
is caused by the observation that at any given time $t$\ both $\varphi $ and 
$\frac{\partial \varphi }{\partial x^{0}}$ may independently have arbitrary
values and signs. \ It is important because of the following. \ For the sake
of generality and comparison, we include the mass $m^{2}$ term into both
equations (32),(33) \ thus converting them into the Klein-Gordon (K-G)
equations [$28$], page 32, 
\begin{eqnarray}
\left( \square +m^{2}\right) \varphi &=&0,  \TCItag{39} \\
\left( \square +m^{2}\right) \varphi ^{\ast } &=&0.  \TCItag{40}
\end{eqnarray}%
Since the time of Schr\"{o}dinger's discovery \ of the equation bearing his
name, the K-G equation was considered as the relativistic analog of Schr\"{o}%
dinger's equation. In analogy with the non-relativistic case, by multiplying
(39) by $\varphi $ and (40) by $\varphi ^{\ast }$ and subtracting (40) from
(39) we are repeating the same steps as for the non- relativistic Schr\"{o}%
dinger equation in order to obtain the continuity equation (37). \ In the
nonrelativistic case (that is for the Schr\"{o}dinger equation) this
procedure yields: $j_{0}=\rho =\psi ^{\ast }\psi $ and $\vec{j}=\frac{\hbar 
}{2mi}(\psi ^{\ast }\mathbf{\nabla }\psi -\psi \mathbf{\nabla }\psi ^{\ast
}).$ The continuity equation in the nonrelativistic case is in its standard
form: $\frac{\partial }{\partial t}\rho +\mathbf{\nabla }\cdot \vec{j}=0.$
To compare this result with equations (37), (38) we must multiply (37) by $%
\frac{\hbar }{2m}.$ Then, we obtain respectively $J_{0}=\frac{i\hbar }{2m}%
(\varphi ^{\ast }\frac{\partial \varphi }{\partial x^{0}}-\frac{\partial
\varphi ^{\ast }}{\partial x^{0}}\varphi )$ and $J_{k}=\frac{\hbar i}{2m}%
(\varphi ^{\ast }\frac{\partial \varphi }{\partial x^{k}}-\frac{\partial
\varphi ^{\ast }}{\partial x^{k}}\varphi )=-$ $j_{k}$ , $k=1,2,3$. Thus,
even though (up to a sign) the currents $\vec{J}$ and $\vec{j}$ \ coincide,
the densities $j_{0}$ and $J_{0}$ are noticeably different. This fact
matters when the time-dependent problems are discussed for the K-G fields.
Because of the nonpositivity of $J_{0}$ the full time-dependent K-G equation
was discarded (historically) from consideration\ as the relativistic analog
of the Schr\"{o}dinger equation. In the time-independent case the situation
is not as dramatic. Specifically, suppose that the field $\varphi $ in (39)
can be written as $\varphi (\mathbf{x},t)=\psi (\mathbf{x})\exp (-i\omega t)$
then, the K-G equation is converted into the Helmholtz equation 
\begin{equation}
\left( \mathbf{\nabla }^{2}+\tilde{m}^{2}\right) \psi =0,\text{ }\omega
^{2}-m^{2}=\tilde{m}^{2}.  \tag{41}
\end{equation}%
Accordingly, the field $\varphi ^{\ast }$ can be written now as $\varphi
^{\ast }(\mathbf{x},t)=\psi (\mathbf{x})\exp (i\omega t)$. In view of these
results, $J_{0}$ now acquires the form, $J_{0}=\frac{\hbar \omega }{m}(\psi
^{\ast }\psi ).$ At the same time\ $J_{k}=0.$ This result allows us to study
all kinds of stationary K-G equations [$33$] -all making physical sense. The
obtained result raises the following question. In section 3 we demonstrated
that the continuity equation (13) is the same as the energy-momentum
conservation equation (36). At the same time, the continuity equation (10)
for the photon is the same as the continuity equation (13) for the complex
scalar field. This means that this equation can be used instead of the
continuity equation (37) for development of the time-dependent quantum
mechanical formalism for the complex scalar K-G field. It is appropriate at
this point to mention that historically the complex massive scalar field was
used in the Yukawa theory of strong interactions where it is known as $\pi -$
meson field. By extending the results of Harish-Chandra [29] for this field
written in the language of Duffin-Kemmer formalism, Tokouoka [30] studied in
detail the meson-nucleon interactions. Much more recent results are
discussed, for example, in [34] and references therein. Since the meson
-nucleon interactions are described nowadays with help of quantum
chromodynamics (QCD), the complex scalar field acts effectively as the
abelianized version of the non-Abelian Yang-Mills gauge field. More on this
is discussed in the next section where we shall also explain how to get rid
of the mass term in the K-G equation. According to [33], page 99, every
spinor component of the Dirac equation with nonzero mass is satisfying the
massive K-G equation. \ In the zero mass limit this fact creates the
equivalence class between the massless K-G and Dirac equations. Apparently,
components of equations for higher spin particles, e.g. spin-2 gravitons,
also belong to the same equivalence D-K class [10, 28, 33]. In this section
we still need to demonstrate how the results of sections 3 and 4 are
connected with the non-relativistic time-independent Schr\"{o}dinger
equation (TISE). In the next section we shall present evidence that such an
equation also belongs to the same equivalence class\ (in the sense of
Hadamard) as the rest of basic equations.

To connect the results of sections 3 and 4 with TISE, just described
treatment of the stationary K-G equation is the most helpful. After the mass
term is eliminated in this equation (read the next section) and the constant 
$c$ is restored, the K-G equation is converted into the wave equation (32)
(or (33)). This fact allows us to use the results of Schr\"{o}dinger's
second foundational paper on quantum mechanics [35], pages 13-40. In (32) we
replace the speed of light $c$ with $u=c/n,$ where $n=n(x,y,z)$ is the
effective refractive index. Next, we use the same ansatz $\varphi (\mathbf{x}%
,t)=\psi (\mathbf{x})\exp (-i\omega t)$ for the wave function. Also, we
replace the dispersion relation $\omega =ck$ for the electromagnetic waves
in the vacuum by $\omega =uk$ and use the de Broglie-type relation $%
k=p/\hbar $ along with the fact that for the mechanical system $E=\mathbf{p}%
^{2}/2m$ +$V$. Thus, we obtain $\mathbf{p}^{2}$ $=2m(E-V).$ With help of
these results the TISE follows from the wave equation (32) under conditions
just described. Thus, we obtain, 
\begin{equation}
\lbrack -\frac{\hbar ^{2}}{2m}\nabla ^{2}+V-E]\psi =0.  \tag{42}
\end{equation}%
The normalized result for $J_{0}$ can now be used immediately so that the
continuity equation (37) works in this case. Nevertheless, the question
remains: what to do with the continuity equation (36)? This equation is used
for development of quantum mechanics of photons. Can it be used in the
nonrelativistic case for the stationary Schr\"{o}dinger equation? Equations
(32)-(36) suggest that this could be possible. Nothing meaningful is
obtained if we act straightforwardly, though. Indeed, by using the K-G
ansatz for the wave function in (34) and (35) and restoring $c$ in these
equations yields:%
\begin{eqnarray}
T^{00} &=&\left( \frac{\omega }{c}\right) ^{2}\psi ^{\ast }\psi +\mathbf{%
\nabla }\psi \cdot \mathbf{\nabla }\psi ^{\ast },  \TCItag{43} \\
T^{0i} &=&0.  \TCItag{44}
\end{eqnarray}%
The result obtained for $T^{00}$ is different from that for $J_{0}.$ Unlike
[14], we are not going to look for arguments in favor of $\mathbf{\nabla }%
\psi \cdot \mathbf{\nabla }\psi ^{\ast }\simeq \left( \frac{\omega }{c}%
\right) ^{2}$ relation$.$This was done already in section 3, where the
results were obtained in the regime of geometrical optics. Instead, now we
are going to use the original Schr\"{o}dinger's ideas again. This time, we
are going to use the methods developed in his 1st paper on quantum mechanics
[35], pages 1-12, as well as from the already discussed 2nd paper. In
particular, we again replace $c$ by $u$ in (43) and then replace $\left( 
\frac{\omega }{u}\right) ^{2}$ by $\left( \frac{p}{\hbar }\right) ^{2}.$\ \
Next, by looking at (43) we require%
\[
\mathbf{\nabla }\psi \cdot \mathbf{\nabla }\psi ^{\ast }=\left( \frac{p}{%
\hbar }\right) ^{2}\psi ^{\ast }\psi 
\]%
and use $\ \left( \frac{p}{\hbar }\right) ^{2}=\frac{2m(E-V)}{^{\hbar ^{2}}}%
. $ \ These results can be rewritten in terms of the H-J equation if we
temporarily use only the real valued functions: $\psi (\mathbf{x})=\psi
^{\ast }(\mathbf{x})$%
\begin{equation}
\left( \nabla \psi \right) ^{2}=\frac{2m(E-V)}{^{\hbar ^{2}}}\psi ^{2}. 
\tag{45}
\end{equation}%
Notice that this H-J equation contains explicitly Planck's constant $\hbar $
while the H-J equation used in the semiclassical WKB calculations is $\hbar
- $ independent by design. Nevertheless, (45) coincides exactly with the
equation ($1^{\prime \prime })$ of Schr\"{o}dinger's 1st paper on quantum
mechanics, [35], pages 1-12. \ This difference has a profound effect on the
rest of Schr\"{o}dinger's calculations. It allows\ him and, hence us, to
restore the stationary Schr\"{o}dinger equation without any approximations.
For this purpose, Schr\"{o}dinger introduces the functional $J[\psi ]$ 
\begin{equation}
J[\psi ]=\frac{1}{2}\dint_{\Delta }d^{3}x[\left( \nabla \psi \right) ^{2}-%
\frac{2m(E-V)}{^{\hbar ^{2}}}\psi ^{2}]  \tag{46}
\end{equation}%
which he is minimizing under the subsidiary condition%
\begin{equation}
\dint_{\Delta }d^{3}x\psi ^{2}=1  \tag{47}
\end{equation}%
thus producing the stationary Schr\"{o}dinger equation (42). Therefore, the
continuity equations (36) and (37) \ both can be used in conjunction with
the nonrelativistic stationary Schr\"{o}dinger's equation (42).

\bigskip

\bigskip

\textbf{5. \ \ Chladni patterns and knotty wavefunction configurations}

\ \ \ \ \ \ 

To begin, we would like to discuss now the following issue: How presence
(absence) of masses is affecting topics discussed in previous sections?
Surprisingly, the link between the discussed topics and the masses can be
established with help of the Huygens' principle. This principle can be
discussed purely mechanically with methods of contact geometry and topology
as it was done, for example, by Arnol'd [$36$]. Reader's friendly basics of
contact geometry and topology are provided in our book [$37$]. The same
principle could be discussed using the theory of partial differential
equations. It is beautifully discussed, for example, in books by Hilbert and
Courant [$38$], Luneburg [22], G\"{u}nter [$39$] and later, by many others.
For the sake of space, we shall not go into mathematical details of Huygens'
principle in this paper. Details are given in our paper [$40$]. We only
mention that \ equations (32),(33) do obey the Huygens' principle so that 
\textsl{all equations which can be reduced to} (32),(33)\ \textsl{do obey
this principle}. According to Hadamard \ whose results are presented in
[39], only 3 operations are allowed for conversion of a given PDE to the
"trivial" PDE equation(s) (32),(33).

Specifically, the essence of \textsl{Huygens' equivalence principle} lies in
the following.

Let L[$\phi $] be the Huygens-equivalent operator (that is the operator
which is written as in equations (32),(33)) and let \~{L}[$\phi $] be
another operator (and equation) to be investigated. Then, the operators L[$%
\phi $] and \~{L}[$\phi $] are Huygens-equivalent if:

a) \~{L}[$\phi $] can be obtained from L[$\phi $] by the nonsingular
transformations of independent variables.

b) \~{L}[$\phi $]=$\lambda ^{-1}$L[$\lambda \phi $] for some positive,
smooth function $\lambda $ of independent variables.

c) \~{L}[$\phi $]=$\rho $L[$\phi $] for some positive smooth function $\rho $
of independent variables.

With these rules established, let us consider as an example an invertible
sequence of transformations from the d'Alembert equation (32) to the massive
K-G equation. Let us use the transformation \~{L}[$\phi $]=$\lambda ^{-1}$L[$%
\lambda \phi $] with $\lambda =e^{\alpha t},$ with $\alpha $ being some
constant. Then, we obtain:%
\[
e^{-\alpha t}\{[\frac{\partial ^{2}}{\partial t^{2}}e^{\alpha t}\phi
]-[\nabla ^{2}e^{\alpha t}\phi ]\}=\frac{\partial ^{2}}{\partial t^{2}}\phi
+2a\frac{\partial }{\partial t}\phi -\nabla ^{2}\phi +a^{2}\phi =0. 
\]%
Next, let $\lambda _{1}=e^{ibx}$ and $\lambda _{2}=e^{-icx}.$ Upon
substitution of these factors into the previous equation, while using the
rule b), we obtain after some calculation: 
\[
\frac{\partial ^{2}}{\partial t^{2}}\phi +2a\frac{\partial }{\partial t}\phi
-\nabla ^{2}\phi +2ib\frac{\partial }{\partial x}\phi -2ic\frac{\partial }{%
\partial x}\phi +(a^{2}-b^{2}-c^{2})\phi =0. 
\]%
If now $b=c$ and $a^{2}=2b^{2},$ we obtain the telegrapher equation%
\[
\frac{\partial ^{2}}{\partial t^{2}}\phi +2a\frac{\partial }{\partial t}\phi
-\nabla ^{2}\phi =0. 
\]%
The K-G equation is obtained now upon substitution $\phi =e^{mt}\psi $ into
telegrapher's equation with subsequent replacement of $a=-2m$ and $m$ by $im$%
. Thus, the K-G and d'Alembert equations are Huygens -equivalent. If this is
so, it would be possible to use the twistor formalism [32],[41] for the K-G
equation.

The question arises: Is the stationary Schr\"{o}dinger equation
Huygens-equivalent to the d'Alembert equation? \ In 1935 Fock initiated
study of this problem (having different goals in mind, though). \ While
studying the (accidental) degeneracy of the stationary Schr\"{o}dinger
equation for the hydrogen atom, he converted this equation into the integral
equation looking very similar to the Poisson integral in the theory of \
functions of one complex variable. Recall, that harmonic functions (that is
functions satisfying the Laplace equation) inside the circle (and, thus,
also in any domain which can be conformally mapped into circle) can be
presented via the Poisson integral. Fock initiated study of what is known
now as dynamical symmetry groups. These groups are allowing us to solve
quantum mechanical problems group-theoretically thus by-passing uses of the
Schr\"{o}dinger equation. This direction of research has grown into a large
field nowadays [$42$]. Fock's work is presented in Vol.1, pages 400-410 of [$%
42$]. Group-theoretic analysis had lead Fock to conclusion that the bound
states of hydrogen atom should be studied in 4-dimensional Euclidean space
while scattering states should be studied in 3+1 dimensional hyperbolic
(Lobachevski) space. \ Study of the Euclidean 4-dimensional version of the
Schr\"{o}dinger equation (for the bound states) was reduced by Fock to the
study of solutions of 4-dimensional Laplacian. The description of the
scattering states was only outlined by Fock. He suggested that this case
requires study of solutions of the d'Alembertian. From here is a connection
with the Huygens principle. This suggestion was waiting for its solution for
31 years. In 1966 it was finally solved by Itzykson and Bander [$43$].
Although Itzykson and Bander [$43$] obtained the d'Alembertian for
scattering states thus making the stationary Schr\"{o}dinger equation
Huygens-equivalent to the d'Alembertian equation, \ the results of [$43$]
happened to be very cumbersome. The transparency of results for both bound
and scattering states of hydrogen atom was achieved \ only in 2008 in the
paper by Frenkel and Libine [$44$]. Using quaternions in 4 dimensions
Frenkel and Libine extended the Poisson formula for the harmonic functions-
from 2 to 4 dimensions. This allowed them to extend easily the obtained
results from Euclidean 4 dimensional to Minkowski 3+1 space.

With these results presented, now we are able to explain the quantum
mechanical significance of null fields. These fields were known since 1787,
when Chladni, a German physicist, \ studied nodal lines of a vibrating metal
plate by stroking this plate covered by sand with a violin bow [$45$]. After
studying a variety of nodal patterns systematically, he wrote a book in
1802, where all these patterns were systematized. Remarkably, the 1802
edition of Chladni book was translated into English and published in 2015
[46]. Nowadays, Chladni patterns can be seen on You Tube [$47$] and, more
scientifically, in plasmonic charge density waves [$48$], etc.

With help of results of previous sections we are in a position now to
provide an explanation of the connection between Chladni's patterns, null
fields, knots/links and quantum mechanics. To set up notations, following [$%
49$],we begin with the description of two dimensional Chaladni problems. \
It is convenient to think about a given Riemannian surface $\Sigma $ with
metric $g_{\Sigma }$ as a vibrating membrane with $u(\mathbf{x},t)$, ($%
\mathbf{x}\in \Sigma )$ being a displacement\ at time $t$ of the membrane
from its original position. The function $u$ is a solution to the wave
equation: 
\begin{equation}
\frac{\partial ^{2}}{\partial t^{2}}u=\nabla _{\Sigma }^{2}u.  \tag{48}
\end{equation}%
By representing solution in the form $u(\mathbf{x},t)=v(t)w(\mathbf{x})$,
the above equation splits into two equations, e.g. 
\begin{equation}
\frac{\partial ^{2}}{\partial t^{2}}v=\lambda v  \tag{49}
\end{equation}%
and%
\begin{equation}
\nabla _{\Sigma }^{2}w=\lambda w.  \tag{50}
\end{equation}%
The (null) \textsl{zero set}, $\Xi (w):=\{\mathbf{x\in \Sigma :}$ $w\mathbf{%
(x)=}0\},$ is called \textsl{nodal} (Chladni) set. The definitions just
described need to be supplemented with the boundary (e.g. Dirichlet or
Neumann) conditions so that the above nodal problem is known as the \textsl{%
fixed membrane problem}. In the case when the membrane is a closed surface
the above problem is known as\textsl{\ free membrane problem. }The first%
\textsl{\ }detailed calculation of Chladni patterns was\textsl{\ }made by
Poisson in 1829. The chronology of subsequent developments along with
detailed numerous examples can be found in the encyclopedic book [$50$] by
Rayleigh, ch.rs 9 and 10. The same results were recently reproduced in the
book [$51$]. At much more advanced level Chladni patters were studied by
Cheng\textsl{\ }[$52$] who proved that for an arbitrary smooth Riemannian
surface ($\Sigma ,g_{\Sigma })$ the nodal set is a collection of immersed
closed curves. Cheng's result is remarkable since later the analogous
results were obtained in 3 dimensions in [$53$]. This time, since the nodal
curves are closed, \ they can generate knots and links, of any complexity.

The results of [53]\ can be reobtained with help of results obtained in this
work superimposed with [41,54,55]. Following [56], we write%
\begin{equation}
\mathbf{F}(\mathbf{r},t)=\mathbf{F}_{+}+\mathbf{F}_{-}.\text{ \ }\mathbf{F}%
_{\pm }(\mathbf{r},t)=\dint\limits_{0}^{\infty }d\omega e^{\mp i\omega t}%
\mathbf{F}_{\pm \omega }(\mathbf{r}).  \tag{51}
\end{equation}%
In such notations (7),(9) can be rewritten as

\begin{eqnarray}
\mathbf{\nabla }\cdot \mathbf{F}_{\pm } &=&0,\text{ }  \TCItag{52} \\
\nabla \times \mathbf{F}_{\omega } &=&k\mathbf{F}_{\omega },  \TCItag{53}
\end{eqnarray}%
with $k=\omega /c.$ In plasma physics (53) is known as the "force-free"
equation while in hydrodynamics it is known as "Beltrami" equation. It was
discussed in detail in our book [$37$] in the context of methods of contact
geometry/topology while in [$54$] and [$55$] these \ methods were used for
generation of all kinds of knots and links. Applying the operator $\mathbf{%
\nabla }\times $ \ to (53) while taking into account (52) results in the
following equation 
\begin{equation}
\nabla ^{2}\mathbf{F}_{\omega }+k^{2}\mathbf{F}_{\omega }=0,  \tag{54}
\end{equation}%
to be compared with (50).\ This\ (vector) version of the Helmholtz equation
is known as Chandrasekhar-Kendall (CK) equation [$37$]. These authors
noticed that every solution of (53) is solution of (54) but the converse is
not true. This happens to be of fundamental importance for our tasks. In [$%
37 $], page 30, it was stated that the solution of (53) is a composition of
fields of 3 types: a) solenoidal (52), b) lamellar $\mathbf{F}_{\omega
}\cdot \left( \nabla \times \mathbf{F}_{-\omega }\right) =0,$c) Beltrami $%
\mathbf{F}_{\omega }\times \left( \nabla \times \mathbf{F}_{-\omega }\right)
=0.$ The null fields used in creation of "linked and closed beams of light" [%
$41$] are lamellar. In terms of notations set up in [56] they are defined as 
\begin{equation}
\mathbf{F}_{\omega }\cdot \mathbf{F}_{-\omega }=0.  \tag{55}
\end{equation}%
As explained in [37], the same classification of fields exists in physics of
liquid crystals. In [37], pages 32-34, it was demonstrated that the
Faddeev-Skyrme (F-S) knot/link generating model [57] is of the liquid
crystalline origin. It also can be looked upon as originating from the
Abelian reduction of the NAYM fields to be discussed further in Section 6.
Therefore, all results of [41] as well as of [$54,55$] are compatible with
those originating from the F-S model [57]. Note added in proof (below)
provides additional information. Now we can study\ the lamellar (null)
solutions of the C-K equation. Being logically guided by [53,56] our results
differ in detail from results of these works. This difference permits us to
inject new physics absent in these references.

We begin with (55). It can be rewritten as 
\begin{equation}
\left\vert \mathbf{E}_{\omega }\right\vert ^{2}-\left\vert \mathbf{H}%
_{\omega }\right\vert ^{2}+2(\mathbf{E}_{\omega }\cdot \mathbf{\bar{H}}%
_{\omega }+\mathbf{\bar{E}}_{\omega }\cdot \mathbf{H}_{\omega })=0.  \tag{56}
\end{equation}%
This \ equation is surely satisfied if $\left\vert \mathbf{E}_{\omega
}\right\vert ^{2}=\left\vert \mathbf{H}_{\omega }\right\vert ^{2}$ and $%
\mathbf{E}_{\omega }\cdot \mathbf{\bar{H}}_{\omega }+\mathbf{\bar{E}}%
_{\omega }\cdot \mathbf{H}_{\omega }=0.$Next, without loss of generality and
following [56], it is permissible to assume that $\mathbf{E}_{\omega }=%
\mathbf{\bar{E}}_{\omega }$ and $\mathbf{H}_{\omega }=\mathbf{\bar{H}}%
_{\omega }.$ This then leads to $\mathbf{E}_{\omega }\cdot \mathbf{H}%
_{\omega }=0.$ Should we avoid use of time Fourier transforms defined in
(51), just presented two (null) equations (that is $\left\vert \mathbf{E}%
_{\omega }\right\vert ^{2}=\left\vert \mathbf{H}_{\omega }\right\vert ^{2}$
and $\mathbf{E}_{\omega }\cdot \mathbf{H}_{\omega }=0)$ would be sufficient
for generation of all kinds of torus knots and Hopf links evolving in time [$%
41,58$]. These conditions play the central role in works by Ranada [15$%
],[16],$[58].

These results do not let us make a connection with Chladni patterns yet and
with physics associated with these patterns. \ This will be done in this and
the next section. In this section using\ [59] we consider the following
remarkable identity%
\begin{equation}
(\nabla ^{2}+k^{2}\mathbf{)(r\cdot v)}=2\nabla \cdot \mathbf{v}+\mathbf{%
r\cdot }(\nabla ^{2}+k^{2}\mathbf{)v.}  \tag{57}
\end{equation}%
Here $\mathbf{v}$ is either $\mathbf{E}_{\omega }$ or $\mathbf{H}_{\omega }$%
. The scalar $\mathbf{(r\cdot F}_{\omega }\mathbf{)}$ is convenient to
rewrite in the notations of [53], i.e. $\mathbf{(r\cdot F}_{\omega }\mathbf{%
)=}u=u_{1}+iu_{2}.$ Evidently, in view of (19), the scalar $u$ can be
identified with $V$(\textbf{x},t) defined in (14) so that $\mathbf{E}%
_{\omega }$ and $\mathbf{H}_{\omega }$ can be recovered if needed.
Fortunately, this will not be necessary. By combining $(54)$ and $(57)$ we
obtain the Chladni-type (that is (50)) equations 
\begin{equation}
(\nabla ^{2}+k^{2}\mathbf{)}u_{_{k}1}=0\text{ and }(\nabla ^{2}+k^{2}\mathbf{%
)}u_{_{k}2}=0.  \tag{58}
\end{equation}%
These equations were introduced and discussed in [53] using purely
mathematical arguments. Since here the same equations we just reobtained
with help of physical arguments, this allows us to extend the results of [$%
53 $].

First, we notice that both equations have \textsl{the same eigenvalue} $%
\lambda _{1}=\lambda _{2}=k^{2}$. Having the same eigenvalue (of
multiplicity 2) the wavefunctions $u_{_{k}1}$ and $u_{_{k}2}$ are \textsl{%
not the same though} as demonstrated in [$53$] and reaffirmed below.
Although presence of the $i$-factor is essential, the difference goes beyond
this fact. This difference should not be confused with the degeneracy
concept in standard quantum mechanics. The presence of eigenvalues having
multiplicities is responsible for the effects of entanglements (more
accurately- the self-entanglements in the present case). This is explained
in detail in our book [$37$], pages 386-395.

Now we are in the position to demonstrate that: a) quantum mechanical
self-entanglement is equivalent to the entanglement in the topological
knot-theoretic sense; b) presence of eigenvalues with multiplicity larger
than one makes Schr\"{o}dinger's and Heisenberg's interpretation of quantum
mechanics not equivalent. The last statement follows immediately from the
detailed Heisenberg-style calculations presented in [60]. Thus, we are only
left with explanation of a).

Following [$54$], let us consider the force-free equation (53), where
temporarily we replace $\mathbf{F}_{\omega }$ by $\mathbf{v}$ as in (57).
Then, by applying to both sides of (53) the operator div and \ by assuming
that $k=const=\kappa (x,y,z)$ we obtain div($\kappa \mathbf{v)}=\mathbf{v}%
\cdot \nabla \kappa =0.$ Let $\mathbf{r}(t)=\{x(t),y(t),z(t)\}$ be some
trajectory on the surface $const=\kappa (x,y,z).$ In such a case $\frac{d}{dt%
}\kappa \{x(t),y(t),z(t)\}=v_{x}\kappa _{x}+v_{y}\kappa _{y}+v_{z}\kappa
_{z}=\mathbf{v}\cdot \nabla \kappa =0.$ This means that the "velocity" $%
\mathbf{v}$ is always tangential to the surface $const=\kappa (x,y,z).$
Since the vector field $\mathbf{v}$ is being assumed nowhere vanishing, the
surface $const=\kappa (x,y,z)$ can only be a torus T$^{2}.$ The field lines
of $\mathbf{v}$ on T$^{2}$ should be closed if $const$ is rational number.
Thus we just demonstrated that the force-free equation (53) supplies us with
the condition of existence of all possible torus knots for all rational $%
\kappa ^{\prime }s$ . Starting with works by Ranada [15,16], such torus
knots were explicitly designed both in [54,55$]$ and [41].

Now it remains to demonstrate that Chladni-type equations (50),(58) lead to
the 3-dimensional Chladni patterns associated with these equations. Since
this was done already in [$53$], our task is only to supply some physics to
the results of [$53$]. Since the surface $const=\kappa (x,y,z)$ is T$^{2}$,
while the solid torus is defined by \^{T}$^{2}=$ $D^{2}$ $\times S^{1},$
with $D^{2}$ being a disc, it is convenient to introduce a cylindrical
system of coordinates and to consider the Neumann-type boundary value
problem\ for the Helmholtz equations (58) written in cylindrical
coordinates. This task is facilitated by the accumulated knowledge about
circular waveguides in electrodynamics.\ The solutions $u_{_{k}1}=J_{1}(\rho 
\sqrt{\tilde{\lambda}})\cos \varphi $ and \ $u_{_{k}2}$ =$J_{1}(\rho \sqrt{%
\tilde{\lambda}})\sin \varphi $\ are discussed in [53]. Here $J_{1}(x)$ is
the standard Bessel function and $\tilde{\lambda}$ adjusted (with account of
cylindrical symmetry ) eigenvalues which are \ just the TE and TM -type
solutions \ known for circular waveguides [61].\ For small $\rho ^{\prime }s$
they are represented by \ \ $u_{_{k}1}\simeq \frac{\sqrt{\tilde{\lambda}}}{2}%
\rho \cos \varphi $ and \ \ $u_{_{k}2}\simeq \frac{\sqrt{\tilde{\lambda}}}{2}%
\rho \sin \varphi $\ . \ Both functions become zero for $\rho =0.$ Thus, the
individual nodal Chladni sets are respectively given by $u_{_{k}1}^{-1}(0)$\
\ and $u_{_{k}2}^{-1}(0),$\ while the Chladni centerline\ $S^{1\text{ }}$
for the solid torus\ is determined by the transversality condition: $%
u(0)=u_{_{k}1}^{-1}(0)\cap $\ $u_{_{k}2}^{-1}(0).$ It is indeed the
transversality condition since we can plot both $\rho \cos \varphi $ and $%
\rho \sin \varphi $ on\ the complex plane \textbf{C} so that they are
transversal to each other [62].The transversality condition needed for the
validity of Thom's isotopy theorem [53,62] (assuring that the obtained
results are stable with respect to possible perturbations). These
perturbations will occur because by definition every knot is an embedding of
\ $S^{1}$ into $S^{3}$ $=\mathbf{R}^{3}\cup \{\infty \}.$ The validity of
Thom's theorem is required to assure that the embedding of solid tori \^{T}$%
^{2}$ into $S^{3}$ could be done in such a way that it shall produce knots
of any complexity as long as they are not wild. \ Although [53] claims that
all Chladni knotted sets can be obtained this way, physically this requires
more explanations. These are presented in the next two sections.\bigskip

\bigskip

\textbf{6. \ Chladni patterns in the sky, or how topological nature of masses%
}

\ \ \ \textbf{\bigskip\ in the Universe is linked with axions of dark matter
\bigskip \bigskip \bigskip }

6.1. Summary of basic facts\bigskip

At this point our readers might ask a question: How the obtained results are
related to the real currents and charges present in the electromagnetic
field? \ Recall, that Bohm[6,18] was concerned exactly with this issue when
he compared \ the quantum mechanical and optical formalisms. Since the
currents originate from moving charges, the above question should be
restated accordingly. This is one of the topics which will lead us to the
discussion of dark matter.

In \textsf{[}$54,55$] we developed further results by Ranada[$15,16,41,58$].
By skillfully using the electric-magnetic duality for the\ source-free
electromagnetic fields Ranada obtained stable localized solutions for the
electromagnetic fields. In [$54,55$] they were reinterpreted respectively as
Dirac monopoles, electrons and dyons (particles possessing simultaneously
both the electric and magnetic charges). Topologically all these objects are
represented \ either by torus knots or by interlinked (Hopf) rings made of
closed electric (for the electrical monopole being interpreted as stable
electron), or magnetic (the Dirac monopole) \ lines and, for dyons, modelled
by two linked electric and two linked magnetic closed lines. All these
links/knots are pass-equivalent to the trefoil knots which are the simplest
torus knots. More complicated objects are also possible [$55$] but
dynamically they are not stable [55,$63$]. Their lifetimes are finite, as it
happens experimentally.\ All these stable objects (electric and magnetic
monopoles, dyons) are made of null fields [58]. \ Even though they were
discussed in Section 5, they will be further discussed below, in this
section. Ref. [$41$] \ also describes knotted structures which are not null.
These were not discussed by Ranada and they are not of instanton origin [54$%
,64$]. Since the electromagnetic fields are Abelian gauge fields their
description should be made by instanton -type calculations normally used for
description of the non-Abelian gauge fields. These calculations make the
existence of non-null structures [41] both physically and mathematically
questionable. Therefore, they are not discussed further in this work.

The question remains: How do the masses enter into this, purely topological,
picture? This issue was touched upon already by Ranada. In [16] he
demonstrated that his electromagnetic knots do have momentum, charge and
spin. \textsl{That is they have particle-like properties}! These results by
Ranada's were elaborated further in [65,66]. \textsl{Furthermore, Ranada's
knotted structures are made out of \ two complex scalar fields }[$15$]. 
\textsl{These are the same fields as those described in previous sections in
connection with quantum mechanics}! Unlike our works [$54,55$], Ranada's
knots were not obtained as result of instanton calculations. This
circumstance prevented Ranada from seeing general trends- patterns/analogies
between the electromagnetism, hydrodynamics of incompressible fluids,
contact mechanics, the Abelian and non-Abelian gauge fields and gravity. All
these analogies are discussed in our works [$54,55$, and in our book [37]. \
Based on known results from knot theory, we argued that complements of knots
and links make the ambient Minkowski space-time curved in the same way as it
is curved by masses in Einsteinian gravity. That is, in compliance with
general relativity, for knots/links the familiar equation 
\begin{equation}
Spacetime\text{ }Curvature=Matter  \tag{59}
\end{equation}%
should also work. This can be easily understood by studying the dynamics of
point-like masses on the trice punctured sphere. The three punctures make
dynamics on the sphere hyperbolic [$67$]. Accordingly, knots/links make the
fundamental group of space (or space-time) nontrivial thus causing the
effects of curvature. For knots/links the ambient space-times are in one-to
one accord with topologies (fundamental groups) of \ these knots/links.
Therefore, when written mathematically, (59) represents Einstein's equations
of general relativity. In knot theory there are no characteristic scales for
these knots/links (unless they are considered as made out of some physical
material). This property is indicative of conformal invariance. It is
possible to develop theory of gravity also with emphasis on conformal
invariance [68]. Exactly for this reason, knotted and linked objects of
microscopic sizes can be interpreted as particles but knotted/linked
structures at larger scales are being interpreted as space-time
(topological) defects [69]. Since all physical masses should be positive, we
have to keep in mind that not all types of knots/ links are acting like
masses. This physical\ restriction on masses excludes from consideration all
types of hyperbolic knots/ links of which the figure 8 knot is the most
elementary. Incidentally, mathematical methods described in [54,55] for
generation of knots/links tell us that creation of hyperbolic knots/links is
possible only in the presence of boundaries. Details can be found in [70].
In [54,55] all knots/links were generated dynamically. Thus, cosmologically
it is physically plausible that massive links/knots do not require existence
of boundaries for their generation. Introduction of finite size masses into
general relativity as well as of finite size charges into theory of
non-Abelian Yang-Mills (NAYM) fields is associated with the fundamental
technical (theoretical) difficulties. These are described in [37], page 97,
and references therein. In the case of NAYM theories the problem of charges\
is by-passed by treating only the \textsl{source-free} NAYM fields. Thus, if
we are thinking about the NAYM fields as just "more complicated" AYM fields
(represented by the pair of complex source-free fields), then it should not
come out as a surprise that the method of Abelian reduction [71] applied to
the NAYM fields permits us to obtain the Dirac monopoles as exactly those,
known in the AYM theory. These were obtained by Ranada without any reference
\ to the method of Abelian reduction. Ranada obtained these results without
uses of the instanton methods as well while Trautman [$64$] and Kholodenko [$%
54$] obtained these (Dirac) monopoles by instanton methods from the AYM
theory by standard instanton methods [71]. Historically,\ these methods were
used first for obtaining the non-Abelian t'Hooft-Polyakov monopoles. How the
Dirac monopole is obtainable from the t'Hooft-Polyakov monopole using method
of Abelian reduction is explained in detail in [71], page 174. Since the
gravity can be rewritten in the language of NAYM gauge fields as it was
first noticed by Utiyama [71,72] already in 1956, it is only natural that
Robinson, in 1961, was able to find a place for the source-free Maxwell's
null fields\ in gravity[73],[74].

With the background just provided we are now in the position to talk briefly
about the connections between the quantum mechanics and the dark matter.
Very mysteriously the quantum mechanics relates to the dark matter from its
inception. Max Planck's 1900 law of blackbody radiation happens to be
directly connected with the dark matter. This happens because of the
following. 1.The theory of cosmic microwave background is based on the
theory of blackbody radiation. It is fully experimentally supported by this
theory [75]. 2. The experimental data providing information about the dark
matter are taken from the data on cosmic microwave background [76]. 3.The
axions, to be defined below, and in the next section, are directly connected
with the dark matter [77]. 4.Quantization of the source-free electromagnetic
fields makes bosons from these fields. This causes us to think about
superfluidity, the Bose condensation, etc. But, as is well known, the
chemical potential of Plankian photons is zero. Therefore, for Plankian
photons there is no Bose condensation, etc. The situation is not at all as
simple as it looks based on information from students' textbooks. Without
any references to knots, links and foundations of quantum mechanics, such a
conclusion was reached by astrophysicists [78-80].

Thus, it remains to demonstrate that the conclusions of astrophysicists are
strongly backed up by the fundamental laws/principles of quantum mechanics
and electrodynamics. This then will help us to strengthen the arguments \ by
astrophysicists by accounting for knotty nature of the Maxwellian
electrodynamics[15,16,54,55,63,81]. The existing quantization for
electromagnetic fields does not contain any information about knots/links.
Therefore, even though the links between the dark matter and cosmic
microwave background are hinting toward the connections with the Bose
condensation, superfluidity, etc., they leave outside the role of
knots/links in these studies.\bigskip

6.2. Restoring Kelvin's/Thompson's vortex atoms with help

\ \ \ \ \ \ of quantization recommendations by Einstein\bigskip

Results presented in Sections 2-5 suffer from several serious drawbacks,
even though they are fully sufficient for development of Bohmian mechanics
[7]. On one hand, we demonstrated that optical and quantum mechanical
formalisms can be reunited rigorously. On another, such a reunification is
possible only for the source-free Maxwellian fields. This immediately raises
a question: where is the place of mass, charge, angular momentum and spin in
such designed quantum mechanics? The answer thus far was as follows. When
Ranada used for his electromagnetic knots/links source-free scalar fields,
he demonstrated [16] that these knots/links do act as if they have mass,
charge. angular momentum and spin. This is all fine, but these particle-like
objects have life on their own apparently totally disconnected from the
designed quantum mechanical formalism. \ 

To repair this deficiency, it is appropriate to recall some comments by
Einstein in his two lectures at the Institute for Advanced Study in the late
1940's [82], page 383. Einstein noticed that in contemporary quantum theory
we first develop the theory of electrons via the Schr\"{o}dinger equation
and work out its consequences for atomic spectra and chemical bonding with
great success. Then, we develop the theory of free quantized EM field
independently, and discuss it as a separate thing. Only at the end do we,
almost as an afterthought, decide to couple them together by introducing a
phenomenological coupling constant "e" and call the result "Quantum
Electrodynamics". Einstein suggested the way out of this logic. He said:" 
\textbf{I feel that it is} \textbf{a delusion to think of the electrons and
the fields as two physically different, independent entities. Since neither
cannot exist without the other, there is only one reality to be described,} 
\textbf{which happens to have two different aspects; and theory ought to
recognize this from the start instead of doing things twice}." \ Einstein
himself worked on this problem and wrote a paper on this subject in 1919. In
doing so he had Poincar$e^{\prime }$ as his predecessor and Pauli as his
successor [83,84]. Einstein results are nicely summarized in [84], pages
795, 796. His \textsl{purely electromagnetic} \textsl{particle} (PEP) is
made out of source-free Maxwellian fields obeying the Beltrami equation, our
equation (53), that is of source-free electromagnetic fields (1). As it was
shown in [54,55]\ all electromagnetic knots/links are obtained (generated)
with help of this equation are of nonhypebolic nature. Therefore, equation
(59) implies that such PEP's are massive nonhyperbolic knots. In the
presence of boundaries the formation of hyperbolic knots is possible,
however, as discussed in detail in [55] and, differently in [81]. But the
negativity of masses is unphysical and, therefore, such (hyperbolic) knots
must be discarded from consideration. Important additional connections
between knot/link topology and particle masses were developed in [63].

\textsl{Thus, Einstein obtained the correct answer, equation (53), but had
not linked it with the quantum mechanics and with knot/link topology}. \
Such linkages are presented in this paper. Fortunately, there is room,
still, for development further of these results thanks, for instance, to the
series of papers by Sallhofer [85,86] and references therein. No connections
with knots/links was made in his papers, though. We would like now to sketch
the results of [85,86]. \textsl{By doing so, we shall complete the
fundamentals of Einstein quantization \ program }\ and, as by product, we
effectively shall restore (at the more advanced level) seemingly outdated
Kelvin's(Thomson's) theory of vortex atoms [87].

Going back to the section 4 we notice that the complete correspondence
between the source-free Maxwellian electrodynamics and Schrodinger's quantum
mechanics had been achieved due to replacement of the speed of light $c$ by
the speed $u=c/n$ \ in the medium with effective refractive index $n\
=n(x,y,z)$. The same idea can be seen present in the works by Sallhofer
[85,86]. Specifically, instead of our system of Maxwell's equations (1), he
begins with the analogous system: 
\begin{eqnarray}
\mathbf{\nabla }\cdot \varepsilon \mathbf{E} &=&0,\mathbf{\nabla }\times 
\mathbf{E}=-\frac{\mu }{c}\frac{\partial \mathbf{H}}{\partial t},  \nonumber
\\
\mathbf{\nabla }\cdot \mu \mathbf{H} &=&0,\mathbf{\nabla }\times \mathbf{H}=%
\frac{\varepsilon }{c}\frac{\partial \mathbf{E}}{\partial t},\text{ } 
\TCItag{60} \\
\mathbf{\nabla }\cdot \mathbf{E} &=&0,\mathbf{\nabla }\cdot \mathbf{H}=0. 
\nonumber
\end{eqnarray}%
Here the magnetic and electric permittivities $\mu /c$ and $\varepsilon /c$
could be made as functions \ of spatial coordinates. They are playing the
same role as previously defined effective refractive index $n$. The last two
equations emphasize the source -free nature of the electromagnetic fields 
\textbf{E} and \textbf{H}.

Next, introduce the set of Pauli matrices $\mathbf{\vec{\sigma}=(}\sigma _{1}%
\mathbf{,}\sigma _{2}\mathbf{,}\sigma _{3}\mathbf{)}$ in a usual way: 
\begin{equation}
\sigma _{1}=\left( 
\begin{array}{cc}
0 & 1 \\ 
1 & 0%
\end{array}%
\right) ,\sigma _{2}=\left( 
\begin{array}{cc}
0 & -i \\ 
i & 0%
\end{array}%
\right) ,\sigma _{3}=\left( 
\begin{array}{cc}
1 & 0 \\ 
0 & -1%
\end{array}%
\right) .  \tag{61}
\end{equation}%
For any nonzero vector \textbf{A} the following identity holds: 
\begin{equation}
\left( \mathbf{\sigma }\cdot \mathbf{\nabla }\right) (\mathbf{\sigma }\cdot 
\mathbf{A})=\mathbf{1\nabla }\cdot \mathbf{A+}i\mathbf{\sigma }\cdot (%
\mathbf{\nabla }\times \mathbf{A).}  \tag{62}
\end{equation}%
Multiply (via scalar multiplication) the first and the second rows of (60)
by $\mathbf{\sigma }$ and apply the matrix identity (62). The result comes
out as 
\begin{eqnarray}
\left( \mathbf{\sigma }\cdot \mathbf{\nabla }\right) (\mathbf{\sigma }\cdot 
\mathbf{H})-\frac{\varepsilon }{c}\frac{\partial }{\partial t}(i\mathbf{%
\sigma }\cdot \mathbf{E}) &=&0,  \TCItag{63} \\
\left( \mathbf{\sigma }\cdot \mathbf{\nabla }\right) (i\mathbf{\sigma }\cdot 
\mathbf{E})-\frac{\mu }{c}\frac{\partial }{\partial t}(\mathbf{\sigma }\cdot 
\mathbf{H}) &=&0.  \nonumber
\end{eqnarray}%
Using the standard Dirac matrices 
\begin{equation}
\mathbf{\gamma }=\left( 
\begin{array}{cc}
0 & \mathbf{\sigma } \\ 
\mathbf{\sigma } & 0%
\end{array}%
\right)  \tag{64}
\end{equation}%
the above system of equations can be rewritten in a familiar form of the
Dirac equation:%
\begin{equation}
\left[ \mathbf{\gamma \cdot \nabla -}\left( 
\begin{array}{cc}
\varepsilon \mathbf{1} & \mathbf{0} \\ 
\mathbf{0} & \mu \mathbf{1}%
\end{array}%
\right) \frac{1}{c}\frac{\partial }{\partial t}\right] \Psi =0,\text{ } 
\tag{65}
\end{equation}%
\begin{equation}
\Psi =\left( 
\begin{array}{c}
(i\mathbf{\sigma }\cdot \mathbf{E}) \\ 
(\mathbf{\sigma }\cdot \mathbf{H})%
\end{array}%
\right) =\left( 
\begin{array}{cc}
iE_{3} & i(E_{1}-E_{2}) \\ 
i(E_{1}+iE_{2}) & -iE_{3} \\ 
H_{3} & H_{1}-iH_{2} \\ 
H_{1}+iH_{2} & -H_{3}%
\end{array}%
\right) .  \tag{66}
\end{equation}%
Substitution of $\Psi =\psi \exp (-i\omega t)$ into (65) produces%
\begin{equation}
\left[ \mathbf{\gamma \cdot \nabla +}i\frac{\omega }{c}\left( 
\begin{array}{cc}
\varepsilon \mathbf{1} & \mathbf{0} \\ 
\mathbf{0} & \mu \mathbf{1}%
\end{array}%
\right) \right] \Psi =0.  \tag{67}
\end{equation}%
The obtained (Dirac) equation is the relativistic analog of (TISE) equation
(42). Following the same logic as in the nonrelativistic case, the analog of
equation (42) can now be written as follows%
\begin{equation}
\left[ \mathbf{\gamma \cdot \nabla +}i\frac{\omega }{c}\left( 
\begin{array}{cc}
(1-\frac{V-mc^{2}}{\hbar \omega })\mathbf{1} & \mathbf{0} \\ 
\mathbf{0} & (1-\frac{V+mc^{2}}{\hbar \omega })\mathbf{1}%
\end{array}%
\right) \right] \Psi =0.  \tag{68}
\end{equation}%
The detailed calculations leading to known Dirac spectrum for the
relativistic electron can be found in [88]. It should be noted that the idea
of relating the Dirac and Maxwell's equations can be traced back to 1931. In
this year two fundamental papers were published in a sequel. One, [89]-by
Oppenheimer and, another, [90]-by Laporte and Uhlenbeck. The results just
presented, in effect are abbreviated versions of just cited papers. Other,
similar papers will be discussed in the next section. The references [85,86]
were used instead only because they are immediately linked with the Beltrami
equation (53). \ Since this equation links\ the Maxwell and Dirac equations
with knots/links, we just executed Einstein's quantization program.\bigskip

6.3. Chladni patterns in the sky caused by axions of dark matter\bigskip

It is believed that some pseudo-scalar particles, called "\textsl{axions}"
[91],[92], are directly responsible for measurable effects attributed to the
dark matter. Incidentally, the same particles are also believed to play an
essential role in the properties of topological insulators [92],[93]. In
view of results presented in Sections 5 and 6, we would like to make several
comments about the axion electrodynamics following, in part, [91].

We begin with the observation that source-free Maxwell's electrodynamics
possesses a special kind of symmetry - the electric-magnetic duality. That
is to say, equations (1) remain form-invariant with respect to
transformations (duality rotations)%
\begin{equation}
\left[ 
\begin{array}{c}
\mathbf{E}^{\prime } \\ 
\mathbf{H}^{\prime }%
\end{array}%
\right] =\left( 
\begin{array}{cc}
\cos \xi & \sin \xi \\ 
-\sin \xi & \cos \xi%
\end{array}%
\right) \left[ 
\begin{array}{c}
\mathbf{E} \\ 
\mathbf{H}%
\end{array}%
\right] \text{.}  \tag{69}
\end{equation}%
Here $\xi $ is an arbitrary angle. The presence of sources and/or axions
formally destroys this symmetry. Nevertheless, it will be eventually
restored in the way consistent with the results of section 5. The axions can
be introduced as follows. The four-dimensional electromagnetic Lagrangian $%
\mathcal{L}_{0}$ is given in the standard form as 
\begin{equation}
\mathcal{L}_{0}=-\frac{1}{4\mu _{0}}F^{\mu \nu }F_{\mu \nu }-A_{\mu
}J_{e}^{\mu },  \tag{70}
\end{equation}%
where, as usual, $F^{\mu \nu }=\partial ^{\mu }A^{\nu }-\partial ^{\nu
}A^{\mu },$ and the electric current $J_{e}^{\mu }=(c\rho _{e},\mathbf{J}%
_{e}).$ Suppose, that there are magnetic monopoles, then one can introduce
the magnetic charge density $\rho _{m}$ and the magnetic current $\mathbf{J}%
_{m}$ so that $J_{m}^{\mu }=(c\rho _{m},\mathbf{J}_{m}).$ Maxwell's
equations accounting for the electric and magnetic charges can be written
now as%
\begin{eqnarray}
\partial _{\mu }F^{\mu \nu } &=&\mu _{0}J_{e}^{\mu },  \nonumber \\
\partial _{\mu }\tilde{F}^{\mu \nu } &=&\mu _{0}J_{m}^{\mu }/c,  \TCItag{71}
\\
\tilde{F}^{\mu \nu } &=&\frac{1}{2}\epsilon ^{\mu \nu \sigma \rho }F_{\sigma
\rho }.  \nonumber
\end{eqnarray}%
Clearly, in the presence of magnetic monopoles along with electric charges
Maxwell's electrodynamics will regain the duality again. The situation
changes when the Lagrangian $\mathcal{L}_{0}$ (with electric and magnetic
charges) is extended by adding to it the axion-like interaction term%
\begin{equation}
\mathcal{L}_{\theta }=-\frac{\kappa }{c\mu _{0}}\theta (x)\mathbf{E\cdot
H\equiv }\frac{\theta (x)\kappa }{4\mu _{0}}\tilde{F}^{\mu \nu }F_{\mu \nu }.
\tag{72}
\end{equation}%
Here, $\theta =\theta (x)$ is the pseudoscalar field representing axions.
The total Lagrangian becomes: 
\begin{equation}
\mathcal{L}^{T}=\mathcal{L}_{0}+\mathcal{L}_{\theta }+\mathcal{L}_{a}=-\frac{%
1}{4\mu _{0}}F^{\mu \nu }F_{\mu \nu }+\frac{\theta (x)\kappa }{4\mu _{0}}%
F_{\mu \nu }\tilde{F}^{\mu \nu }-A_{\mu }J_{e}^{\mu }+\mathcal{L}_{a} 
\tag{73}
\end{equation}%
where,%
\begin{equation}
\mathcal{L}_{a}=\frac{1}{2}[\partial ^{\mu }\theta \partial _{\mu }\theta
-m^{2}\theta ^{2}].  \tag{74}
\end{equation}

\bigskip The minimization of $\mathcal{L}^{T}$ leads to the following set of
equations (in terms of \textbf{E} and \textbf{H} variables)%
\begin{eqnarray}
\nabla \cdot (\mathbf{E}-c\kappa \theta \mathbf{H}) &=&\rho _{e}/\varepsilon
_{0},  \nonumber \\
\nabla \cdot (c\mathbf{H}+\kappa \theta \mathbf{E}) &=&c\mu _{0}\rho _{m}, 
\nonumber \\
\nabla \times (c\mathbf{H}+\kappa \theta \mathbf{E)}\mathbf{=\partial }
&&_{t}(\mathbf{E}-c\kappa \theta \mathbf{H)/}c+c\mu _{0}\mathbf{J}_{e}, 
\TCItag{75} \\
\nabla \times (\mathbf{E}-c\kappa \theta \mathbf{H)}\mathbf{=-\partial }
&&_{t}(c\mathbf{H}+c\kappa \theta \mathbf{H)/}c-\mu _{0}\mathbf{J}_{m}, 
\nonumber \\
\square \theta &=&-\frac{\kappa }{\mu _{0}c}\mathbf{E\cdot H-}m^{2}\theta . 
\nonumber
\end{eqnarray}%
The system of equations (75) possesses the electric -magnetic duality
symmetry. Using (69) and rotating the fields \textbf{E} and \textbf{H} it is
possible to find such an angle $\tilde{\xi}$ that $\mathbf{J}_{m}$ is
eliminated. Thus, our arguments become independent upon the existence of
Dirac monopoles. Working with such calibrated fields, and applying the curl
operator to the 3rd and 4th equations in (75), while selecting to work with
the source-free electromagnetic fields ($\mathbf{J}_{e}=0),$ we end up with
the following system of equations%
\begin{eqnarray}
\nabla \cdot \mathbf{\hat{E}} &=&0,  \nonumber \\
\nabla \cdot \mathbf{\hat{H}} &=&0,  \nonumber \\
\square \mathbf{\hat{E}}\mathbf{=0,} &&  \TCItag{76} \\
\square \mathbf{\hat{H}}\mathbf{=0,} &&  \nonumber \\
\left( \square +m^{2}\right) \theta &=&0.  \nonumber
\end{eqnarray}%
Here we put $c=1$, $\mathbf{\hat{E}}=\mathbf{E}-\kappa \theta \mathbf{H,\hat{%
H}=H}+\kappa \theta \mathbf{E,}$ and had selected only the null
electromagnetic fields for which $\mathbf{E\cdot H=}0\mathbf{.}$ It is clear
then that the first four equations are exactly the same as equations
(52),(53) and (54). In view of results of Section 5, in which we discussed
the Huygens equivalence, as well as in view of equation (57), the last of
equations of (76) is the same as previously derived equation $(\nabla
^{2}+k^{2}\mathbf{)(r\cdot v)=0,}$ while $\nabla \cdot \mathbf{v=}0$ in (57)
is the same as the first two of equations in (76). \ Thus, we just
demonstrated that the system of equations (76) coincides exactly with the
system of equations presented at the end of Section 5. Thus, axions are
describing knotty Chladni (dark matter) patterns in the sky. \bigskip

\textbf{7. \ Attempt at synthesis: inclusion of chirality \bigskip }

7.1. General comments\medskip

Recall, that the chirality originates from the lack of spatial reflection
symmetry of the system in question. What does this subject has to do with
what was said above ? In this section we \ make an attempt to complete
Einstein's program of quantization outlined in the previous section. This
means that Einstein's \textsl{purely electromagnetic} \textsl{particle}
(PEP) should be made out of source-free Maxwellian fields obeying the
Beltrami equation, our equation (53). As described in Section 5, this
equation is equivalent\ to the system of source-free Maxwell's equations
(1). The Beltrami equation (53) is one of the major equations in contact
geometry [37]. And, beginning with our work [54], we demonstrated in detail
that the Beltrami equation is involved in description of all kinds of torus
knots. It is known [94] that all torus knots are chiral. The simplest chiral
torus knot is trefoil knot. The chirality associated with Beltrami equation
can be spotted already at the lowest level of instanton calculations [54,
95]. The relevance of chirality of torus knots to elementary particle
physics was recognized already by Sakharov in late 1960ies [96]. Since the
Beltrami equation was used in previous sections, we now need to explain why
uses of this equation still require more details.

We begin by posing a question: \textsl{If in sections 2-4 and 6 we were able
to map the source-free Maxwell's equations into the Schr\"{o}dinger (or
Dirac) equation, was this mapping surjective}, \textsl{injective or bijective%
}? \ Fortunately, this question was studied already in [97-99]. Imposition
of the requirement of local gauge invariance on the Schr\"{o}dinger (or
Dirac) equation and requirement of relativistic covariance of the emerging
source-free Maxwell's equations as result of this imposition, makes such a
mapping bijective.

The obtained answer still requires explanations. First, mathematically
results of [97-99] can be classified only as a proof of existence. They need
to be contrasted with the constructive results of sections 2-6. In these
sections we built this mapping explicitly but injectivily. Some practical
applications resulting from this injectivity are described in [100]. Second,
it is time now to demonstrate that the mapping is indeed bijective.\medskip

7.2. The proof of bijectivity of Maxwell-Dirac mapping\medskip

Very fortunately, this task was completed for the massless Dirac equation.
To our knowledge there are no results for the massive Dirac equation with
one exception [101] known to us. If $m$ is the mass parameter in the Dirac
equation, then results formally obtainable in $m\rightarrow 0$ limit are
associated with neutrino physics [102]. \ From the very beginning of
invention of neutrino theory by Pauli, it was known that, like photon,
neutrino is also neutral. Thus, in the limit $m\rightarrow 0$ the charged
Dirac particle becomes neutral. There is no difficulty to understand this
result if $m=0$. Nature, however, is more intricate. It\ prepared surprises
for us. First, the mass of neutrino is small but nonzero. This fact has no
explanation within the scope of the Standard Model. Second, there are three
types of neutrinos-all having small but different masses. There are no
explanations within the Standard Model of why these masses must be different
or why these masses should be small but nonzero. \ Third, while the Dirac
electron has nonzero electric charge, the massive neutrino(s) is neutral
[102]. Because the neutrino is neutral, it interacts with the
electromagnetic field quantum mechanically through radiation corrections
only. Just stated bijectivity [101] permits us to treat the Dirac mass as an
adjustable parameter. The absence of charge for the \textbf{massive}
neutrino presents much harder theoretical obstacles for implementation of
the Dirac-Maxwell bijectivity. This circumstance puts the whole\ Einstein's
quantization program into jeopardy. The situation\ is repairable to a some
extent by noticing that both the neutrinos ($\nu _{e},\nu _{\mu },\nu _{\tau
})$ and the leptons ($e$, $\mu ,\tau )$ associated with them are fermions.
Here we are dealing with the peculiar situation of having massless photons
(bosons) and massive neutrinos (fermions) both as \ neutral particles.
Furthermore, both leptons and neutrinos are chiral particles (e.g. read
below), both have \textbf{the same} chirality (this is experimental fact,
there are no theory explaining this ). Thus, by using the Dirac mass $m$ as
parameter in the Dirac equation we are are confronted with the following 
\textsl{problems}:

a) When treated as a parameter, at what stage the limiting $m\rightarrow 0$
process loses its continuity, that is, when the massive electron $e$ (or $%
\mu ,\tau $ leptons$)$ abruptly loses its charge and becomes chargeless
massive neutrino? b) Being chargeless, the neutrinos obtained from the
massive Dirac-like leptons will remain the Dirac-like or the Majorana-like
fermions?

It is well known [102] that the Majorana -like fermions are neutral by
design. Thus, it appears that, all neutrinos should be of Majorana type. The
experiment is not showing this directly, however, because of the phenomena
of neutrino mixing and neutrino oscillations.\ This circumstance makes
implementation of Einstein's quantization program very difficult. The
situation is repairable to a some extent, nevertheless, as it will be
explained below. After these comments, we are ready to demonstrate in some
detail the Maxwell-Dirac and the Maxwell-Majorana bijectivity.

Our readers might object at this point of exposition. If this is so, what is
wrong with the results presented in subsection 6.2? The answer is: These
results are injective only. The bijective treatment in [101] is not readily
suitable for treatment of Majorana neutrinos. The bijective correspondence
discussed in [97-99] cannot be used for the case of neutral Majorana
neutrinos. Thus, more work is needed in the future.

To begin, following [33], page100, we notice that every component $\psi
_{\sigma },$ $\sigma =0\doteqdot 3,$of the spinor $\psi $ of the massive
Dirac equation, is satisfying the massive K-G equation. From Section 5 we
know that, using the Hadamard transformation, the massive K-G can be
reversibly transformed into the D'Alembert equation. The massless K-G (that
is the D'Alembert) equation is obtainable from the massless Dirac equation ($%
\hbar =1,c=1)$%
\begin{equation}
i\dsum\limits_{i=0}^{3}\gamma ^{i}\frac{\partial }{\partial x^{i}}\psi =0. 
\tag{77}
\end{equation}%
Here the Dirac matrices are given in their standard form [28]%
\begin{equation}
\gamma ^{0}=\left( 
\begin{array}{cc}
\mathbf{I} & \mathbf{0} \\ 
\mathbf{0} & -\mathbf{I}%
\end{array}%
\right) ,\gamma ^{k}=\left( 
\begin{array}{cc}
\mathbf{0} & \sigma ^{k} \\ 
-\sigma ^{k} & \mathbf{0}%
\end{array}%
\right) ,k=1\doteqdot 3,\gamma ^{\mu }\gamma ^{\nu }+\gamma ^{\nu }\gamma
^{\mu }=2g^{\mu \nu },\mu ,\nu =0\doteqdot 3,  \tag{78}
\end{equation}%
and the Pauli matrices are also given in their standard form as%
\begin{equation}
\sigma ^{1}=\left( 
\begin{array}{cc}
0 & 1 \\ 
1 & 0%
\end{array}%
\right) ,\sigma ^{2}=\left( 
\begin{array}{cc}
0 & -i \\ 
i & 0%
\end{array}%
\right) ,\sigma ^{3}=\left( 
\begin{array}{cc}
1 & 0 \\ 
0 & -1%
\end{array}%
\right) .  \tag{79}
\end{equation}%
Using these standard notations and choosing%
\begin{equation}
\psi =\left( 
\begin{array}{c}
iH^{3} \\ 
iH^{1}-H^{2} \\ 
E^{3} \\ 
iE^{2}+E^{1}%
\end{array}%
\right)  \tag{80}
\end{equation}%
in equation (77) converts this (Dirac) equation into the system of Maxwell's
equations (1) [103]. The wave function (80) is not the only one converting
the massless Dirac equation into the system of source-free Maxwell's
equations. Following [103] we write down the set of all wave functions
accomplishing the same task. These are: 
\begin{eqnarray}
\chi _{1} &=&\left( 
\begin{array}{c}
iH^{3} \\ 
iH^{1}-H^{2} \\ 
E^{3} \\ 
iE^{2}+E^{1}%
\end{array}%
\right) ,\chi _{2}=\left( 
\begin{array}{c}
-iE^{3} \\ 
-iE^{1}+E^{2} \\ 
H^{3} \\ 
iH^{2}+H^{1}%
\end{array}%
\right) ,\chi _{3}=\left( 
\begin{array}{c}
E^{2}+iE^{1} \\ 
-iE^{3} \\ 
-H^{1}-iH^{2} \\ 
H^{3}%
\end{array}%
\right) ,  \nonumber \\
\chi _{4} &=&\left( 
\begin{array}{c}
iH^{1}+H^{2} \\ 
-iH^{3} \\ 
-iE^{2}+E^{1} \\ 
-E^{3}%
\end{array}%
\right) ,\chi _{5}=\left( 
\begin{array}{c}
-iH^{3} \\ 
-H^{1}-iH^{2} \\ 
iE^{3} \\ 
-iE^{2}+iE^{1}%
\end{array}%
\right) ,\chi _{6}=\left( 
\begin{array}{c}
iE^{3} \\ 
iE^{1}+E^{2} \\ 
iH^{3} \\ 
-iH^{2}+iH^{1}%
\end{array}%
\right) ,  \nonumber \\
\chi _{7} &=&\left( 
\begin{array}{c}
iE^{2}-E^{1} \\ 
E^{3} \\ 
-iH^{1}-H^{2} \\ 
iH^{3}%
\end{array}%
\right) ,\chi _{8}=\left( 
\begin{array}{c}
+iH^{2}-H^{1} \\ 
H^{3} \\ 
E^{2}+iE^{1} \\ 
-iE^{3}%
\end{array}%
\right) .  \TCItag{81}
\end{eqnarray}%
Just described wave functions realate to each other group-theoretically.
Details are provided in [103].

\medskip

\bigskip 7.3. \ Some facts about chirality\medskip

The 4-components function $\psi $ in (77) is splitable into two
two-component (Weyl) equations typically used in the theory of massless
neutrino [28], [102]. This is being achieved via introduction of chirality
operators. In notations of [28] these operators can be defined as follows.
First, let $\gamma _{5}=-i\gamma _{0}\gamma _{1}\gamma _{2}\gamma _{3},$
then the projection operators are defined via $P_{L}=\frac{1}{2}(1-\gamma
_{5}),P_{R}=\frac{1}{2}(1+\gamma _{5})$ so that we get%
\begin{equation}
\gamma _{5}=-\left( 
\begin{array}{cc}
\mathbf{0} & \mathbf{I} \\ 
\mathbf{I} & \mathbf{0}%
\end{array}%
\right) ,  \tag{82}
\end{equation}%
and%
\begin{equation}
P_{L}P_{R}=P_{R}P_{L}=0,P_{L}+P_{R}=1,P_{L}^{2}=P_{L},P_{R}^{2}=P_{R}. 
\tag{83}
\end{equation}%
If $\psi $ is the Dirac spinor, then:%
\begin{equation}
\psi _{L}=P_{L}\psi ,\psi _{R}=P_{R}\psi ,P_{L}\psi _{R}=P_{R}\psi _{L}=0, 
\tag{84}
\end{equation}%
and 
\begin{equation}
\gamma _{5}\psi _{L},_{R}=\pm \psi _{L},_{R}.  \tag{85}
\end{equation}%
Clearly, because of (85), it makes sense to call $\gamma _{5}$ as the 
\textsl{chirality operator}. The eigenvalues $\pm 1$ are the chirality
eigenvalues and $\psi _{L},_{R}$ are the chiral projections. Using these
results, the Dirac equation can be rewritten\ now in the manifestly chiral
form. Indeed, since 
\begin{equation}
\psi =\left( 
\begin{array}{c}
\psi _{a} \\ 
\psi _{b}%
\end{array}%
\right) ,\text{ with }\psi _{a}=\left( 
\begin{array}{c}
\psi _{1} \\ 
\psi _{2}%
\end{array}%
\right) ,\psi _{b}=\left( 
\begin{array}{c}
\psi _{3} \\ 
\psi _{2}%
\end{array}%
\right)  \tag{86}
\end{equation}%
and, using (82), we can write: 
\begin{equation}
\psi _{L},_{R}=P_{L,R}\psi =\frac{1}{2}(1\pm \gamma _{5})\psi =\frac{1}{2}%
\left( 
\begin{array}{cc}
I & \mp I \\ 
\mp I & I%
\end{array}%
\right) \left( 
\begin{array}{c}
\psi _{a} \\ 
\psi _{b}%
\end{array}%
\right) .  \tag{87}
\end{equation}%
Thus, 
\begin{equation}
\psi _{L}=\frac{1}{2}\left( 
\begin{array}{c}
\psi _{a}-\psi _{b} \\ 
-\psi _{a}+\psi _{b}%
\end{array}%
\right) =\left( 
\begin{array}{c}
\phi \\ 
-\phi%
\end{array}%
\right) ,\text{ where }\phi =\frac{1}{2}(\psi _{a}-\psi _{b})=\frac{1}{2}%
\left( 
\begin{array}{c}
\psi _{1}-\psi _{3} \\ 
\psi _{2}-\psi _{4}%
\end{array}%
\right) ,  \tag{88}
\end{equation}%
\begin{equation}
\text{and, }\psi _{R}=\frac{1}{2}\left( 
\begin{array}{c}
\psi _{a}+\psi _{b} \\ 
\psi _{a}+\psi _{b}%
\end{array}%
\right) =\left( 
\begin{array}{c}
\chi \\ 
\chi%
\end{array}%
\right) ,\text{ where }\chi =\frac{1}{2}(\psi _{a}+\psi _{b})=\frac{1}{2}%
\left( 
\begin{array}{c}
\psi _{1}+\psi _{3} \\ 
\psi _{2}+\psi _{4}%
\end{array}%
\right) .  \tag{89}
\end{equation}%
With help of these results equation (77) can now be rewritten as 
\begin{eqnarray}
E\phi &=&-\mathbf{\sigma }\cdot \mathbf{p}\phi ,  \TCItag{90} \\
E\chi &=&+\mathbf{\sigma }\cdot \mathbf{p}\chi .  \nonumber
\end{eqnarray}%
Obtained result is prompting us to introduce the helicity operator $\mathcal{%
H}:$%
\begin{equation}
\mathcal{H=}\frac{\mathbf{\sigma }\cdot \mathbf{p}}{\left\vert \mathbf{p}%
\right\vert }.  \tag{91}
\end{equation}%
Thus, $\psi _{L}$ is the eigenspinor representing the helicity $\mathcal{H}%
=+1$ for particles and $\mathcal{H}=-1$ for antiparticles. Accordingly, $%
\psi _{R}$ is the eigenspinor representing the helicity $\mathcal{H}=-1$ for
particles and $\mathcal{H}=+1$ for antiparticles. When $m>0$, the chirality
eigenspinors $\psi _{R}$ and $\psi _{L}$ no longer \ describe particles with
fixed helicity. In this case the helicity is no longer a good quantum number.

The two-component theory of Dirac-type electron neutrino $\nu _{e}$ implies
that the spinor $\psi _{\nu }$ representing neutrino $\nu $ and
participating in the weak interactions is \textbf{always} in the form:%
\begin{equation}
\psi _{\nu _{e}}=P_{L}\psi .  \tag{92}
\end{equation}%
The Dirac electron associated with $\nu _{e}$ is described by the left
handed spinor.\medskip

7.4. \ Some facts about the charge conjugation, parity, time reversal
\medskip and

\ \ \ \ \ \ \ the Majorana condition\medskip

Our readers might notice already that the massive Dirac equation does not
carry explicitly the information about the charge $\mp q$ of
electron/positron. The simplest way to detect the presence of a charge is to
turn on the external electromagnetic field $A_{\mu }$. That is to study the
equation:%
\begin{equation}
(i\dsum\limits_{i=0}^{3}\gamma ^{i}\frac{\partial }{\partial x^{i}}\psi \mp
q\gamma ^{\mu }A_{\mu }-m)=0.  \tag{93}
\end{equation}%
For the massive Majorana neutrino the charge $\mp q$ is zero. Moreover, the
bijective Maxwell-Dirac (actually, the Maxwell-Majoranain the present case)
isomorphism [97-99] breaks down instantly as soon as the external
electromagnetic field is turned on. Thus, the major problem is the following.

\textsl{If Einstein's quantization program involves knotted/linked
structures representing particles, then a) how these structures interact
with the electromagnetic field and, b) how stable these structures are with
or without presence of electromagnetic field?}

An attempt to answer these questions is presented in Section 8. In the
meantime, we return to the topics of this subsection. The adjoint $\bar{\psi}
$ to Dirac spinors is defined by%
\begin{equation}
\bar{\psi}(x)\equiv \psi ^{\dagger }(x)\gamma ^{0}.  \tag{94}
\end{equation}%
The charge conjugation $\mathit{C}$ is defined as%
\begin{eqnarray}
\psi ^{C}(x) &=&\xi _{C}C\bar{\psi}^{T}(x)=-\xi _{C}\gamma ^{0}C\psi ^{\star
}(x),  \nonumber \\
C\gamma _{\mu }^{T}C^{-1} &=&-\gamma _{\mu ,}C^{\dagger
}=C^{-1},C^{T}=-C,C(\gamma ^{5})^{T}C^{-1}=\gamma ^{5}.  \nonumber \\
\xi _{C}C\bar{\psi}^{T}(x) &\rightarrow &^{C}\left\vert \xi _{C}\right\vert
^{2}\psi (x),\left\vert \xi _{C}\right\vert ^{2}=1.  \TCItag{95}
\end{eqnarray}%
Under the parity \textit{P} transformation (space mirror image inversion) 
\textit{x}$^{\mu }=(x^{0},\vec{x})\rightarrow ^{P}x_{P}^{\mu }=(x^{0},-\vec{x%
}).$ The spinor field $\psi (x)$ transforms as [102], page 52, 
\begin{equation}
\psi ^{P}(x_{P})=\xi _{P}\gamma ^{0}\psi (x),\xi _{P}=\pm 1,\pm i.  \tag{96}
\end{equation}%
Under the time reversal \textit{T} transformation \textit{x}$^{\mu }=(x^{0},%
\vec{x})\rightarrow ^{T}x_{T}^{\mu }=(-x^{0},\vec{x}).$The spinor field $%
\psi (x)$ transforms as [102], page 55, 
\begin{equation}
\psi ^{T}(x_{T})=\xi _{T}\gamma ^{0}\gamma ^{5}C\bar{\psi}^{T}(x)=\xi
_{T}\gamma ^{5}C\psi ^{\star }(x),\left\vert \xi _{T}\right\vert ^{2}=1. 
\tag{97}
\end{equation}%
With these definitions, \textsl{the Majorana condition} is given by%
\begin{equation}
\psi ^{C}(x)=\psi (x).  \tag{98}
\end{equation}%
The physics of this condition is easily checkable on example of the
electromagnetic current $J=q\bar{\psi}(x)\gamma ^{\mu }\psi (x).$ Indeed, we
have 
\[
\bar{\psi}(x)\gamma ^{\mu }\psi (x)=-\psi (x)^{T}C^{\dagger }\gamma ^{\mu }C%
\bar{\psi}^{T}(x)=\bar{\psi}(x)C\left( \gamma ^{\mu }\right) ^{T}C^{\dagger
}\psi (x)=-\bar{\psi}(x)\gamma ^{\mu }\psi (x)=0. 
\]%
Using (95) and (98) we obtain,%
\begin{equation}
\psi (x)=-\xi _{C}\gamma ^{0}C\psi ^{\star }(x)  \tag{99}
\end{equation}%
and, because [102], page 48,%
\begin{equation}
C=-i\left( 
\begin{array}{cc}
\mathbf{0} & \sigma ^{2} \\ 
-\sigma ^{2} & \mathbf{0}%
\end{array}%
\right) \text{ so that }\gamma ^{0}C=-i\left( 
\begin{array}{cc}
\mathbf{0} & \sigma ^{2} \\ 
\sigma ^{2} & \mathbf{0}%
\end{array}%
\right) .  \tag{100}
\end{equation}%
By selecting $\xi _{C}=i$ the equation (99) acquires the following form%
\begin{equation}
\psi (x)\equiv \left( 
\begin{array}{c}
\phi _{1} \\ 
\phi _{2}%
\end{array}%
\right) =\left( 
\begin{array}{cc}
\mathbf{0} & \sigma ^{2} \\ 
\sigma ^{2} & \mathbf{0}%
\end{array}%
\right) \left( 
\begin{array}{c}
\phi _{1}^{\ast } \\ 
\phi _{2}^{\ast }%
\end{array}%
\right) =\left( 
\begin{array}{c}
\sigma ^{2}\phi _{2}^{\ast } \\ 
\sigma ^{2}\phi _{1}^{\ast }%
\end{array}%
\right) ,\phi _{1}=\left( 
\begin{array}{c}
\varphi _{11} \\ 
\varphi _{12}%
\end{array}%
\right) ,\phi _{2}=\left( 
\begin{array}{c}
\varphi _{21} \\ 
\varphi _{22}%
\end{array}%
\right) .  \tag{101}
\end{equation}%
In view of the definition of charge conjugation (95), this operation
commutes with the Dirac operator defined in (77). Therefore, its \ action on
the equation (101) produces again the set of Maxwell's source-free
equations. Evidently, the equation (101) is just the duality rotation (69).
The presented result "apparently" completes Einstein's quantization program.
The explanation of quotation marks will be given in Section 8. To finish
this section, we need to explain the physical meaning of the duality
rotation in the present case. To do so we need to return to [103] and to
point out that the duality \ rotation is one of the symmetries of Maxwell
(or massless Dirac) equations. As it is known, with each symmetry associated
the conservation law. Accordingly, the uncovered rotation symmetries\ lead
to conservation laws carrying useful information about the system. At the
physical level of rigor, such (duality) symmetries were discussed in detail
in [104]. The discovered new invariants found their macroscopic uses in
studies of electromagnetically chiral media [105], page 7. How the
macroscpic chiral Maxwell's equations originate microscopically in the
chiral media is nicely explained in [106]. Macroscopically, in the absence
of chirality, we have 
\begin{equation}
\mathbf{D}=\varepsilon \mathbf{E,B=\mu H.}  \tag{102}
\end{equation}%
In the presence of chrality we have instead, [106], page 66, the
Drude-Born-Fedorov (DBF) equations 
\begin{eqnarray}
\mathbf{D} &=&\varepsilon (\mathbf{E}+\beta \mathbf{\nabla }\times \mathbf{E}%
),  \nonumber \\
\mathbf{B} &=&\mu (\mathbf{H+}\beta \mathbf{\nabla }\times \mathbf{H).} 
\TCItag{103}
\end{eqnarray}%
The effects of chirality in this formalism are being controlled by the
parameter $\beta .$ There is yet another set of equations, [106], page 69,%
\begin{eqnarray}
\mathbf{D} &=&\hat{\varepsilon}\mathbf{E+}i\xi \mathbf{B,}  \nonumber \\
\mathbf{H} &=&\left( 1/\hat{\mu}\right) \mathbf{B}+i\eta \mathbf{E.} 
\TCItag{104}
\end{eqnarray}%
These are known as chiral constitutive equations (ChC eq.s). The second ones
might be related to the first ones but are being considered as more
fundamental. The chirality built into neutrino formalism just described can
be used to recover the ChC equations [107].\bigskip

\textbf{8. Completion of the Einstein quantization program\bigskip }

8.1. Conformal invariance, Lie sphere geometry, Dupin

\ \ \ \ \ \ cyclides and Chladni patterns

\ \ \ \ \ \ 

Bateman and Cunnigham discovered in 1910 that Maxwell's equations (even with
sources) are invariant under the action of conformal group SO(4,2) of
Minkowski space. A summary of their achievements is presented in [108]. From
this reference we also find a link to the paper by Gross [109]. He proved
that solutions of source-free Maxwell's equations provide unitary
representations of the conformal group of Minkowski space and then extended
these results to other massless relativistic equations using the
Bargmann-Wigner [110] description of particles with discrete spin. In [108]
these results were considerably simplified culminating with a proof that any
zero-mass, discrete spin representation of the Poincar$e^{\prime }$ group
admits an unitary representation of the conformal group. The group SO(4, 2)
plays the central role in the Lie sphere geometry. A concise introduction to
this topic is given in our work [40], section 7. In the field of our study,
the Lie sphere geometry reveals itself in the form of \textsl{Dupin cyclides}%
. These are invariants of the conformal group SO(4, 2), that is of the Lie
sphere geometry [111],[112]. Within the context of hyperbolic (wave-like)
equations the Dupin cyclides were discovered by Friedlander in 1946. Within
the context of quantum mechanics they were discovered in our work [40].
Brief but informative information about Dupin cyclides is contained in
[113]. From this reference it follows that the simplest Dupin cyclides are
planes, spheres, cylinders, cones and tori. In [114] the method of designing
more complicated Dupin cyclides\ from cylinders, tori and cones by applying
the M\"{o}bius inversion is described in detail.

In the next subsection we shall discuss some results coming from the use of
the AdS-CFT correspondence. To our knowledge, no mention exists in physics
literature about the connections between the AdS/CFT and the Lie sphere
geometry. A collection of mathematically rigorous results on this topic is
provided in [115]. Here, by combining some results from our work [40], and
from [115], we provide the absolute essentials. We begin with the \textit{%
Poincare' disc} model \textbf{D}$^{2}$ of the hyperbolic space \textbf{H}$%
^{2}.$ Geodesics in this model are made of horocycles.\textit{\ }These are
the circular segments whose both ends lie at the circular boundary\textit{\ }%
$\mathit{S}_{\infty }^{1}$\textit{\ }of \textbf{D}$^{2}.$ The boundary is
considered as a "spatial infinity." By appropriately choosing constants $%
a,b,c,d$ \ in the M\"{o}bius transformation%
\[
f(z)=(az+b)/(cz+d),z\in \mathbf{C}=R^{2}\cup \{\infty \}, 
\]%
the disc model \textbf{D}$^{2}$ can be transformed into the Poincar$%
e^{\prime }$ half plane model of hyperbolic space \textbf{H}$^{2}.$ This
two-dimensional model is generalizable to higher dimensions, where it is
known as the \textit{hyperbolic ball model}. Thus, the two dimensional
combination (\textbf{H}$^{2},$\textit{S}$_{\infty }^{1})$ is being replaced
by (\textbf{H}$^{n+1},$\textit{S}$_{\infty }^{n})\ $\ in spaces of higher
dimensions. The idea of AdS/CFT can be seen already in 2 dimensions. In it,
the deformations of \textit{S}$_{\infty }^{1}$ (of the boundary) leads to
the Virasoro algebra and, hence, to 2-dimensional conformal field theories.
By analogy with 2 dimensions, it was expected in higher dimensions that the
hyperbolic-like behavior in anti-de Sitter spacetime (AdS) is affected by
(linked with) the conformal field theory (CFT) residing at its boundary. The
Mostow rigidity theorem makes the deformations of conformal 3-sphere $%
\mathit{S}_{\infty }^{3}$ impossible to perform. To bypass this difficulty
[115] the hyperbolic space \textbf{H}$^{n}$ is replaced by the anti-de
Sitter space-the Einstein space of constant negative curvature. This space
is obtained from the vacuum Einstein equations with added (negative)
cosmological constant. By doing so, the hyperbolic sphere at infinity 
\textit{S}$_{\infty }^{n}$ is replaced by the space of conformally flat
solutions Ein$^{n,1}$of Einstein's equations. Being conformally flat, these
are conformally equivalent to the Minkowski spacetime. That is the spacetime
described by conformal symmetry group SO(4, 2). The objects living in such
spacetimes are invariants of the Lie sphere geometry [115],[116].

Being armed with such information, we would like to reobtain the results of
Bateman, Cunningham and Ranada [15,16] from the point of view of Dupin
cyclides. In Section 5 we demonstrated that the presence of Beltrami
equations (53) implies the existence of Chladni patterns in the form of
torus knots. By definition, the torus knots are living on surfaces of
toruses, that is on Dupin cyclides. Since equations (53) are equivalent to
the set of source -free Maxwellian equations (1) and since the Dupin
cyclides are invariants of the conformal group SO(4,2), we just proved the
main results of Bateman and Cunningham. Since in his works [15,16] Ranada
was also using the source-free Maxwell's equations, just obtained results
are consistent with those by Ranada. Few additional comments are required,
though. They will be supplied in the rest of this section. The results, just
described, lead to Chladni patterns, while Ranada's knots are \textsl{real}
(not empty spaces). The question arises: Is there a way to correct this
situation? The answer is:"Yes, there is". The correction was made in our
works [54], and in ([55], Sections 1-4 and Appendix B). The obtained results
still admit different and profoundly important interpretation\medskip .

8.2. \ Statistical mechanics of the Dirac equation, differential geometry

\ \ \ \ \ \ \ of elastic curves and Kirchhoff's elastic rods

\ \ \ \ \ \ \ \ 

Hundreds upon hundreds of papers and many books had been written on the
topic of what is (the classical analog of) the electron. The latest
representative books are [117-120]. While the number of papers, even the
most recent ones, is overwhelmingly large, the contribution into this topic
made by this author differs from all of these since our purpose was to apply
the Dirac results (Dirac equation) to polymer solutions, e.g., read
[121-124]. Polymers typically are modelled as random walks on the lattice.
The Dirac polymer chains (semiflexible polymers) differ from Schr\"{o}%
dinger's (fully flexible) polymer chains by the energetical requirement for
the each consecutive step of the random walk. There are 2 options for the
consecutive lattice step: a) to continue to go straight, b) to go sideways,
or even completely back. If a) and b) are not regulated energetically, then
we are dealing with the classical random walk. If going sideways is
energetically controlled, we are dealing with the semiflexible chains. The
connection between such \textsl{biased} random walk and the Dirac equation
for the first time was independently studied in [125] and [126]$,$ problem
2.6. In the case of semiflexible polymers there is one-to-one correspondence
between the rigidity of these polymers and the mass $m$ of the Dirac
particle: for $m=0$ the random walk degenerates into straight line (the
neutrino path), while for $m\rightarrow \infty $ we are dealing with the
fully flexible polymers, that is with the nonrelativistic (Schr\"{o}%
dinger-like) limit of the Dirac propagator. To facilitate our readers'
understanding, of what follows, we would like now to provide some needed
facts from polymer physics and differential geometry.

Specifically, the random walks on the lattice are characterized by the
end-to-end distribution (moment generating) function $G_{0}(\mathbf{p}%
,N)=\int d^{d}re^{i\mathbf{p}\cdot \mathbf{r}}G_{0}(\mathbf{r},N)$. For the
random (Gaussian chains) walks of $N$ steps on the, say, the square/cubic
lattice, whose local step is $l,$ typically we need to consider the walk of $%
N$ steps starting at the origin \textbf{0} and ending at some distance 
\textbf{r} from the origin. In such a case the distribution function $G_{0}(%
\mathbf{p},N)$ is known to be: $G_{0}(\mathbf{p},N)=\exp (-\mathbf{p}%
^{2}lN/2d).$Here $d$ is the lattice dimensionality. With such a generating
function it is of interest to calculate the averages (the moments), e.g. \ $<%
\mathbf{R}^{2}>=\int d^{d}r\mathbf{r}^{2}G_{0}(\mathbf{r},N)$, etc. Using
the generating function $G_{0}(\mathbf{p},N)$ in 3 dimensions we obtain: $<%
\mathbf{R}^{2}>=Nl.$ More physically interesting is to evaluate the
experimentally measurable scattering function $S_{0}(\mathbf{p,}N)$. For the
fully flexible (Gaussian) polymer chains, it is given by $S_{0}(\mathbf{p,}%
N)=\frac{1}{N^{2}}\dint\limits_{0}^{N}d\tau \dint\limits_{0}^{N}d\tau
^{\prime }G_{0}(\mathbf{p},\left\vert \tau -\tau ^{\prime }\right\vert )=%
\frac{2}{x^{2}}(x-1+e^{x}),$ $x=(\mathbf{p}^{2}/2d)Nl$ . To extend this
result to semiflexible chains, we notice that the Laplace transform $G_{0}(%
\mathbf{p},s)$ of $G_{0}(\mathbf{p},N)$ is known to be (in the appropriately
chosen system of units) given by $G_{0}(\mathbf{p},s)=(\mathbf{p}%
^{2}+s)^{-1},$ where $s$ is the Laplace variable conjugate to $N$. Our
readers immediately recognize in $G_{0}(\mathbf{p},s)$ the Green's function
for the massive Klein -Gordon (K-G) propagator (the Euclidean version). We
can obtain the Dirac propagator from the K-G propagator following Dirac's
logic. This then will enable us to obtain the (Dirac) generating function $%
G_{D}(\mathbf{p},N)$ for the semiflexible polymers and, therefore, the
scattering function $S_{D}(\mathbf{p,}N)$. Such an approach is logical but
is not physically illuminating. This happens because the statistical
properties of both the Gaussian and the Dirac polymer chains can be computer
simulated. Besides, they can be experimentally measured by analyzing the
light (or neutron) scattering data for $S_{D}(\mathbf{p,}N)$. This
circumstance leads us to study the\ transition -from discrete to continuous
limit, for various models of polymer chains relying on results of computer
simulations and the light scattering experiments. \ Both are based on these
models. Although for the Gaussian chains there is only one (random walk)
model, there is \textsl{countable infinity} of models [127] for semiflexible
chains and the \textsl{Dirac chain is just one model out of this infinity.}
\ Other than Dirac, in polymer physics the most popular is the Kratky-Porod
(K-P) model [128]. It is based on applications of Euler's elastica to
polymer physics. Details are given below. It will be compared with the Dirac
model because: a) it will help us to extend the limits of the Maxwell-Dirac
bijectivity, b) it will help us to model the Dirac fermions on computers in
terms of random walks with rigidity. Before making such a comparison,
several results are helpful to discuss. For instance, the Dirac propagator-
the solution of problem 2.6. from Feynman-Hibbs book [126] ( presented in
detail in [121]) is given by 
\begin{equation}
\frac{1}{2}G_{D}(\mathbf{p},N)=\cosh (mEN)+\frac{1}{E}\sinh (mEN),E^{2}=1-%
\frac{\mathbf{p}^{2}}{m^{2}}.  \tag{105}
\end{equation}%
Its Laplace transform is obtained as. 
\begin{equation}
\frac{1}{2}G_{D}(\mathbf{p},N)=\frac{s+m}{\mathbf{p}^{2}+s^{2}-m^{2}}. 
\tag{106}
\end{equation}%
In the limit $m\rightarrow 0$ this result can be rewritten as 
\begin{equation}
\frac{1}{2}TrG_{D}(\mathbf{p},N)=Tr\frac{i\mathbf{\vec{\sigma}}\cdot \mathbf{%
p+}s}{(i\mathbf{\vec{\sigma}}\cdot \mathbf{p+}s)(-i\mathbf{\vec{\sigma}}%
\cdot \mathbf{p+}s)}.  \tag{107}
\end{equation}%
Here the Pauli matrices $\mathbf{\vec{\sigma}=(}\sigma _{1}\mathbf{,}\sigma
_{2}\mathbf{,}\sigma _{3}\mathbf{)}$ are the same as in (61). The trace
operation $Tr$ reflects the averaging over directions of the ends of the
polymer chain [121]. Evidently, the Dirac propagator $G_{D}(\mathbf{p},s)=(-i%
\mathbf{\vec{\sigma}}\cdot \mathbf{p+}s)^{-1}$ is replacing now the
Klein-Gordon propagator $G_{0}(\mathbf{p},s)=(\mathbf{p}^{2}+s)^{-1}$
introduced previously for description of Gaussian chains. Since the effects
of rigidity are contained in the mass parameter $m,$ just presented result
should be slightly modified to account for the parameter $m$. For this
purpose, following our work [123], \ we have to use the Euclideanized
version of the 3+1 dimensional Dirac propagator, \ that is we have to
replace\{$\gamma ^{\mu },\gamma ^{\nu }\}=-2g^{\mu \nu }$ \ (where $g^{\mu
\nu }$ is the diagonal Minkowski matrix with \{1,-1,-1,-1\} on the diagonal)
by its Euclidean version \{$\gamma ^{\mu },\gamma ^{\nu }\}=-2\delta ^{\mu
\nu }.$This leads us to 
\begin{equation}
G_{D}(p\mathbf{,}m)=\frac{-i}{\gamma ^{\mu }p_{\mu }+m}=i\frac{\gamma ^{\mu
}p_{\mu }-m}{-p^{2}+m^{2}},\text{ }p^{2}=\mathbf{p}^{2}+s^{2}.  \tag{108}
\end{equation}%
This happens because instead of the Fourier transforms (used in the high
energy physics) in the polymer physics we must use the Laplace transforms.
To bring this result in correspondence with (106) it is sufficient to make
yet another replacement: $m\rightarrow im$ and to take the trace$.$ This
produces: 
\begin{equation}
TrS_{D}(\mathbf{p,}m)=\frac{m}{\mathbf{p}^{2}+s^{2}-m^{2}}.  \tag{109}
\end{equation}%
By additional change of \ $m$ variable equations (106) and (109) can be made
to coincide [123].The inverse Laplace transforming this result leads to $%
G_{D}(\mathbf{p,\left\vert \tau -\tau ^{\prime }\right\vert ,}m)$. This
result should be used instead of its Gaussian/Schr\"{o}dinger version $G_{0}(%
\mathbf{p},\left\vert \tau -\tau ^{\prime }\right\vert )$ in claculations of
the scattering function. The results of such calculation are presented in
[123]. They were compared against results coming from all known models of
polymer chains. The scattering function for the Dirac chain happens to fit
ideally both the numerical (Monte Carlo) and real experimental data. It was
already included in the standard toolkit manual (for polymer
experimentalists) distributed world-wide [129] by the Paul Scherrer
Institute (Switzerland). In this manual and, therefore in the polymer
literature, it is known as "Kholodenko worm" (because semiflexible polymers
are called "wormlike" in polymer community) or "Kholodenko-Dirac". \ 

Being armed with these results, we are now ready to talk about how results
just discussed can be reproduced differential-geometrically. Since this
topic was discussed in detail in our works [37, 54, 55,121], this fact
allows us to be brief. \ The task is reduced to the discussion of some path
integrals equivalently representing the Dirac propagator. \ By analogy with
results, just described, we begin with the path integral for the K-G
propagator $G_{0}(\mathbf{p},s)=(\mathbf{p}^{2}+s)^{-1}.$ Such path integral
and its calculation is described in detail in the book by Polyakov [130],
pages 151-169. Equivalently, one can use [131]. The end-to-end distribution
function, that is the K-G propagator, is formally given by: 
\begin{equation}
G(r,m^{2})=\dint \frac{\mathfrak{D}[\mathbf{r}(\tau )]}{\mathfrak{D}[f(\tau
)]}\exp \{-\mu _{0}i\dint\nolimits_{0}^{1}d\tau \sqrt{<\mathbf{v}\cdot 
\mathbf{v}>}\},\mu _{0}=m^{2}\sim s,\text{ }\mathbf{v}=\frac{d\mathbf{r}}{%
d\tau }.  \tag{110}
\end{equation}%
The exponent in (110) is nothing but the length of the curve. By design, it
is written in the reparametrization-invariant form. That is, it is invariant
under the transformations of the type: 
\begin{equation}
\mathbf{r}(\tau )\rightarrow \mathbf{r}(f(\tau )),f(0)=0,f(1)=1,\frac{df}{%
d\tau }>0.  \tag{111}
\end{equation}%
This result has its origins in differential geometry and reflects the fact
that the total length of the curve is invariant scalar. \ In the case of
semiflexible polymers, the following (Kratky-Porod) path integral is
typically being used [128] 
\begin{equation}
G(\mathbf{v}(N),\mathbf{v}(0))=\dint\nolimits_{\mathbf{v}=\mathbf{v}(0)}^{%
\mathbf{v}=\mathbf{v}(N)}\mathfrak{D}[\mathbf{v}(\tau )]\exp \{-\frac{\eta }{%
2}\dint\nolimits_{0}^{N}d\tau <\mathbf{t}\cdot \mathbf{t}>\}\dprod\limits_{%
\tau }\delta (\mathbf{v}^{2}(\tau )-1),\mathbf{t}=\frac{d\mathbf{v}}{d\tau }
\tag{112}
\end{equation}%
in polymer physics community. Here the exponent is not reparametrization
invariant. It represents Euler's elastica [132]. As before, here $N$ is the
number of monomeric units in polymer chain, $\eta $ is the rigidity
parameter (related to the mass $m$ of Dirac propagator). The constraint $%
\mathbf{v}^{2}(\tau )=1$ is reflectring the differential-geometric
requirement for natural parametrization of the curve. If the natural
parametrization is used in (110), then the exponent in (110) becomes a
number and the path integral is seemingly unnecessary. In (112) the
expression in the exponent under "time" integral is the square of the local
curvature of the curve as a function of trajectory parameter $\tau $.
Mathematically, the path integral (112) describes a Brownian motion
(diffusion) on the sphere (of unit radius). Therefore, for short "times" the
Gaussian result 
\begin{equation}
<[\mathbf{v}(\tau )-\mathbf{v}(0)]^{2}>=\frac{2}{\eta }\tau   \tag{113}
\end{equation}%
follows since for short times motion on the sphere and on the plane are the
same. This result is consistent with the (already presented) result: $<%
\mathbf{R}^{2}>=Nl$, $N\rightarrow \tau .$ Accordingly, the statistical
average in (113) is performed with $G_{0}(\mathbf{r},N)$ (up to a change of
scale ) as before. Since $\mathbf{v}^{2}(\tau )=1$ we can rewrite (113) as 
\[
<2-2\mathbf{v}(\tau )\cdot \mathbf{v}(0)>=\frac{2}{\eta }\tau 
\]%
yielding upon exponentiation 
\begin{equation}
<\mathbf{v}(\tau )\cdot \mathbf{v}(0)>=\exp (-\frac{\tau }{\eta }). 
\tag{114}
\end{equation}%
A closed analytical form for the Kratky-Porod propagator (112) is the same
as for the rigid rotator in quantum mechanics, that is:%
\begin{equation}
G(\mathbf{v}(N),\mathbf{v}(0))=\dsum\limits_{l=0}^{\infty }\exp
\{-(l(l+1)N\}\dsum\limits_{m=-l}^{l}Y_{lm}(\theta ,\phi )Y_{lm}^{\ast
}(\theta _{0},\phi _{0}).  \tag{115}
\end{equation}%
Not only technically to work with such a propagator is more difficult
[128],[134] \ than with the Dirac propagator but, since every 3 dimensional
curve is characterized by its curvature and torsion, and the torsion is
absent in the exponent of the path integral (112), it is surprising that the
comparison between the scattering functions $S_{D}(\mathbf{p,}N)$ (exact
analytic result [123]) for the Dirac chains and that $S_{K-P}(\mathbf{p,}N)$
for the Kratky-Porod chains (the exact clalculations for $S_{K-P}(\mathbf{p,}%
N)$ are not available, [128],[134]) produce closely similar results [135].
Since the discretized models of the Dirac propagator [126],[127] involve
study of the biased random walks and since the random walks on the sphere,
equation (112), still cannot be considered as biased walks, the question
arises: is it possible to improve the Kratky-Porod propagator, equation
(112), so that the improved propagator becomes that for the Dirac chains? \
This improvement, evidently, should involve inclusion of some kind of bias
terms in the exponent in (112). Without any reference to the equation (105),
such an improvement was indeed attempted in [136] with the result, equation
(3) of [136], indeed coinciding with our (105). Provided that constants in
(3) are appropriately reinterpreted, as it is done in detail in (Section
8.4.3 of our book [37]), the complete agreement between equation (3) of
[136] and (105) is achieved. Just obtained results permit us to make some
important comments. These are presented in the next two subsections.\medskip 

8.3. Differential geometry of elastic curves and Kirchhoff's elastic
rods\medskip .

\ \ \ \ \ \ Emergence of the elastic torus knots\medskip

From differential geometry it follows that the bias in equation (3) of [136]
makes the polymer conformation 3 dimensional. \ This happens because we must
take into account that [121]\medskip :

\textbf{Theorem 1.} \textit{A curve }$\vec{\gamma}(s)$ \textit{in three
dimensional space is fully }

\ \ \ \ \ \ \ \ \ \ \ \ \ \ \ \ \ \ \ \ \ \textit{determined by its
curvature }$\mathfrak{R}$\textit{(}$s)$\textit{\ and torsion }$\mathfrak{T}$%
\textit{(}$s).\medskip $

\textbf{Theorem 2. }\textit{The torsion} $\mathfrak{T}$\textit{(}$s)$ 
\textit{of planar curves is zero}.\medskip \medskip

These results are correct only in flat space. If the\ curve is embedded into
space of nonzero curvature, the above theorems must be amended [137]. Also,
they must be amended if the curve is replaced by a beam of, say, circular
cross-section $\alpha .$ More complications will follow if the cross-section
is represented by some scalar function $\alpha (s)$ of its location $s$
along the curve $\vec{\gamma}(s)$. Complications associated with such
thickening of a curve are described by the theory of Kirchhoff rods [138].
Developments of rod theory depends upon the condition: rod is stretchable or
nonstrechable. If we are dealing with nonstrechable rods (open or closed),
it is convenient to introduce the concept of \textit{centerline}. It is a
unit speed curve along the axis of the rod. In the context of semiflexible
polymers the presence of this centerline is accounted by the constraint $%
\dprod\limits_{s}\ \delta (\mathbf{v}^{2}(s)-1)$ in the path integral, as it
was done for the Kratky-Porod model (112). The treatment of polymer chains
of finite thicknesses is described in the book by Yamakawa [128]. In this
work we shall rely, however, on rigorous mathematics results. We begin
with\medskip

\textbf{Theorem 3.} [139] \textit{Every torus knot type is realized by a
smooth}

\ \ \ \ \ \ \ \ \ \ \ \ \ \ \ \ \ \ \ \ \ \textit{\ closed elastic rod
centerline.\medskip }

If $\vec{\gamma}(s)$is the centerline of a uniform symmetric (that is $%
\alpha $ is $s-$ independent) Kirchhoff elastic rod, then it is providing an
extremum of the functional

$\mathcal{F}$[$\vec{\gamma}(s)]$ defined by%
\begin{equation}
\mathcal{F}[\vec{\gamma}(s)]=\lambda _{1}\dint\nolimits_{\gamma }ds+\lambda
_{2}\dint\nolimits_{\gamma }ds\mathfrak{T}(s)+\lambda
_{3}\dint\nolimits_{\gamma }ds\mathfrak{R}^{2}(s),\text{ with }\lambda
_{3}\neq 0.  \tag{116}
\end{equation}%
Here $\lambda _{1},\lambda _{2},\lambda _{3}$ are some constants (Lagrangian
multipliers) and $\mathfrak{R}$ and $\mathfrak{T}$ are the curvature and
torsion defined in the Serret -Frenet (S-F) equations given by%
\begin{equation}
\frac{d\vec{\gamma}}{ds}=\mathbf{e}_{1},\frac{d\mathbf{e}_{1}}{ds}=\mathfrak{%
R}\mathbf{e}_{2},\frac{d\mathbf{e}_{2}}{ds}=-\mathfrak{R}\mathbf{e}_{1}+%
\mathfrak{T}\mathbf{e}_{3},\frac{d\mathbf{e}_{3}}{ds}=-\mathfrak{T}\mathbf{e}%
_{2},  \tag{117}
\end{equation}%
while the mutually orthogonal vectors\textbf{\ e}$_{1}$, \textbf{e}$_{2}$, 
\textbf{e}$_{3}$\textbf{\ }compose the \ S-F moving frame. The following
theorem is of major physical importance (to be explained below and in the
next subsection)\bigskip

\textbf{Theorem 4.}\textit{\ An elastic rod centreline of non-constant
curvature can}

\ \ \ \ \ \ \ \ \ \ \ \ \ \ \ \ \ \ \ \ \textit{intersect itself only at the
origin of natural system of }

\ \ \ \ \ \ \ \ \ \ \ \ \ \ \ \ \ \ \ \ \textit{cylindrical coordinates. \
If this does not occur, and }

\ \ \ \ \ \ \ \ \ \ \ \ \ \ \ \ \ \ \ \ \textit{the centerline is closed,
then it is embedded and lies}

\ \ \ \ \ \ \ \ \ \ \ \ \ \ \ \ \ \ \ \textit{\ on embedded torus of
revolution.\medskip }

\textit{\bigskip }

\textit{\medskip }Building on information supplied by this theorem, the next
theorem makes things perfect for applications\medskip .

\textbf{Theorem \ 5}. \textit{Given any relatively prime integers k,n such
that}

\ \ \ \ \ \ \ \ \ \ \ \ \ \ \ \ \ \ \ \ \ \ $\left\vert k/n\right\vert <1/2,$%
\textit{there exists a unique smooth closed elastic}

\ \ \ \ \ \ \ \ \ \ \ \ \ \ \ \ \ \ \ \ \ \ \textit{rod of constant torsion
with the knot type of a (k,n) }

\ \ \ \ \ \ \ \ \ \ \ \ \ \ \ \ \ \ \ \ \ \ \textit{torus knot.\medskip }

The results presented in this subsection allow us to reach the following
conclusions.

1.The source-free set of Maxwell equations (1) can be rewritten in the form
of the Beltrami equation. This was achieved in our works [54], and in([55], 
Sections 1-4 and Appendix B). These results allow us to bypass uses of the
Fourier transformed form of the Beltrami equation (53). In our book [37],
page 3, it was explained that the Beltrami equation is just the London
equation of superconductivity. This observation leads naturally to the
discussion of knotted vortex filaments.

2. By repeating arguments of Section 5, that is by applying the operator $%
div $ to both sides of the Beltrami equation\textbf{\ }$\nabla \times 
\mathbf{F=}k\mathbf{F}$ and by assuming that $k=k(x,y,z)$ we obtain $\mathbf{%
F}\cdot \mathbf{\nabla k=0.}$ Let now $\vec{\gamma}(s)=\{x(s),y(s),z(s)\}$
be a trajectory on the surface $k(x,y,z)=k=$constant. This then implies $%
\frac{d}{ds}\kappa \{x(s),y(s),z(s)\}=v_{x}\kappa _{x}+v_{y}\kappa
_{y}+v_{z}\kappa _{z}=\mathbf{v}\cdot \nabla \kappa =0.$ Now we have to take
into consideration that $\mathbf{F}=\mathbf{v}$ in our case. Thus, the
"velocity" $\mathbf{v}$ is always tangential to the surface $const=\kappa
(x,y,z).$We also have to assume that the vector field $\mathbf{v}$ is
nowhere vanishing on the surface $const=\kappa (x,y,z).$ This is possible
only if the surface is torus T$^{2}$. The field lines of \textbf{v} on T$%
^{2} $ should be closed if $const$ is rational number. This condition is
essential for the closed filament $\vec{\gamma}(s)$ to be a torus knot.

3. Facts just stated are entirely compatible with Theorems 3.--5. Therefore,
the presented summary allows us to develop results of our next, the final,
subsection of this section.\medskip \medskip

8.4. From Kirchhoff to Dirac: a solution of the electron problem

\ \ \ \ \ \ by methods of projective relativity\medskip

In subsection 8.2. we mentioned that the amount of papers and books about
the nature of electron is overwhelmingly large. This means only that the
proposed models of electron were always incomplete in some way or another.
It is being hoped, that presented below results may close the existing gaps
thus making the microscopic picture of electron complete. More on this topic
is also presented in the next section. \ 

The following results are based on the isomorphism proven by Kirchhoff
between the description of static and dynamics of rods with thickness and
statics and dynamics of rigid bodies [138]. Our uses of Kirchhoff
isomorphism are based on utilization of Theorems 3.-5. of previous
subsection. They serve to demonstrate the usefulness of (Kirchhoff)
functional (116) in quantum mechanics. Based on just described connection
between the statistical description of semiflexible polymers with Dirac
propagators, it follows that the functional \ (116) is present in the
exponent of the path integral for the Dirac propagator. Closed paths
obtainable by minimization of this functional are described by the theorems
of previous subsection. These are the torus knots. Accordingly, they are
also solutions of the Beltrami equation and, therefore, are localized
solutions of the sorceless set of Maxwell's equations (1). \textsl{Thus, the
task of establishing the Maxwell - Dirac correspondence is to be completed
in this subsection because of this crucial property of the Maxwell-Dirac
correspondence. }This aspect of correspondence is inseparably linked with
clarification of the meaning of electron as quantum mechanical particle.

To begin, we shall use some results from our paper [121]. In particular, we
begin with presentation of the system of S-F equations (117) in the form
known in the dynamics of rigid body[36], that is we write: 
\begin{equation}
\frac{de_{i}}{ds}=\dsum\limits_{J=1}^{3}\omega _{ij}e_{j}  \tag{118}
\end{equation}%
where,%
\begin{equation}
\omega _{ij}=\left( 
\begin{array}{ccc}
0 & \mathfrak{R} & 0 \\ 
-\mathfrak{R} & 0 & \mathfrak{T} \\ 
0 & -\mathfrak{T} & 0%
\end{array}%
\right) ,  \tag{119}
\end{equation}%
and $e_{i}\cdot e_{j}=\delta _{ij}.$ According to Arnol'd [36], only in 3
dimensions the operations $\times $ and $\wedge $ are equivalent. This
observation makes the \ S-F vectors $e_{i}$ act as elements of the Clifford
algebra. That is, the property $e_{1}\wedge e_{2}=-e_{2}\wedge e_{1}$ allows
us to write the defining anticommutator of the Clifford algebra as 
\begin{equation}
\{e_{i},e_{j}\}_{+}=2\delta _{ij}.  \tag{120}
\end{equation}%
Incidentally, the same conclusion was reached independently by Martin [140].
The set of equations (118) can be equivalently rewritten as 
\begin{equation}
\frac{de_{i}}{ds}=\dsum\limits_{i,j}^{3}\varepsilon _{ijk}B_{j}e_{k}, 
\tag{121}
\end{equation}%
where the components of "magnetic" field \textbf{B }are defined now as $%
B_{1}=-\mathfrak{T},B_{2}=0,B_{3}=-\mathfrak{R}.$ Having in mind Feynman's
path integral method of obtaining quantum mechanical results, it is
instructive to reobtain equation (121) variationally from the action
functional $\mathcal{A}(t_{f},t_{i})$ defined in [141] as%
\begin{equation}
\mathcal{A}(s_{f},s_{i})=\frac{i}{2}\dint\nolimits_{s_{i}}^{s_{f}}ds[\tilde{%
\omega}^{kl}e_{k}\dot{e}_{l}-H(\{e_{i}\})],\text{ \ }\tilde{\omega}%
^{kl}=\delta ^{kl}.  \tag{122}
\end{equation}%
In (122) the Hamiltonian is given by%
\begin{equation}
H(\{e_{i}\})=-\dsum\limits_{i,j,k=1}^{3}\varepsilon _{ijk}B_{i}e_{j}e_{k} 
\tag{123}
\end{equation}%
The action $\mathcal{A}(t_{f},t_{i})$ has the imaginary $i$ in front since
the quantum mechanical path integrals come with $i$ in front of the action
functional [126]. The S-F equation (118) (or (121)) does not contain
imaginary part, e.g. look at the K-P path integral (112). Suppose that the
equation (121) is solved. Then, by multiplying (121) from the left by $e_{i}$%
, summing over the index $i$ and remembering about the noncommutativity of $%
e_{i}$ , $\dot{e}_{i}\equiv \frac{de_{i}}{ds}$ and, also of the antisymmetry
of the Kronecker symbol $\varepsilon _{ijk},$ we obtain:%
\begin{equation}
\mathcal{A}(s_{f},s_{i})=i\dsum\nolimits_{i=1}^{3}\dint%
\nolimits_{s_{i}}^{s_{f}}dse_{i}\dot{e}_{i}.  \tag{124}
\end{equation}%
In view of (117) we realize that the vector $\mathbf{e}_{1}$ is directed
along the rod's centerline so that the mutually perpendicular vectors $%
\mathbf{e}_{2}$ and $\mathbf{e}_{3}$ are rotating in the plane perpendicular
to $\mathbf{e}_{1},$while $\mathbf{e}_{1\text{ \ }}$is moving along the
centerline$.$ \ Just obtained result should be made compatible with the
action in the exponent in equation (110) since the total action is the sum
of both actions coming from (110) and (124). In the equation (110) the
worldline is "living " in the 4-dimensional Minkowski space-time while in
(124), the wordline is "living" in 3-dimensional Euclidean space. \ To make
these actions compatible with each other, first of all, we notice that both
actions are reparametrization-invariant. Therefore, we can put the time
limits $s_{i}=0,s_{f}=1$ in (124) in accord with (110). Then we must study
the extension of the S-F equations to 4 dimensions. It happens, that: a) the
S-F equations exist in spaces of any dimensionality and of any signature
[142], b) although the equivalence $\times $ and $\wedge $ is lost, \ the
noncommutativity property survies, especially \ upon quantization [140], c)
the problem about the dynamics\ of a charged particle in the presence of
joint electric and magnetic fields is described by the S-F equations in
4-dimensional Minkowski space [143] (this observation makes the topic of
this subsection compatible with that of the previous subsection). In
Minkowski 4-space let the vector \textbf{\r{e}}$_{0}$be directed along the
4-dimensional world line (4-velocity), then the vectors $\mathbf{e}%
_{i},i=1\div 3$ must rotate in the hyperplane perpendicular to \textbf{\r{e}}%
$_{0}.$Therefore, \ to account for this fact, \ the constraint%
\begin{equation}
\eta _{\mu \nu }e^{\mu }\dot{x}^{\nu }=0  \tag{125}
\end{equation}%
should be imposed. Following [141], we choose $\eta _{\mu \nu
}=diag\{-1,+1,+1,+1\}.$At the same time, owing to [140], we know that the
vectors $e^{\mu }$ must be treated noncommutatively, e.g. see equation (132)
below. To move forward, it is helpful to utilize some facts by studying the
action $S[x(\tau )]$ in the exponent of (110) 
\begin{equation}
S[x(\tau )]=-m\dint\nolimits_{0}^{1}d\tau \dsum\limits_{\mu =0}^{3}\sqrt{%
\dot{x}^{\mu }\dot{x}_{\mu }}\equiv \dint\nolimits_{0}^{1}d\tau \mathcal{L}%
[x(\tau )].  \tag{126}
\end{equation}%
Since%
\begin{equation}
\frac{m\dot{x}_{\mu }(\tau )}{\sqrt{\dot{x}^{\mu }\dot{x}_{\mu }}}=p_{\mu } 
\tag{127}
\end{equation}%
the Hamiltonian for this action is zero. Indeed 
\begin{equation}
H=p_{\mu }\dot{x}_{\mu }+m\sqrt{\dot{x}^{\mu }\dot{x}_{\mu }}=0.  \tag{128}
\end{equation}%
At the same time,%
\begin{equation}
p_{\mu }p^{\mu }+m^{2}=0.  \tag{129}
\end{equation}%
This equation is yet another constraint. For compatibility of (125), (127)
and (129) it is useful to rewrite the constraint (125) as 
\begin{equation}
\eta _{\mu \nu }e^{\mu }p^{\nu }=0  \tag{130}
\end{equation}%
while the action (126) to rewrite as%
\begin{equation}
S[x(\tau ),p(\tau )]=\dint\nolimits_{0}^{1}d\tau \lbrack p_{\mu }\dot{x}%
_{\mu }-\frac{e}{2}(p_{\mu }p^{\mu }+m^{2})].  \tag{131}
\end{equation}%
Next, for compatibility with (131), we elevate the action (124) to 4
dimensions. This is permissible in view of [143]. Accordingly, the 3
elements Clifford algebra (120) has now acquired the 4th element. To account
for Minkowski space, we have to replace (120) by 
\begin{equation}
\{e_{\mu },e_{\nu }\}_{+}=2\eta _{\mu \nu }.  \tag{132}
\end{equation}%
With this anticommutator, the noncommutative elements $e_{\mu }$ \ become,
in fact, the Dirac the gamma matrices [144]. However, already from the
information presented in Section 7 we know that there is also an independent
"chiral" gamma matrix $\gamma _{5}$ so that the dimensionality of space of $%
e_{\mu }^{\prime }s$ (or $\gamma _{\mu }^{\prime }s)$ is \textbf{five. }The
question immediately emerges: Is there any additional physics (other than in
Section 7) behind this 5th dimension? Surprisingly, already in 1919 this
question was independently raised by Kaluza who published his findings in
1921. His results were elevated to the quantum level by Klein in 1926 [145].
Both these authors were concerned with usefulness of the 5th dimension for
unification of gravity and electromagnetism. Unfortunately, further
refinements demonstrated that many results of the Kaluza-Klein theory,
especially the predicted mass of the electron, are in dramatic disagreement
with the experiment [146]. Quite independently, many mathematicians and
physicists such as Pauli, Dirac, Schr\"{o}dinger, had chosen another path to
the fifth dimension (e.g. \ study of dS and AdS spaces, mentioned in
subsection 8.1, etc.) and the process is still ongoing. The initial stages
of this alternative developments are beautifully summarized in the book by
O.Veblen. This direction of research is known in literature as "Projective
theory of relativity"[147]. The latest important developments are presented
in [148] and summarized in [149]. For the purposes of this work, we only
need to use the results of reference [150] by Dirac. It is mentioned as
Ref.12 in [149]. Although [150] is the precursor of Dirac's theory of
constraints [151], it has its own great merit, especially since it utilizes
Eisenhart's theory of contact transformations [152]- the major theme of our
book [37].

We just have discussed in (128) the situation, when the dynamical system has
a Lagrangian but not a Hamiltonian. What information does this fact
provides? Dirac studied this phenomenon in his book [151] on treatment of
dynamical systems with constraints. The situation when both the Lagrangian
and the Hamiltonian are zero is much more exotic. This case is studied in
[150]. By studying the examples from this reference not only we shall
understand the origins of 5th dimension in our problem, but also we shall
learn how to deal with such situations when they occur. We begin with study
of the lightcone equation:

\begin{equation}
c^{2}ds^{2}=c^{2}dt^{2}-\dsum\nolimits_{i=1}^{3}dx_{i}^{2}.  \tag{133}
\end{equation}%
It is related to the action (126). Choosing the signature of Minkowski
space-time in accord with that, in Dirac's paper, we obtain the Lagrangian%
\begin{equation}
\mathcal{L=-}mc(c^{2}\dot{x}_{0}^{2}-\dsum\nolimits_{i=1}^{3}\dot{x}%
_{i}^{2})^{\frac{1}{2}}.  \tag{134}
\end{equation}%
Suppose now that the particle has zero rest mass, $m=0.$ In such a case the
Lagrangian $\mathcal{L}$ is zero. That is:%
\begin{equation}
\mathcal{L=}c^{2}\dot{x}_{0}^{2}-\dsum\nolimits_{i=1}^{3}\dot{x}_{i}^{2}=0. 
\tag{135}
\end{equation}%
This is surely the case to be studied in projective geometry. That is since
the multiplication by some constant $\lambda $ of the equation (135) changes
nothing, we are dealing with\ the equivalence class made of infinity of
Lagrangians which differ from each other by the actual value of $\lambda $ .
But, since such $\lambda $ containing expression represents the Lagrangian
(the equivalence class of), the associated momenta are obtained as usual.
That is: 
\begin{eqnarray}
p_{0} &=&\lambda \frac{\partial \mathcal{L}}{\partial \dot{x}_{0}}=2\lambda
c^{2}\dot{x}_{0},  \TCItag{136} \\
p_{r} &=&\lambda \frac{\partial \mathcal{L}}{\partial \dot{x}_{r}}=-2\lambda 
\dot{x}_{r},r=1,2,3.  \nonumber
\end{eqnarray}%
By employing (135) combined with the equations (136) produces, after some
algebra, the Hamiltonian which is also zero,%
\begin{equation}
H=p_{0}^{2}-c^{2}\dsum\nolimits_{r=1}^{3}\dot{x}_{r}^{2}=0.  \tag{137}
\end{equation}%
Let us now look at the equation (133) as an equation in projective space.
That is, it is more convenient now to rewrite (133) as 
\begin{equation}
c^{2}ds^{2}-c^{2}dt^{2}+\dsum\nolimits_{r=1}^{3}dx_{r}^{2}=0.  \tag{138}
\end{equation}%
This homogenous equation is manifestly 5-dimensional. By keeping this fact
in mind, let us consider its generalization yielding another Lagrangian
equal to zero in five-dimensional projective space. That is, this time we
are discussing the classical electrodynamics in the projective space. We have%
\begin{equation}
\mathcal{L=-}mc(c^{2}\dot{x}_{0}^{2}-\dsum\nolimits_{r=1}^{3}\dot{x}_{r})^{%
\frac{1}{2}}+e/c\dsum\nolimits_{r=1}^{3}A_{r}\dot{x}_{r}-eA_{0}\dot{x}_{0}-%
\dot{x}_{5}=0.  \tag{139}
\end{equation}

When $A_{0}$ and $A_{r}$ are zero, to be in agreement with (135), we have to
assume that $\dot{x}_{5}$ is proportional to $m$. \ Then, (139) is becoming
zero in accord with (138).Following the same steps as \ in (136), we now
obtain:%
\begin{eqnarray}
p_{0} &=&\lambda \frac{\partial \mathcal{L}}{\partial \dot{x}_{0}}=-\lambda
(mc^{2}\frac{dx_{0}}{ds}+eA_{0}),  \nonumber \\
p_{r} &=&\lambda \frac{\partial \mathcal{L}}{\partial \dot{x}_{r}}=\lambda (m%
\frac{dx_{r}}{ds}+eA_{r}),  \nonumber \\
p_{5} &=&\lambda \frac{\partial \mathcal{L}}{\partial \dot{x}_{5}}=-\lambda .
\TCItag{140}
\end{eqnarray}%
\qquad By keeping in mind how the Hamiltonian (137) was obtained, and by
repeating the same steps in the present case, we obtain yet another zero
Hamiltonian: 
\begin{equation}
(p_{0}-eA_{0}p_{5})^{2}-c^{2}\dsum%
\nolimits_{r=1}^{3}[p_{r}+e/cA_{r}p_{5}]^{2}-m^{2}c^{4}p_{5}^{2}=0. 
\tag{141}
\end{equation}%
Again, let now $A_{0}$ and $A_{r}$ be zero$.$ Then, if we put $c=1,$ we
reobtain the constraint equation (129). This time it can be reinterpreted as
homogenous equation in the projective space \textbf{RP}$^{4}$. \ Without
loss of generality, in this space we can fix, say, the fifth coordinate in 
\textbf{R}$^{5}$ and reinterpret it as the mass $m$ in accord with (139).
This provides us with the mass shell condition leading to the massive
Klein-Gordon equation at the quantum level. The constraint (130) is
compatible with the constraint (129) under the conditions just described. To
demonstrate this we have to place the constraint (130) into \textbf{RP}$%
^{4}. $ Then, to avoid mistakes, following [141], we relabel $e_{i}$ as $\xi
_{i}$ so that the anticommutator (120) is rewritten as 
\begin{equation}
\{\xi _{\mu },\xi _{\nu }\}=\hbar \eta _{\mu \nu },\text{ }\{\xi _{5},\xi
_{5}\}=\hbar ,  \tag{142}
\end{equation}%
provided that 
\begin{equation}
\xi _{\mu }=\left( \hbar /2\right) ^{\frac{1}{2}}\gamma _{5}\gamma _{\mu }%
\text{ ,\ \ }\xi _{5}=\left( \hbar /2\right) ^{\frac{1}{2}}\gamma _{5},\text{
}\{\gamma _{\mu },\gamma _{\nu }\}=-2\eta _{\mu \nu }.  \tag{143}
\end{equation}%
At the quantum level the constraints (129) and (130) are transformed into
the following equations%
\begin{equation}
\left( \hat{p}^{2}+m^{2}\right) \psi =0,  \tag{144}
\end{equation}%
\begin{equation}
\left( \hat{p}^{\mu }\xi _{\mu }+m\xi _{5}\right) \psi =0.  \tag{145}
\end{equation}%
By multiplying (145) by $\xi _{5}$ from the left and, by taking into account
(142) and (143), the standard massive Dirac equation is obtained,%
\begin{equation}
(\hat{p}^{\mu }\gamma _{\mu }+m)\psi =0.  \tag{146}
\end{equation}%
Finally, we can assemble the total action in the path integral for the\
Dirac propagator. It is given by:%
\begin{equation}
S[x(\tau ),\xi (\tau ),e,\chi ]=\dint\limits_{0}^{1}d\tau \lbrack p_{\mu }%
\dot{x}_{\mu }-\frac{e}{2}(p_{\mu }p^{\mu }+m^{2})+i\xi _{\mu }\dot{\xi}%
^{\mu }+i\xi _{5}\dot{\xi}^{5}+i\chi (p_{\mu }\xi ^{\mu }+m\xi ^{5})]. 
\tag{147}
\end{equation}%
The same action (up to signature) can be found the the book by Polyakov
[130], page 224. By writing the action in exactly the same form as \ on page
224, we have to use the equation (131) to obtain: 
\begin{equation}
\frac{\delta S[x,p]}{\delta p_{\mu }}=\dot{x}_{\mu }-ep_{\mu }=0,\text{ } 
\tag{148}
\end{equation}%
and to use it back in (131), implying:%
\begin{equation}
S[x]=\dint\limits_{0}^{1}d\tau \frac{1}{2}[e^{-1}\dot{x}^{\mu }\dot{x}_{\mu
}-em^{2}].  \tag{149}
\end{equation}%
When this result is substituted back to (147), we obtain the action, 
\begin{equation}
S[x(\tau ),\xi (\tau ),e,\chi ]=\dint\limits_{0}^{1}d\tau \{\frac{1}{2}%
[e^{-1}\dot{x}^{\mu }\dot{x}_{\mu }-em^{2}]+i\xi _{\mu }\dot{\xi}^{\mu
}+i\xi _{5}\dot{\xi}^{5}+i\chi (p_{\mu }\xi ^{\mu }+m\xi ^{5})\},  \tag{150}
\end{equation}%
in the form given by Polyakov. While Polyakov uses the concepts of
supersymmetry in his calculations, in this work we used the concepts of
projective relativity to arrive at the same results. Only through uses of
projective relativity it is possible to connect the differential-geometric
results of previous subsection with the quantum mechanical results of this
subsection. Moreover, mathematically the equation (149) is consistent with
the equation (124) since they both "live" in the same space. In both cases
the final total (combined) action used in the path integral contains
dynamical variables defined on orbits determined from the respective
equations of motion. These are obtained variationally from the respective
actions given by equations (122) and (131). \medskip Their combination is
consistent with the action functional (116) used for description of rigid
rods as explained in Section 8.2.\medskip

\textbf{9. Instead of a discussion\medskip }

9.1. Work by Varlamov\medskip

We just demonstrated the place of electron in the Maxwell-Dirac
correspondence. Thanks to series of works [153-155] by Varlamov it is
possible to find the place of the obtained results inside the Standard Model
of particle physics. The Standard Model consists of 17 fundamental
particles. Only two of these -- the electron and the photon, were known 100
years ago. Already at this level of knowledge they represented two major
groups of particles: the fermions and the bosons. The fermions are the
building blocks of matter. The known today 12 fermions are split into six
quarks and six leptons. The electron is a lepton. Protons and neutrons are
not a part of the Standard Model because they are made of quarks. All
heavier particles comprising the whole matter are made of quarks and
leptons. Only five bosons (including photon) are responsible for all the
interactions between the matter (leptons and quarks). They carry three of
the four fundamental forces in nature: the strong force, the weak force and
the electromagnetism. The fourth force is gravity. Varlamov's achievement
lies in demonstrating, using rigorous mathematics, that the mass spectrum of
all mesons (made of two quarks) and baryons (made of three quarks) is
possible to obtain with help of just one formula containing the mass of
electron $m_{e}$ as an input. The formula%
\begin{equation}
m^{s}=m_{e}(l+\frac{1}{2})(\tilde{l}+\frac{1}{2}),s=\left\vert l-\tilde{l}%
\right\vert  \tag{151}
\end{equation}%
fits the experimental data with the margin of errors less than or at most
1\%. The formula is obtained as some Lorentz group representation
characterized by the pair $(l,\tilde{l})$ in which $l$ and $\tilde{l}$ are
nonnegative numbers ($m,n$) or ($\frac{m}{2},$ $n)$,or ($m$,$\frac{n}{2}),$%
or ($\frac{m}{2},\frac{n}{2}).$ In this case, if there is a division by 2,
the respective numbers are odd.

In subsection 8.1.we introduced the Dupin cyclides. These are invariants of
the Lie sphere geometry. The M\"{o}bius group is a subgroup of the Lie
sphere geometry isomorphic to the Lorentz group [40]. Therefore, all
Varlamov's results are transferable to the results under umbrella of the Lie
sphere geometry. Thus, the Dupin cyclides are also invariants of the Lorentz
group as was established by Friedlander already in 1946 [40]. Since
topologically the electron is a torus knot-the trefoil (as results of the
previous section indicate), living on the simplest Dupin cyclide-the torus,
the rest of the short-lived particles must also be living on the Dupin
cyclides and all of them have the trefoil knot (that is electron) as their
precursor. As is known from knot theory [55],[63], all knots are obtainable
either from trefoils or from figure 8 knots. But the figure 8 knot is the
hyperbolic knot. Therefore, it cannot serve as the precursor since its
topological mass is a negative number. Thus, it must be excluded from
consideration for reasons explained in subsection 6.1. \medskip Varlamov's
results have their predecessor-work by Asim Barut [$156$]. The Abstract of [$%
156$] reads as follows:" The complex machinery of gauge theories and QCD is
not the only possible avenue to understand the physics of elementary
particles. There is another simple and more direct approach based on
absolutely stable particles as constituents of all other particles.
Mathematically, the ensuing group structure is identical to the quark-lepton
model, but the dynamics and the underlying physics is different. How all
forces of elementary particles are already dynamically unified under
electromagnetism is discussed". This paper by Barut came out at about the
same time as the first papers by Ranada [15],[16]. Based on the classical
training, Barut was aware of the linearity of classical electrodynamics.\
Ranada's results fundamentally changed this perception. Knots and links are
known to be the observables of the non-Abelian Chern-Simons field theory [$%
157$] derivable from the non-Abelian gauge field theory [$158$]. Thus,
having\ an electron as the simplest torus knot (Secions 7 and 8) lying in
the foundation of mass spectrum of mesons and baryons is consistent with
ideas of Barut and with fundamentals of knot theory[96]. To make Barut's
ideas consistent with modern ideas of quantum chromodynamics (that is to
account for knots) requires more works. These can be initiated based on the
information provided in the Note added in proof (below). How the mass tower
is built knot-theoretically is discussed in our work [63].\bigskip\ 

9.2. From projective relativity to knotty geometrodynamics\medskip

It is worth mentioning that the results of Lie sphere geometry have their
origins also in projective geometry[40]. They, as well as the results of
projective relativity (touched upon in subsection 8.4) serve well for the
Minkowski and curved spaces of Minkowski signature. The purpose of both the
Kaluza-Klein theory and the theory of projective relativity is to provide
logically consistent way to unify gravity and electromagnetism. The
alternative logically consistent way to unify gravity and electromagnetism
was made by Rainich in 1925 [1$59$]. His results were rediscovered by Misner
and Wheeler [$160$] and the ongoing development had acquired the name
"geometrodynamics" [$161$].

We would like to provide some details explaining why we believe, that this
approach to unification of gravity and electromagnetism is the most
promising. First, as noticed already in our book [37], page 97, the existing
non-Abelian gauge theories, and also gravity viewed as one of these
non-Abelian gauge theories [71],[72], are having profound difficulties in
including the extended geometrical objects into formalism of these theories.
Chronologically Rainich was not aware of this difficulties but, as if he was
able to foresee these difficulties, in his theory [$159$] he discussed only 
\textsl{source-free Maxwellian fields }selfconsistently\textsl{\ }creating
the gravitational background. That is to say, in Einstein's field equations
(59) the source-free electromagnetism is placed on the right hand side in\
these equations as the source of gravity. In their paper [$160$] the authors
claimed that the dynamics of source-free electric and magnetic fields is
described in terms of "rate of change of curvature of pure Riemannian
geometry, and nothing more", page 530. On the same page we find: " ...the
electromagnetic field leaves an imprint upon the metric that is so
characteristic, that from that imprint one can read back to find out all
that one needs to know about the electromagnetic field". The converse is
also true, e.g. read Note added in proof ( below).\ The authors were
concerned about the meaning of charges in Rainich theory. As we know now,
the answer was provided by Ranada [15],[16], and was elaborated by many
authors afterwards, including the author of this paper. Furthermore, as it
is demonstrated in this paper, \textsl{only source-free electromagnetic
fields make the probabilistic and optical interpretations of quantum} 
\textsl{mechanics equivalent because de Broigle waves have no sources}.
Because the Maxwell-Dirac isomorphism is the major theme of this work, the
topological and Diracian nature of Ranada's charges, provide the answer to
the question about the charges in geometrodynamics: all charges, without
exception, are of topological origin. These are either the individual
nonhyperbolic knots or the nonhyperbolic links. Rainich's, Misner's,
Wheeler's "\textsl{Already Unified Field Theory}" (AUFT) had one significant
drawback, though. It had not found inside itself a place for the null
electromagnetic fields. However, as we extensively discussed in this work,
these are essential for the existence of Ranada's knotty charges.
Furthermore, the stability of these knotty structures (or their ability to
disintegrate) is also\ linked with the existence of null fields[58], [81],[$%
162$]. Moreover, study of stability of nonhyperbolic knots/links requires
applications of contact geometry and topology [37],[$163$]. Recall
(subsection 8.4.), that this discipline was used not only in our works on
Randa's knots [55],[56] but also in Dirac's work [151] on
Lagrangian/Hamiltonian dynamics in projective spaces[152]. This usage of
contact geometry and topology is not some artefact of the theory because the
source-free Maxwellian field equations can be rewritten as the Beltrami
field equations. This is demonstrated in Appendix B of our paper [55].
Furthermore, because of this, based on the results of this work, we can
state that the Beltrami field equations under appropriate circumstances
(whenever the Maxwell-Dirac isomorphism is working) can be translated
one-to-one into Dirac equation thus providing yet another outlet for
connecting the contact geometry and topology with quantum mechanics. In
quantum mechanics the Beltrami equation \ is known as the London equation of
conventional (low temperature) superconductivity [37], page 3. Going back to
AUFT, the place for null fields in AUFT was found in a beautiful paper by
Geroch [$164$]. \ Subsequently, many authors had simplified and improved his
results, e.g., read [$165$].\medskip\ To make this presentation complete, we
must add yet another piece of information. Ranada's knots/links were of
electromagnetic origin. Since in AUFT gravity and electromagnetisn are
presented on equal footing, the question arises: are there exist
Randada-type knots and links of gravitational origin? Fortunately, this
topic (without reference to AUFT) was considered and solved positively. In [$%
166\mathbf{][}167$] gravitational hopfions (knots and links) mathematically
identical to those obtained by Ranada were obtained and described. \ In [37]
we already introduced and studied \ the Faddeev-Niemi hopfions. These
hopfions significantly develop results of 1962 work by Skyrme [$168$]
entitled "A unified field theory of mesons and baryons". Thus, with these
results presented and supplemented with the results of Note added in proof
(read below), we are coming back to the updated result by Varlamov (151) and
to the paper by Barut [156].\bigskip

9.3. Cosmological significance/meaning of the obtained results\medskip

9.3.1. Dirac-Kerr-Newman electron\medskip

Recently Burinskii in a series of papers, e.g. read [$169$], have
demonstrated that "the Dirac equation can be naturally incorporated into the
Kerr-Schild formalism \ as a master equation controlling the Kerr-Newman
geometry" (describing the Kerr-Newman solution \ of Einstein equations for
the charged rotating black hole)."As \ a result, \textbf{the Dirac electron
acquires an} \textbf{extended space-time structure} \textbf{of the
Kerr-Newman geometry}: a singular ring of Compton size and \ a twistorial
polarization of the gravitational and electromagnetic fields. The behavior
of this Dirac-Kerr-Newman system in weak and slowly changing electromagnetic
fields\textbf{\ is determined \ by the wave function of the Dirac equation
and is indistinguishable \ from the behavior of the Dirac electron. }The
wave function of the Dirac equation plays in this model the role of an
"order parameter" which controls the dynamics, spin polarization and the
twistorial nature of space-time". This information is useful to supplement
with results by Dalhuisen [74],[170] connecting twistors with
electromagnetic knots. In addition, in [171],[172] the connection between
twistors, geometric algebra and cyclides is discussed.\medskip

9.3.2. Geometrodynamics of neutrino, Majorana fermions and knot
theory\medskip

Rainich's, Misner's, Wheeler's AUFT had inspired others to think about the
development of analogous unified field theory of gravitation and neutrino,
e.g. read [$173$], [174]. This idea seemed meaningful since till the
discovery of neutrino oscillations [102] everybody believed that neutrino is
massless. By analogy with the photon, the massless neutrino does not carry
charge but it is a fermion while the photon is boson. Nevertheless, by
analogy with superconductivity in which Cooper pairs are bosons, the
neutrino theory of light was created [$175$]. After discoveries of neutrino
masses this neutrino theory of light was abandoned. However, it is commonly
believed (based on experimental data) that neutinos are the Majorana
fermions, that is these are neutral fermions with very small masses [102].
Since they are neutral, uses of the concepts of gauge invariance [97] for
establishing the analog of the Dirac-Maxwell correspondence cannot be used.
Nevertheless, for the \textbf{nonexisting} massless neutrinos described by
the massless Majorana equation very recently the correspondence analogous to
the Dirac-Maxwell was found [$176$]. In view of results of Section 5., the
presence of mass in the Majorana equation can be eliminated so that the
correspondence found in [$176$] can be extended to the massive case. Because
of this, one can think about the relationship between the Majorana fermions
and knots. This thought encounters many difficulties,though. First, there is
no analog of the Varlamov's formula (129) for the Majorana fermion, second,
the path integrals for the Dirac and Majorana electrons are not the same,
e.g., see [$177$]. Therefore, the differential-geometric results of sections
8.3.,8.4. cannot be used. The results of knot theory still can be used
though on Maxwell's side of the Majorana-Maxwell correspondence because
source-free set of Maxwell's equations is exactly equivalent to the Beltrami
equation as explained in this work. On the Majorana side the connections
with knot theory are known in 2 dimensions [$178$] thus far. It happens
because the knot-theoretic Arf invariant [63],[179] is connected with the 2D
Ising model. And the solution of this model in 2 dimensions involves uses of
the massive Dirac fermions [130]. Using the logic behind equations (86)-(89)
(see Section 7.3) Shankar in his book [180], page 153, demonstrated that the
action for 2D Dirac fermion can be represented by the sum of actions for two
decoupled Majorana fermions: 
\begin{equation}
\dint dxd\tau \bar{\Psi}(\eth +m)\Psi =\frac{1}{2}\dint dxd\tau \lbrack \bar{%
\eta}(\eth +m)\eta +\bar{\chi}(\eth +m)\chi ].  \tag{152}
\end{equation}%
Here $\Psi =\frac{1}{\sqrt{2}}(\chi +i\eta ),\bar{\Psi}=\frac{1}{\sqrt{2}}(%
\bar{\chi}+i\bar{\eta}),\bar{\chi}=\chi ^{T}\sigma _{2},\bar{\eta}=\eta
^{T}\sigma _{2},$ and, $\eth =\sigma _{2}\partial _{\tau }+\sigma
_{1}\partial _{x}$. The same result is presented in the book by Mussardo\ in
[181], page 337, in somewhat more focused form.

\textbf{\bigskip }

\bigskip

\textbf{Note added in proof}. When this work was completed, the author came
across the following 3 papers. These are: a) "Torus knots as Hopfions" by
M.Kobayashi and M.Nitta \textit{Phys.Lett.B} \textbf{2014, }728\textbf{, }%
314-318; b) "Helical buckling of Skyrme-Faddeev solitons", by D.Foster and
D.Harland,\textit{\ Proc.Roy.Soc. A} \textbf{2012}, 468, 3172-3190; c)"
Spacetime metric from linear electrodynamics" by Yu.Obukhov and F.Hehl, 
\textit{Phys.Lett.B} \textbf{1999}, 458, 466-470. Listed papers provide
strong additional support to\ the results of Sections 8 and 9. Finally, the
paper " The topological origin of quantum randomness" by S.Heusler,
P.Schlummer and M.Ubben, \textit{Symmetry} \textbf{2021}, 13, 581, provides
a very interesting complementary view of the results presented in this
paper.\bigskip

\textbf{Funding}: This research received no external funding

\bigskip

\textbf{Acknowledgements}: The author acknowledges many stimulating

conversations with Professor Louis H.Kauffman

(University of Illinois at Chicago) as well as very helpful critique of

one of the referees. It resulted in author's rewriting of Section 9.

\bigskip

\bigskip \textbf{Conflict of Interest}: The author declares no conflict of
interest

\bigskip

\textbf{References}

\bigskip

1. \ R.Feynman, R.;Leighton, R.: Sands,M. \textit{The Feynman Lectures on
Physics},

\ \ \ \ \ Vol.3; Basic Books, New York, NY,USA, 2011.

2. \ Bach, R.;Pope, D.; Liou, S-H.; Batelaan, H. Controlled double-slit

\ \ \ \ \ electron diffraction. \textit{New J.Phys}.\textbf{2013}, 15,
033018.

3. \ Zhou, H.; Pereault, W.; Mukherjee, N.; Zare\ R. Quantum mechanical

\ \ \ \ \ double slit for molecular scattering. Science \textbf{2021}, 374.
960-964.

4. \ Arndt, M.; Nairz,O.; Vos-Andreae, J.; Keller, C.; Zouw,G.; Zellinger, A.

\ \ \ \ \ Wave-particle duality of C 60 molecules. \textit{Nature} \textbf{%
1999}, 401, 680.

5. \ Eibenberger, S.; Gerlich, S.; Arndt, M.; Mayor,M.;T\"{u}xen, J. Matter

\ \ \ \ --wave interference of particles selected from a molecular library

\ \ \ \ \ with masses exceeding 10 000 amu.\ {\Large \ }\textit{Phys. Chem.
Chem. Phys.} \textbf{2013},15,

\ \ \ \ \ 14696-14700.

6. \ Bohm, D. \textit{Quantum Theory}; Dover Publications Inc., New York,

\ \ \ \ \ USA , 1989.

7. \ Sanz,A.; Miret-Art\'{e}s, S.\textit{\ A Trajectory Description of
Quantum}

\ \ \ \ \ \textit{Processes}. I.\textit{\ Fundamentals}; Springer-Verlag,
Berlin, Germany, 2012.

8. \ Leonhardt,U. \textit{Measuring the Quantum State of Light}; Cambridge U.

\ \ \ \ \ Press, Cambridge, UK, 1997.

9.\ \ Fox, M. \textit{Quantum Optics: An Introduction;} Oxford U. Press,

\ \ \ \ \ Oxford,UK, 2006.

10. Berestetskii,V.; Lifshitz, E.;Pitaevskii, L. \textit{Relativistic Quantum%
}

\ \ \ \ \ \textit{Theory}; Pergamon Press, Oxford, UK, 1971.

11.\ Collett, E. Mathematical Formulation of the Interference Laws

\ \ \ \ \ \ of Fresnel and Arago. \textit{Am.J.Phys}. \textbf{1971}, 39,
1483-1495.

12. Kanseri,B.; N.Bisht N.; Kandpal, H. Observation of the Fresnel

\ \ \ \ \ and Arago laws using the Mach-Zehnder interferometer.

\ \ \ \ \ \ \textit{Am. J.Phys.\textbf{2008}. }76, 39-42.

13. Green H.;Wolf, E. A Scalar Representation of Electromagnetic Fields.

\ \ \ \ \ \textit{Proc.Phys.Soc. A}\ \textbf{1953,} 66, 1129-1137.

14. Wolf, E. A Scalar Representation of Electromagnetic Fields : II.

\ \ \ \ \ \textit{Proc.Phys.Soc. A \ \textbf{1959,} }74, 269-280.

15. Ranada, A. A topological theory of the electromagnetic field.

\ \ \ \ \ \textit{Lett. Math. Phys}. \textbf{1989}, 18, 97-106.

16. Ranada, A. Topological electromagnetism. \textit{J.Phys.} A \ \textbf{%
1992},

\ \ \ \ \ 25, 1621-1641.

17. Altun S. Knotted solutions of Maxwell's equations. MS Thesis.

\ \ \ \ \ \ Available online:
https://etd.lib.metu.edu.tr/upload/12623449/index.pdf

18. Bohm D.; Hiley, B. \textit{Undivided Universe}; Rutlege Publ.Co.,

\ \ \ \ \ \ London, UK, 1993.

19. Kobe, D. \textit{Found.Phys.} \textbf{1999}, 29.1203-1231.

20. Raymer R.; Smith,B. SPIE Conference on Optics and Photonics,

\ \ \ \ \ \ Conference number 5866, \textbf{2005}.

21. Raymer, M.; Smith, B. Photon wave functions, wave-packet

\ \ \ \ \ \ quantization of light, and coherence theory. \textit{New J.Phys}%
. \textbf{2007}, 9, 414.

22. Luneburg, R. \textit{Mathematical Theory of Optics}. U.of California
Press,

\ \ \ \ \ \ Berkeley, CA, USA,1966.

23. V. MaslovV.; Fedoryk M., \textit{Semiclassical Approximation in Quantum}

\ \ \ \ \ \ \textit{Mechanics}. Reidel Publ. Co., Boston, MA, USA, 1981.

24. R\"{o}mer, H.\textit{\ Theoretical Optics}. Wiley-VCH, Hoboken, NJ, USA,
2005.

25. Roman, P. A Scalar Representation of Electromagnetic Fields: III.

\ \ \ \ \ \ \textit{Proc.Phys.Soc. A} \textbf{1959}, 74, 281-289.

26. Born, M.;Wolf, E. \textit{Principles of Optics. 15th Edition.}

\ \ \ \ \ \ Cambridge U.Press, Cambridge, UK, 2019.

27. Oughstun, K. \textit{Electromagnetic and Optical Pulse Propagation.}

\ \ \ \ \textit{\ \ }Springer, Nature, AG, Switzerland, 2019.

28. Bogoliubov, N.; Shirkov,D. \textit{Introduction to the Theory of
Quantized}

\ \ \ \ \ \textit{\ Fields}. John Wiley\&Sons, New York, NY, USA, 1976.

29. Harish-Chandra,\ H. The correspondence between the particle and the

\ \ \ \ \ \ wave aspects of the meson and the photon.\textit{\ }

\ \ \ \ \ \ \textit{Proc.Roy.Soc.London, A \textbf{1946,} }186, 502-525.

30. Tokuoka, Z. On the Equivalence of the Particle Formalism and the

\ \ \ \ \ \ Wave Formalism of Meson, II. \textit{Progr.Theor.Phys}. \textbf{%
1953},10, 137-157.

31. Ghose,P.; Majumdar,A.;Guha, S.; J.Sau,\textit{\ J. }Bohmian trajectories
for

\ \ \ \ \ \ photons \textit{Phys.Lett.A} \textbf{2001}, 290, 205-213.

32. Shabat, B. \textit{Introduction to Complex Analysis. Part II}. American

\ \ \ \ \ \ Mathematical Society, Providence, RI, USA, 1992.

33. Greiner, W. \textit{Relativistic Quantum Mechanics. Wave Equations}.

\ \ \ \ \ \ Springer-Verlag, Berlin, Germany, 2000.

34. Rojas, E.;El-Bennich, B.; de Melo J.; Paracha, M. Insights into the

\ \ \ \ \ \ Quark--Gluon Vertex from Lattice QCD and Meson Spectroscopy.

\ \ \ \ \ \ \textit{Few-Body Syst.} \textbf{2015} 56, 639-644.

35. Schr\"{o}dinger, E. \textit{Collected Papers on Wave Mechanics}.Chelsea

\ \ \ \ \ \ Publ.Co., New York, NY, USA, 1978.

36. V. Arnol'd, \textit{Mathematical Methods of Classical Mechanics}.

\ \ \ \ \ \ Springer-Verlag, Berlin, Germany, 1989.

37. Kholodenko, A. \textit{Applications of Contact Geometry and }

\ \ \ \ \ \ \textit{Topology in Physics}. World Scientific, Singapore, 2013.

38. Hilbert D.; Courant, R. \textit{Methods of Mathematical Physics}, Vol.2.

\ \ \ \ \ \ Interscience Publishers, New York,\ \ NY, 1962.

39. P.G\"{u}nter, P. \textit{Huygens Principle and Hyperbolic Equations}.

\ \ \ \ \ \ Academic Press Inc., Boston, MA, USA, 1988.

40. Kholodenko, A.; Kauffman, L. Huygens triviality of the time-independent

\ \ \ \ \ \ Schr\"{o}dinger equation. Applications to atomic and high energy
physics.

\ \ \ \ \ \ \textit{Ann.Phys}.\textbf{2018}, 390,1-59.

41. Arrayas,A.; Bouwmeester,D.; Trueba, J.

\ \ \ \ \ \ Knots in electromagnetism. \textit{Phys.Rep}. \textbf{2017},
667, 1-61.

42. Bohm, A.;Ne'emanY.;Barut, A. \textit{Dynamical Groups and \ Spectrum }

\ \ \ \ \ \ \textit{Generating Algebras}, Vol's 1 \& 2, World Scientific,
Singapore, 1988.

43. Itzykson,C.; Bander, M. Group theory of the hydrogen atom, I and II.%
\textit{\ }

\ \ \ \ \ \ \textit{Rev.Mod.Phys.}\textbf{1966} 38, 330-345 and 346-358.

44. Frenkel,I.; Libine, M. Quaternionic analysis, representation theory and

\ \ \ \ \ \ physics. \textit{Adv.Math}. \textbf{2008}, 218 1806-1877.

45. Jacobson, D.;Nadirashvili, N.;Toth, J.Geometric properties of

\ \ \ \ \ \ eigenfunctions. \textit{Russian Math.Surveys} \textbf{2001}, 56
1085-1105.

46. Chladni,E. \textit{Tretease on Acoustic}, Springer-Verlag, Berlin, 2015.

47. Available online:
https://www.youtube.com/watch?v=OLNFrxgMJ6E\&ab\_channel=

\ \ \ \ \ \ TheRoyalInstitution

48. Gao, Z.;Yin, L.;Fang, W.; Kong,Q.;Fan, C.; Kang, B.;Hu, J-J.

\ \ \ \ \ \ Chen, H-Y.\ Imaging Chladni figure of plasmonic charge density

\ \ \ \ \ \ wave in real space. \textit{ACS Photonics} \textbf{2019}, 6,
2685-2693.

49. Komendarczyk, R. On the contact geometry of nodal sets.

\ \ \ \ \ \ \textit{AMS Transactions} \textbf{2005}, 358, 2399-2413.

50. Rayleigh, J. \textit{Theory of Sound}. Vol.1. Macmillan and Co. Ltd,

\ \ \ \ \ \ London, UK, 1896.

51. Rossing T.; Fletcher, N. \textit{Principles of Vibrations and Sound.}

\ \ \ \ \ \ Springer-Verlag, Berlin, Germany, 2004.

52. Cheng, S-Y. Eigenfunctions and nodal sets. \textit{Comm.Math.Helvetici}

\ \ \ \ \ \ \textbf{1976}, 51, 43-55.

53. Enciso, A.; Hartley,D.;Peralta-Salas, D. Laplace operators with

\ \ \ \ \ \ eigenfunctions whose nodal set is a knot.

\ \ \ \ \ \ \textit{J.Funct.Analysis} \textbf{2016}, 271, 182-200.

54. Kholodenko, A. Optical knots and contact geometry I. From

\ \ \ \ \ \ Arnol'd inequality to Ranada's dyons. \textit{Analysis \& Math.
Phys}.

\ \ \ \ \ \ \textbf{2016,} 6, 163-198.

55. Kholodenko, A. Optical knots and contact geometry II. From

\ \ \ \ \ \ Ranada dyons to transverse and cosmetic knots. \textit{Ann.Phys}.

\ \ \ \ \ \ \textbf{2016}, 371, 77-124.

56. Kaiser, G. Helicity, polarization and Riemann--Silberstein vortices.

\ \ \ \textit{\ \ \ J.Opt. A} \textbf{2004}, 6, S243-S245.

57. Availkable online: http://hopfion.com/faddeev.html

58. Besieris, I.; Shaarawi, A. Hopf-Ran\~{a}da linked and knotted light

\ \ \ \ \ \ beam solution viewed as a null electromagnetic field.

\ \ \ \ \ \ \textit{Optics Lett}. \textbf{2009}, 34 \ 3887-3889.

59. Bouwkamp,C.; Casimir, H. On multipole expansions in the

\ \ \ \ \ \ theory of electromagnetic radiation.\textit{\ Physica} \textbf{%
1954}, 20, 539-554.

60. Kholodenko, A. Heisenberg honeycombs solve Veneziano puzzle.

\ \ \ \ \textit{\ \ Int.Math.Forum} \textbf{2009}, 4, 441-509.

61. https://www.classe.cornell.edu/\symbol{126}%
liepe/webpage/docs/P4456L19.pdf

62. Majthay, A. \textit{Foundations of Catastrophe Theory}. \ Pitman

\ \ \ \ \ \ Publishers, Boston, USA, 1985.

63. Kholodenko, A. Black magic session of concordance: Regge

\ \ \ \ \ \ mass spectrum from Casson's invariant. \textit{Int. J.Mod.Phys.}A

\ \ \ \ \ \ \textbf{2015}, 30, 1550189.

64. Trautman, A. Solutions of the Maxwell and Yang-Mills

\ \ \ \ \ \ equations associated with Hopf fibrings.

\ \ \ \ \ \ \textit{Int.J.Theor.Phys.}\ \textbf{1977}, 16, 561-565.

65. Irvine,W; Bouwmeister,D. Linked and knotted beams of light.

\ \ \ \ \ \ \textit{Nature Physics} \textbf{2008,} 4\textbf{,}716-720\textbf{%
.}

66. Enciso, A.; Peralta-Salas,D.;Torres de Lizaur, F. Helicity is

\ \ \ \ \ \ the only integral invariant of volume-preserving transformations.

\ \ \ \ \ \ \textit{PNAS} \textbf{2016}, 113, 2035-2040.

67. Nag.S. \ \textit{The Complex Analytic Theory of Teichmuller Spaces.}

\ \ \ \ \ Wiley-Interscience, New York, US, 1988.

68. Valiente-Kroon, J.\textit{\ Conformal Methods in General Relativity.}

\ \ \ \ \ \ Cambridge U. Press, Cambridge, UK, 2016.

69. Vilenkin,A.; Shellard, E.\textit{Cosmic Strings and Other Topological}

\ \ \ \ \textit{\ \ Defects. }Cambridge U. Press, Cambridge, UK, 1994.

70. Peralta -Salas,D. Selected topics on the topology of ideal fluid flows.

\ \ \ \ \ \ \textit{Int.J. of Geom.Methods in Mod.Phys}. \textbf{2016}, 13,
Sup.1. 1630012.

71. G\"{o}ckeler, M.; Shuker, T. \textit{Differential Geometry, Gauge }

\ \ \ \ \ \ \textit{Theories and Gravity}.\ Cambr.U. Press, Cambridge, UK,
1987.

72. Utiyama, R. Invariant Theoretical Interpretation of Interaction.

\ \ \ \ \ \ \textit{Phys.Rev}. \textbf{1956}, 101, 1597-1607.

73. Robinson, I. Null Electromagnetic Fields.

\ \ \ \ \ \ \textit{J.Math.Phys}. \textbf{1961}, 2, 290-291.

74. Dalhuisen, J. \textit{The Robinson Congruence in Electrodynamics }

\ \ \ \ \ \ \textit{and General Relativity.}PhD Thesis University of Leiden,

\ \ \ \ \ \ Netherlands, 2014.

75. Bluestone, S. The Planck radiation law: Exercises using the

\ \ \ \ \ \ cosmic background radiation data.\textit{\ J.Chem.Ed\ }\textbf{%
2001}, 78, 215-218.

76. Hu,W. Mapping the dark matter through the cosmic microwave

\ \ \ \ \ \ background damping tail.

\ \ \ \ \ \ \textit{Astrophysical Journal} \textbf{2001}, 557, L79-L83.

77. Available online: https://en.wikipedia.org/wiki/Axion

78. Birkinshaw,M.The Sunaev-Zeldovich effect. \textit{Phys.Reports}

\ \ \ \ \ \ \textbf{1999}, 310, 97-195.

79. Ferreira,E. Ultra-light dark matter.

\ \ \ \ \ \ \textit{Astron.Astrophys}.Rev. \textbf{2021}, 29, 1-167.

80. Khoury J. Dark matter superfluidity.\textit{\ SciPost Phys. Lect.Notes}

\ \ \ \ \textbf{\ \ 2022}, 42, 1-22.

81. Bode, B. Stable knots and links in electromagnetic fields.

\ \ \ \ \ \ \textit{Comm.Math.Phys}.\textbf{\ 2021}, 387, 1757-1770.

82. Jaynes,E. Probability in quantum theory in \textit{Complexity, Entropy}

\ \ \ \ \textit{\ \ and Physics of information. }Ed.W. Zurek\textit{,}

\ \ \ \ \textit{\ \ }CRC Press, London, 1990.

83. Waite,T.; Barut,A.; Zeni, J. The purely electromagnetic electron

\ \ \ \ \ \ re-visited in \textit{Electron Theory and QED. }Ed. J.Dowling%
\textit{. }

\ \ \ \ \ \ Plenum Press, New York, 1966.

84. Rodrigues,W.; Vas, J.From electromagnetism to relativistic

\ \ \ \ \ \ quantum mechanics. Found.Phys.\textbf{1998}, 28, 789-914.

85. Sallhofer, H. Hydrogen in Electrodynamics.I. Preliminary

\ \ \ \ \ \ theories. \textit{Z.Naturforsch} \textbf{\ 1988}, 43a, 139-1043.

86. Sallhofer,H. Elementary derivation of the Dirac Equation. X

\ \ \ \ \ \ Z. \textit{Z.Naturforsch} \textbf{\ 1986}, 41a, 468-470.

87. van Dongen, J. \textit{The Vortex Theory of Atoms. }Ms.Sci.Thesis.

\ \ \ \ \ \ Utrecht University, Netherlands, 2012.

88. Simulik,V. Solutions of the Maxwell equations describing the

\ \ \ \ \ \ spectrum of hydrogen. \textit{Ukrainian Math.Journal }

\ \ \ \ \ \ \textbf{1997}, 49, 1075-1088.

89. Oppenheimer, J. Note on the light quanta and the

\ \ \ \ \ \ electromagnetic field. \textit{Phys.Rev}.\textbf{\ 1931},38,
725-746.

90. Laporte,O.;Uhlenbeck,G. Application of spinor analysis to

\ \ \ \ \ \ the Maxwell and Dirac equations.

\ \ \ \ \ \ \textit{Phys.Rev}.\textbf{1931}, 37, 1380-1397.

91. Visinelli, L. Axion-Electromagnetic waves.

\ \ \ \ \ \ \ \textit{Mod.Phys.Lett.A} \textbf{2013}, 28,1350162.

92. Axion. https://en.wikipedia.org/wiki/Axion

93. Asker, A. Axion electrodynamics and measurable effects in

\ \ \ \ \ \ topological insulators. MS Thesis, Karlstadt University, 2018.

94. Murasugi, K. \textit{Knot Theory and its Applications}. Birkh\"{a}user,

\ \ \ \ \ \ Boston, US, 1996.

95. Chubykalo,A.; Espinosa, A. Self-dual electromagnetic fields.

\ \ \ \ \ \ \textit{Am.J.Phys}. \textbf{2010}, 78, 858-861.

96. Sakharov, A. Topological structure of elementary charges

\ \ \ \ \ \ and CPT symmetry. In \textit{Problems of Theoretical Physics},

\ \ \ \ \ \ pp 242-247. Nauka Publishing, Moscow, Russia, 1972.

97. Kobe, D. Derivation of Maxwell's equations from the local

\ \ \ \ \ \ gauge invariance of quantum mechanics.

\ \ \ \ \textit{\ \ Am.J. Phys.}\textbf{1978}, 46,342-348.

98. Pierce,A. Derivation of Maxwell's equations via the

\ \ \ \ \ \ covariance requirements of the special theory of relativity,

\ \ \ \ \ \ starting with Newton's laws. \textit{J.Appl.Math.and Physics}

\ \ \ \ \ \ \textbf{2019},7,2052-2073.

99. Ramos, J.;Gilmore,R. Derivation of source-free Maxwell and

\ \ \ \ \ \ gravitational radiation equations by group theoretical methods.

\ \textit{\ \ \ \ \ Int.J.Mod.Phys.D} \textbf{2006}, 15, 505-519.

100.Khosravi F. \textit{Unified Spin Electrodynamics of Dirac and }

\ \ \ \ \ \ \ \textit{Maxwell Fields.}PhD thesis. U.of Alberta, Canada, 2020.

101.Simulik, V. Some algebraic properties of Maxwell-Dirac isomorphism.

\ \ \ \ \ \ \ \ \textit{Z.Naturforsch} \textbf{1994}, 49a, 1074-1076.

102.Giunti,C.; Kim,Ch. \textit{Fundamentals of Neutrino Physics and}

\ \ \ \ \ \ \ \ \textit{Astrophysics. }Oxford U.Press,New York, US, 2007.

103. Simulik, V. Connection between the symmetry of the Dirac and

\ \ \ \ \ \ \ \ Maxwell equations. Conservation laws. \textit{%
Theor.\&Math.Physics}

\ \ \ \ \ \ \ \ \textbf{1991},87, 76-85.

104. Bliokh,K.; Bekshaev, A.; Nori, F. Dual electromagnetism: helicity,

\ \ \ \ \ \ \ \ spin , momentum and angular momentum. \textit{New Journal of
Physics}

\ \ \ \ \ \ \ \ \textbf{2013} 15, 033026.

105. Mun,J.;Kim,M.; Yang,Y.;Badloe,T.;Ni,J.; Chen,Y.;Qui,Ch.;Rho,I.

\ \ \ \ \ \ \ \ Electromagnetic chirality: from fundamentals to
nontraditional

\ \ \ \ \ \ \ \ chirooptical phenomena. \textit{Light:Science and
Applications \textbf{2020},}

\ \ \ \ \ \ \ \ 9, 139.

106. Cho,K.\textit{Reconstruction of Macroscopic Maxwell Equations.}

\ \ \ \ \ \ \ \ Springer-Verlag, Berlin, Germany, 2018.

107. Sadykov,S. Maxwell's equations in the Majorana representation

\ \ \ \ \ \ \ \ in a locally transparent isotropic chiral medium. \textit{%
Optics and}

\ \ \ \ \ \ \ \ \textit{Spectroscopy} \textbf{2004}, 97,305-307.

108. Bose,S.;Parker, R. Zero mass representation of Poincar$e^{\prime }$
group

\ \ \ \ \ \ \ \ and conformal invariance. \textit{J.Math.Phys}. \textbf{1969}%
, 10, 812-813.

109. Gross, L. Norm invariance of mass-zero equations under the

\ \ \ \ \ \ \ \ conformal group. \textit{J.Math.Phys.}\textbf{\ 1964}, 5,
687-695

110. Bargmann,V.;Wigner,E. Group theoretical discussion of relativistic

\ \ \ \ \ \ \ \ wave equations.\textit{\ PNAS} \textbf{1948}, 34, 211-223.

111. Cecil,Th. \textit{Notes on Lie sphere geometry and\ cyclides of Dupin}.

\ \ \ \ \ \ \ \ Lecture notes.2020.Department of Mathematics and Computer
Science,

\ \ \ \ \ \ \ \ College of Holly Cross, Worcester, MA, US.

112. Jensen,G.; Musso,E.; Nicolodi,L.\textit{Surfaces in Classical
Geometries.}

\ \ \ \ \ \ \ \ Springer International Publishing, Switzerland, 2016.

113. Ward,R. Progressing waves in flat spacetime and in plane-wave

\ \ \ \ \ \ \ \ \ spacetimes. \textit{Class.Quantum Grav}.\textbf{1987}, 4,
775-778.

114. Schrott,M.; Odehnal,B. Ortho-Circles of Dupin cyclides.

\ \ \ \ \ \ \ \ \textit{J.of Geometry and Graphics }\textbf{2006}, 10, 73-98.

115. Buquard,O.\textit{\ AdS/CFT Correspondence: Einstein Metrics \ and}

\textit{\ \ \ \ \ \ \ \ Their Conformal Boundaries}. European Math.Society,
ETH-Zentrum,

\ \ \ \ \ \ \ \ CH-8092, Zurich, Switzerland.

116.Danciger,J.\textit{Geometric Transitions from Hyperbolic to AdS Geometry}%
.

\ \ \ \ \ \ \ PhD, Department of Mathematics, Stanford University,

\ \ \ \ \ \ \ Stanford, USA, 2011.

117.Simulik,V. \textit{What is the Electron}?Aperion, Montreal, Canada, 2005.

118.Rohrlich, F. \textit{Classical Charged Particles}. World Scientific,

\ \ \ \ \ \ \ Singapore, 2007.

119.Dowling, J. \textit{Electron Theory and Quantum Electrodynamics. 100}

\ \ \ \ \ \ \textit{\ Years Later}. Springer Science+Business Media Inc.,

\ \ \ \ \ \ \ New York, US, 1997.

120.Yaghjian, A. \textit{Relativistic Dynamics of a Charged Sphere. }Springer

\ \ \ \ \ \ \ Science+Business Media Inc., New York, US, 2022.

121.Kholodenko,A. Fermi-Bose transmutation: from semiflexible

\ \ \ \ \ \ \ polymers to superstrings. \textit{Ann.Phys}.\textbf{1990},
202, 186-225.

122.Kholodenko,A. Potts model, Dirac propagator, and conformational

\ \ \ \ \ \ \ statistics of semiflexible polymers.\textit{J.Stat.Phys.}%
\textbf{1991}, 65, 291-316.

123.Kholodenko,A. Analytical calculation of the scattering function

\ \ \ \ \ \ \ for polymers of arbitrary flexibility using the Dirac
propagator.

\ \ \ \ \ \ \ Macromolecules, \textbf{1993}, 26, 4179-4183.

124.Kholodenko,A.; Ballauff, M.; Granados, A., Conformational

\ \ \ \ \ \ \ statistics of semiflexible polymers: comparison between

\ \ \ \ \ \ \ different models. \textit{Physica A }\textbf{1998,} 260,
267-293.

125. Gaveau,B.: Jacobson,T.; Kac, M.; S. Schulman, L.Relativistic

\ \ \ \ \ \ \ Extension of the Analogy between Quantum Mechanics

\ \ \ \ \ \ \ and Brownian Motion. \textit{Phys}. \textit{Rev.Lett}. \textbf{%
1984,}53\textbf{,} 419-422.

126.Feynman,R.; Hibbs, A. \ \ \textit{Quantum Mechanics and Path }

\ \ \ \ \ \ \ \textit{Integrals.} Dover Publications Inc., Mineola, NY, US,
2010.

127.Kholodenko,A. Statistical mechanics of semiflexible polymers:

\ \ \ \ \ \ \ yesterday, today and tomorrow. \textit{Faraday Transactions}

\ \ \ \ \ \ \ \textbf{1995}, 91, 2473-2482.

128.Yamakawa,H.;Yoshizaki,T. \textit{Helical Wormlike Chains in}

\ \ \ \ \ \ \ \textit{\ Polymer Solutions. }Springer-Verlag, Berlin,

\ \ \ \ \ \ \ \ Germany, 2016.

129. Kohlbrecher, J. SASfit: \textit{A program for fitting simple }

\ \ \ \ \ \ \ \ \textit{structural models to small angle scattering data.}
Paul

\ \ \ \ \ \ \ \ Scherrer Institute, Laboratory for neutron scattering

\ \ \ \ \ \ \ \ and imaging. Villigen, Switzerland, 2023.

130. Polyakov, A. \textit{Gauge Fields and Strings.}Harwood Academic

\ \ \ \ \ \ \ \ Publishers\textit{, }New York, US, 1987.

131. Cohen, A.;Moore, G.: Nelson, P. An off shell propagator

\ \ \ \ \ \ \ \ for string theory.\textit{Nucl.Phys. B} \textbf{1986}, 267,
143-157.

132. Matsutani, S.Euler's elastica and beyond. \textit{J.of Geom.Sym.}

\ \ \ \ \ \ \ \ \textit{in Physics} \textbf{2010},17, 45-86.

133. \ Available online: https://en.wikipedia.org/wiki/Worm-like\_chain

134. Koyama,R. Light scattering of stiff chain polymers.

\ \ \ \ \ \ \ \ \textit{J.Phys.Soc.Japan }\textbf{1973}, 34, 1029-1038.

135. P\"{o}tschke, D.; Hickl,P.; Ballauff, M.; Astrand,P-O; Pedersen,

\ \ \ \ \ \ \ \ J. Analysis of \ the conformation of worm-like chains by

\ \ \ \ \ \ \ \ small -angle scattering: Monte -Carlo simulations in

\ \ \ \ \ \ \ \ comparison to analytical theory.

\ \ \ \ \ \ \ \ \textit{Macromol.Theory Simul}. \textbf{2000}, 9, 345-353.

136. Gunn, J.; Warner,M. Giant dielectric response and hairpins in

\ \ \ \ \ \ \ \ polymer nematics. \textit{Phys.Rev.Lett}\textbf{.1987}, 58,
393-396.

137. Biswas, D.;Ghosh, S. Quantum mechanics of particle on a torus

\ \ \ \ \ \ \ \ knot: curvature and torsion effects.\textit{EPL}\textbf{\
2020,}132\textbf{, }10004.

138. O'Reilly,O. \textit{Modeling Nonlinear Problems in the Mechanics }

\ \ \ \ \ \ \ \ \textit{of Strings and Rods}. Springer International
Publishing,

\ \ \ \ \ \ \ \ AG. Cham. Switzerland, 2017.

139. Ivey, Th.; Singer, D. Knot types, homotopies and stability of

\ \ \ \ \ \ \ \ closed elastic rods. \textit{Proc. London Math. Soc}. 
\textbf{1999}, 79, 429--450.

140. Martin, J. Generalized classical dynamics and the classical

\ \ \ \ \ \ \ \ analogue of a Fermi oscillator.

\ \ \ \ \ \ \ \ \textit{Proc.Roy.Soc.London} A \textbf{1959}, 251,536-542.

141. Berezin, F.: Marinov,M. Particle spin dynamics as the

\ \ \ \ \ \ \ \ Grassmann variant of classical mechanics.

\ \ \ \ \ \ \ \ \textit{Ann.Phys}. \textbf{1977}, 104, 336-362.

142. Rashevskii, P. \textit{Riemannian Geometry and Tensor Analysis. }

\ \ \ \ \ \ \ \ "Nauka\textquotedblright , Moscow, USSR, 1967.

143. Forminga, J.; Romero,C. On differential geometry of time-like

\ \ \ \ \ \ \ \ curves in Minkowski spacetime. \textit{Am.J.Phys.} \textbf{%
2006}, 74,1012-1016.

144. Gamma matrices, https://en.wikipedia.org/wiki/Gamma\_matrices

145. Kaluza--Klein theory. Available online:

\ \ \ \ \ \ \ \ https://en.wikipedia.org/wiki/Kaluza\%E2\%80\%93Klein\_theory

146. Ravndal, F. Oskar Klein and the fifth dimension. arXiv:1309.4113

147. Veblen,O. \textit{Projective Relativity Theory. }Published by Julius
Springer,

\ \ \ \ \ \ \ \ Berlin, Germany, 1933.

148. Lessner,G. Unified field theory on the basis of the projective

\ \ \ \ \ \ \ \ theory of relativity. \textit{Phys.Rev.D} \textbf{1982}, 25,
3202-3217.

149. Fauser, B. Projective relativity: Present status and Outlook.

\ \ \ \ \ \ \ \ \textit{General Relativity and Gravitation}, \textbf{2001},
33,875-887.

150. Dirac, P. Homogenous variables in classical dynamics.

\ \ \ \ \ \ \ \ Proc.Camb.Phil.Soc. \textbf{1933,} 29, 389-400.

151. Dirac.P. \textit{Lectures on Quantum Mechanics}.

\ \ \ \ \ \ \ \ Yeshiva University, New York, US, 1964.

152. Eisenhart,L. Contact transformations.

\ \ \ \ \ \ \ \ \textit{Ann.Math}.\textbf{1928-1929}, 30, 211-249.

153.Varlamov, V. Spinor structure and internal symmetries.

\ \ \ \ \ \ \ \ \textit{Int.J.Theor.Phys.} \textbf{2015}, 54,3533-3577.

154. Varlamov,V. Spinor structure and mass spectrum.

\ \ \ \ \ \ \ \ \textit{Int.J.Theor.Phys.}\textbf{2016},55,5008-5045.

155. Varlamov,V. Lorentz group and mass spectrum of

\ \ \ \ \ \ \ \ elementary particles. arXiv:1705.02227

156. Barut,A. Unification based on electromagnetism.

\ \ \ \ \ \ \ \ \textit{Annalen der Physik} \textbf{1986},43, 83-92.

157. Witten, E. Quantum field theory and Jones polynomial.

\ \ \ \ \ \ \ \ \textit{Comm.Math.Phys.}\textbf{1989}, 121, 351-359.

158. Frankel,Th.\textit{The Geometry of Physics.An Introduction}.

\ \ \ \ \ \ \ Cambridge U.Press, Cambridge, UK, 2011.

159. Rainich,G. Electrodynamics in the general relativity theory.

\ \ \ \ \ \ \ \ \textit{AMS\ Transactions} \textbf{1925}, 27, 106-136.

160. Misner,Ch.: Wheeler, J. Classical physics as geometry.

\ \ \ \ \ \ \ \ \textit{Ann.Phys}. \textbf{1957}, 2, 525-603.

161. Geometrodynamics.Available online:

\ \ \ \ \ \ \ \ https://en.wikipedia.org/wiki/Geometrodynamics

162. Kedia,H.;Peralta-Salas,D.; Irvine,W. When knots in light stay

\ \ \ \ \ \ \ \ knotted? \textit{J.Phys.A} \textbf{2018}, 51,025204.

163. Bode,B.; Peralta-Salas,D. The topology of stable

\ \ \ \ \ \ \ \ electromagnetic structures and legendrian fields on

\ \ \ \ \ \ \ \ the 3-sphere. arXiv: 2302.01043

164. Geroch,R.Electromagnetism as aspect of geometry. Already

\ \ \ \ \ \ \ \ unified field theory-the null field case.

\ \ \ \ \ \ \ \ \textit{Ann.Phys.} \textbf{1966}, 36,147-187.

165. Torre,C. The spacetime geometry of a null electromagnetic field.

\ \ \ \ \ \ \ \ \textit{Class.Quatum Gravity}\textbf{\ 2014}, 31, 045022.

166. Smolka,Th.; Jiezirski, J. Simple description of generalized

\ \ \ \ \ \ \ \ electromagnetic and gravitational hopfions.

\ \ \ \ \ \ \ \ \textit{Class.Quantum Grav}.\textbf{2018}, 35, 245019.

167.Tompson,A.;Wikes,A.; Swearngin,J.; Bouwmeister,D.

\ \ \ \ \ \ \ \ Classification of electromagnetic and gravitational hopfions

\ \ \ \ \ \ \ \ by algebraic type.\textit{\ J.Phys.A} \textbf{2015}, 48,
205202.

168. Skyrme,T. A unified field theory of mesons and baryons.

\ \ \ \ \ \ \ \ \textit{Nucl.Phys}.\textbf{1962}, 31, 556-569.

169. Burinskii,A. The Dirac-Kerr-Newman electron. \textit{Gravitation}

\ \ \ \ \ \ \ \ \textit{and Cosmology} \textbf{2008},14, 109-122.

170. Dalhuisen, J.; Bouwmeester,D. Twistors and electromagnetic knots

\ \ \ \ \ \ \ \textit{\ J.Phys.A }\textbf{2012}, 45, 135201.

171.Arcaute,E.; Lasenby, A.; Doran,Ch. Twistors in geometric algebra.

\ \ \ \ \ \ \ \textit{Adv.Appl.Clifford Alg}. \ \textbf{2008}, 18,373-394.

172. Easter,R.; Hitzer, E. Double conformal geometric algebra for

\ \ \ \ \ \ \ \ quadrics and Darboux cyclides. CGI '16: Proceedings of the

\ \ \ \ \ \ \ \ 33rd Computer Graphics International. June \textbf{2016}
Pages 93--96

\ \ \ \ \ \ \ \ https://doi.org/10.1145/2949035.2949059.

173. Inomata,A.; McKinley,W. Geometric theory of neutrinos.

\ \ \ \ \ \ \ \ \textit{Phys.Rev.}\textbf{1965}, 140, B1467-B1473.

174. Collinson,C.; Shaw R. The Rainich condition for neutrino fields.

\ \ \ \ \ \ \ \textit{Int.J.ofTheor.Phys}.\textbf{1972}, 6, 347-357.

175. Neutrino theory of light.Available online:

\ \ \ \ \ \ \ \ https://en.wikipedia.org/wiki/Neutrino\_theory\_of\_light

176. Dennis, M., Tijssen,T.; Morgan, M. On the Majorana representation

\ \ \ \ \ \ \ \ of the optical Dirac equation. \textit{J.Phys.A} \textbf{2023%
}, 56, 024004

177. Greco A. A note on the path integral representation for Majorana

\ \ \ \ \ \ \ \ fermions.\textit{\ J.Phys.A} \textbf{2016}, 49, 155004.

178. Karch, A.;Tong,D.; Turner,C. A web of 2d dualities: Z$_{2}$ gauge

\ \ \ \ \ \ \ \ fields and Arf invariant. \textit{SciPost Phys. }\textbf{2019%
}, 7, 007.

179. Kauffman,L. The Arf invariant of classical knots.

\ \ \ \ \ \ \ \ \textit{Cont.Math.} \textbf{1985}, 44, 101-116.

180. Shankar,R.\textit{\ Quantum Field Theory \ and Condensed Matter}.

\ \ \ \ \ \ \ \ \textit{An introduction}. Cambridge U.Press, Cambridge, UK,
2017.

181. Mussardo,G. \textit{Statistical Field theory: An Introduction to }

\ \ \ \ \ \ \ \ \textit{Exactly Solved Models in Statistical Physics. }%
Oxford U.Press,

\ \ \ \ \ \ \ \ Oxford, UK, 2020.

\ \ \ 

\textit{\bigskip }

\bigskip

\bigskip

\bigskip

\bigskip

\bigskip

\bigskip

\bigskip

\bigskip

\bigskip

\bigskip

\bigskip

\bigskip

\bigskip

\bigskip

\bigskip

\bigskip

\bigskip

\bigskip

\end{document}